\documentclass[letterpaper,11pt]{article}
\pdfoutput=1

\usepackage{jheppub}
\usepackage{multirow}
\usepackage{diagbox}

\usepackage{subcaption}
\usepackage[countmax]{subfloat}
\usepackage{amssymb}
\usepackage{amsmath}
\usepackage{color}
\usepackage{graphicx}
\usepackage{verbatim}
\usepackage{amsthm}
\usepackage{slashed}
\usepackage{hyperref}
\usepackage{makecell}

\usepackage{side}

\preprint{}

\title{
Machine learning of log-likelihood functions in global analysis of parton distributions
}

\author{DianYu Liu$^1$, ChuanLe Sun$^1$, Jun Gao$^{1,2,3}$}

\affiliation{$^1$INPAC, Shanghai Key Laboratory for Particle Physics and Cosmology, School of
Physics and Astronomy, Shanghai Jiao-Tong University, Shanghai 200240, China}
\affiliation{$^2$Key Laboratory for Particle Astrophysics and Cosmology (MOE), Shanghai 200240, China}
\affiliation{$^3$Center for High Energy Physics, Peking University, Beijing 100871, China}

\emailAdd{dianyu.liu@sjtu.edu.cn}
\emailAdd{chlsun60@sjtu.edu.cn}
\emailAdd{jung49@sjtu.edu.cn}

\abstract{
Modern analysis on parton distribution functions (PDFs) requires calculations of the
log-likelihood functions from thousands of experimental data points, and
scans of multi-dimensional parameter space with tens of degrees of freedom.
In conventional analysis the Hessian approximation has been widely used for the
estimation of the PDF uncertainties.
The Lagrange Multiplier (LM) scan while being a more faithful method is less
used due to computational limitations, and is the main focus of this study.
We propose to use Neural Networks (NNs) and machine learning techniques to
model the profile of the log-likelihood functions or cross sections for multi-dimensional parameter
space in order to overcome those limitations which work beyond the quadratic approximations and meanwhile ensures efficient scans
of the full parameter space.
We demonstrate the efficiency of the new approach in the framework of the CT18 global
analysis of PDFs by constructing NNs for various target functions, and performing
LM scans on PDFs and cross sections at hadron colliders.
We further study the impact of the NOMAD dimuon data on constraining PDFs with the new approach,
and find enhanced strange-quark distributions and reduced PDF uncertainties.
Moreover, we show how the approach can be used to constrain new physics
beyond the Standard Model (BSM) by a joint fit of both PDFs and Wilson coefficients of
operators in the SM effective field theory.
}
  

\begin{document}

\maketitle
\section{Introduction}

Precise understanding of the parton structure of the proton is a central topic of QCD~\cite{Gao:2017yyd,Kovarik:2019xvh}.
The parton structure can be described by parton distribution functions (PDFs), which
represent distributions of momentum fractions of the proton carried by quarks and gluons, for instance
in the case of QCD collinear factorization~\cite{Collins:1989gx}.
They are usually determined by fitting to a variety of experimental data, such as data from proton-proton collision, proton-antiproton collision, electron-proton collision, and neutrino–nucleus scatterings.
Besides, there have also been recent developments on calculating PDFs from first principles based on the large momentum effective theory~\cite{Ji:2020ect} and lattice QCD simulations~\cite{LatticeParton:2018gjr}.

Especially, PDFs play important roles in LHC studies. 
For example, PDF uncertainties represent one of the dominant uncertainties in measurements of 
the Higgs boson couplings~\cite{LHCHiggsCrossSectionWorkingGroup:2016ypw}.
Better control of PDF uncertainties are necessary in direct searches for new heavy resonances~\cite{Beenakker:2015rna} and indirect searches for new physics beyond the SM~\cite{Alioli:2017jdo}. 
Furthermore, PDF uncertainties also have a large impact on precision measurements of the SM parameters 
including the strong coupling constant~\cite{ATLAS:2020mee}, the weak mixing angle 
and the W boson mass~\cite{Kaur:2019ndj,ATLAS:2017rzl}.

Modern analysis of PDFs requires calculations of the
log-likelihood functions from thousands of experimental data points, and
scans of multi-dimensional parameter space with tens of degrees of freedom.
There are several groups providing regular updates of PDFs via global fits, 
see Refs.~\cite{Hou:2019efy,Bailey:2020ooq,Ball:2021leu,Alekhin:2018pai,H1:2017bml,Jimenez-Delgado:2014twa,Park:2021kgf,ATLAS:2021vod} 
for recent results on PDF determinations.
The difference between those PDF sets is mainly due to the choice of the  experimental data sets, 
the theoretical calculations used, and the parametrization form of PDFs.
PDF uncertainties can be determined with three methods: 
the Hessian~\cite{Pumplin:2001ct,Martin:2009iq}, Monte Carlo (MC)~\cite{Forte:2002fg}, 
and Lagrange Multiplier (LM)~\cite{Pumplin:2000vx,Stump:2001gu} method.
There also exist recently developed approaches, meta analysis~\cite{Gao:2013bia}, ePump~\cite{Schmidt:2018hvu} and L2 sensitivity~\cite{Wang:2018heo},
on accessing impacts of experimental data on PDFs based on the Hessian method. 
In the Hessian method, the log-likelihood function ($\chi^2$) of a global fit is
approximated with a quadratic form of the PDF parameters at the neighborhood of the global
minimum.
The uncertainties are thus determined through error PDFs along eigenvector
directions, constructed by requiring the increase of the total $\chi^2$ of 1 or of a certain tolerance.
In the MC method, one can obtain the PDF uncertainties from an ensemble of PDF replicas
which are fitted to an ensemble of ``pseudo-data".
Those pseudo-data are generated from the probability distributions related to
the original experimental data sets.
On another hand, for the LM method, PDF uncertainties of an observable can
be determined from the profiled $\chi^2$ as a function of the observable,
without relying on any assumptions about the behavior of the $\chi^2$ at the neighborhood
of the global minimum.
This means PDF uncertainties estimated from the LM method are more robust than those from
the Hessian method.
However, the LM method requires a detailed scan of the PDF parameter space
for every observable studied, which is usually time consuming.

This drawback can be overcome with the help of machine learning (ML).
ML has been widely used in studies of high-energy physics
in recent years.
In many cases, ML is used for classifications such as particle identification 
and event selection in experimental data analysis~\cite{Guest:2018yhq}.
Neural networks (NNs) are also helpful in
regression problems, for example, applications of NNs in the study of PDFs have been
pioneered by the NNPDF collaboration~\cite{NNPDF:2014otw}.
Dependence of PDFs on the momentum fraction are parametrized using NNs, which ensures
a great flexibility~\cite{Forte:2020yip}. 
On another hand, dependence of the $\chi^2$ or any physics quantity, such as the cross section, on PDFs is
complex in general.
NNs offer an opportunity to relate physics quantities to PDFs efficiently. 
One can build NNs with PDFs as input variables to model their PDF dependence.
Compared with traditional methods, NNs can greatly improve efficiencies on generating
predictions for those physics quantities.

With above motivations, in this paper we propose a new approach with which PDF uncertainties can be calculated
efficiently using the LM method with the assistance of NNs.
It takes three steps to achieve this goal. 
First, we construct and train NNs to model the $\chi^2$ of each
individual data set used in the global fit with PDFs.
Second, we construct and train other NNs to associate the physics quantity
to be studied with PDFs.
Finally, we can perform LM scans to determine PDF uncertainties in a
robust way.
The speed of LM scans can be improved by several orders of magnitude due to the
introduction of the NNs.
We demonstrate above idea in the framework of CT18 NNLO global analysis~\cite{Hou:2019efy}
and beyond.
We show how the new approach can help to understand various PDF uncertainties
and the interplay between different data sets in the global fit.
Moreover, we explore several directions beyond CT18 as will be explained below.

Only a few data sets in the CT18 global fit are sensitive to the strange-quark PDFs. 
The dimuon production in neutrino scatterings provides an opportunity to directly constrain strange-quark
distributions in the nucleon.
In recent NOMAD measurements~\cite{NOMAD:2013hbk}, a sample of about $9\times10^6$ events of inclusive charged-current deep-inelastic
scattering (CCDIS), together with about 15344 events of dimuon production, is collected.
The large statistics lead to a better control on various systematic errors and also
an improvement in statistical uncertainties.
We include the NOMAD data in the global fit and evaluate the impact on the PDFs
using the aforementioned approach.

The High Luminosity LHC (HL-LHC) is supposed to accumulate
an integrated luminosity of 3000 fb$^{-1}$ for ATLAS and CMS and of 300 fb$^{-1}$ for
LHCb~\cite{Dainese:2019rgk}.
We take two of those HL-LHC pseudo-data sets constructed in Ref.~\cite{AbdulKhalek:2018rok,AbdulKhalek:2019mps},
the high-mass Drell-Yan data and the forward W/Z production data, and evaluate their impacts on PDFs.
Our projection shows they can largely improve separations of different flavors,
especially for sea quarks.
In the searches for new physics beyond the SM from scatterings involving nucleons, for
instance at HERA or LHC, one key problem is on the degeneracy of PDF variations
and the new physics contributions, especially in cases when similar measurements
are used in both the global fit of PDFs and in the searches of new physics.
Ideally a joint global fit including both PDFs and model parameters of the new physics should be performed, see Refs.~\cite{Carrazza:2019sec,Greljo:2021kvv,Madigan:2021uho,CMS:2021yzl,Iranipour:2022iak} for
examples.
We demonstrate successful application of our approach in such scenario by
a simultaneous fit of both PDFs and the Wilson coefficient of lepton-quark
contact interactions in the SM effective field theory (SMEFT).

The rest of this paper is organized as follows.
In Section~\ref{sec:nn}, we describe the basic setup of our approach, including architectures
of the NNs, PDF parametrizations and experimental data sets considered in the global fit.
In Section~\ref{sec:validation}, we discuss performances of the approach and show that the
accuracy of approximations with NNs are far sufficient for phenomenological studies.
In Section~\ref{sec:LM}, we explain the method of LM scans and discuss several features
of the CT18 analysis based on the new approach.
In Section~\ref{sec:application}, we study the impact of the NOMAD measurements and
of the two pseudo-data of HL-LHC on PDFs, and show a joint fit with
both PDFs and new physics contributions.
Finally, we conclude in Section~\ref{sec:conc}.

\section{Setup of the Neural Network program}\label{sec:nn}
In this section, we give a brief introduction to the setup of our NNs, including the
architectures, the input variables, and the target functions.
We further explain the training processes from the generation of samples to the minimization of the loss function.

\subsection{Basic setup of NNs}
\begin{figure}[htbp]
  \centering
  \includegraphics[width=1\textwidth,clip]{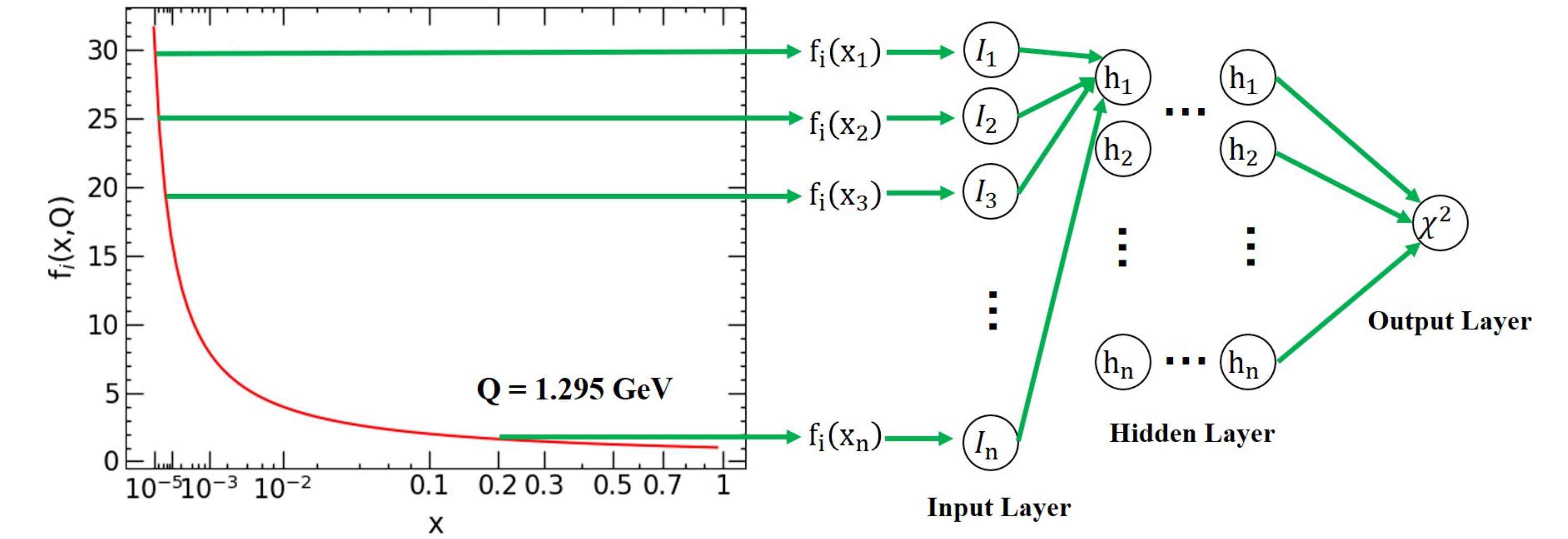}
  \hfill
  \caption{An example of the architecture of NNs in this work, taking $\chi^2$ as the
  target function.}
  \label{Fig:NN_archi}
\end{figure}

The general structure of NNs includes three parts: the input layer, several hidden layers and the output layer.
Each of these layers contains a collection of nodes termed by perceptrons.
There exist various implementations of NNs, and we use Keras~\cite{keras} in this work.
From the NNs built by Keras, PDFs as inputs are associated with either $\chi^2$ or physics quantities as
outputs. 
The log-likelihood function $\chi^2$ quantifies agreements between theory predictions and experimental measurements
for each data set and is calculated according to~\cite{Hou:2019efy}.
The physics quantities considered include cross sections of several benchmark processes at the LHC,
and PDFs or their ratios at different $Q$ values.
An example of the architecture of our NNs is shown in Fig.~\ref{Fig:NN_archi}, in which the inputs
are PDFs at an initial scale and the outputs are the $\chi^2$ of the fit to an experimental data set.
In this figure, the PDFs $f_{i}(x, Q)$ are evaluated at an initial scale of $Q = 1.295$ GeV with $x$ selected among 14
different values, and $i \in \{g, u, d, \bar{u}, \bar{d}, s\}$ runs over all parton flavors. 
We always assume $s=\bar s$ at the initial scale. 
They altogether form the input layer with 84 nodes \{$I_1$, $I_2$ $\ldots$ $I_{84}$\}.
In addition, the differences in setups between NNs for different target functions are
shown in Table~\ref{tab:architecture}.
The choice on the architecture is based on the observation that cross sections or evolved PDFs are
in general non-linear functions of the PDF parameters.
The $\chi^2$ is positive defined and is a sum of various individual terms that depend on
cross sections quadratically, and thus can be approximated by a more complicated architecture as prescribed.  
We include more details on the construction of our NNs in Appendix~\ref{sec:para}.

\begin{table}[htpb]
  \centering
  \resizebox{\textwidth}{16.5mm}{
  \begin{tabular}{c|cccc}
  \hline
  Target & \makecell*[c]{No. of\\hidden layers} & \makecell*[c]{No. of nodes for\\each hidden layer} & \makecell*[c]{Activation functions\\for each layer} & \makecell*[c]{No. of\\total params}\\
  \hline
  $\chi^2$ & 2 & 60,40 & $\tanh$, $(x^{2}+2)$, linear & 7581\\
  \hline
  \makecell*[c]{$\sigma$, $f_{i}(x,Q)$,\\ $f_{i}(x,Q)$/$f_{j}(x,Q)$} & 1 & 40 & $\tanh$, linear & 3441\\
  \hline
  \end{tabular}
  }
  \caption{
    The architecture of NNs in this paper. 
    Structure is set up for either $\chi^2$ or other quantities.}
  \label{tab:architecture}
\end{table}

To construct the $l_{th}$ layer of a NN, we define
\begin{equation}
b_{i}^{(l)}= \left\{\begin{array}{ll}
\sum_{j} w_{i j}^{(l)} I_{j}, & (l = 1) ,\\
\\
\sum_{j} w_{i j}^{(l)} h_{j}^{(l-1)}, & (l > 1) ,
\end{array}\right.
\end{equation}
where $b_{i}^{(l)}$ is the value before the activation of the $i_{th}$ node in the $l_{th}$ layer, $w_{i j}^{(l)}$ is the weight matrix connecting the $(l-1)_{th}$ layer to the $l_{th}$ layer, $I_{j}$ is the value of the $j_{th}$ node in the input layer, and $h_{j}^{(l-1)}$ is the value of the $j_{th}$ node in the $(l-1)_{th}$ layer.
The value of the $i_{th}$ node in the $l_{th}$ layer is then obtained by applying the activation function $t^{(l)}$ on $b_i^{(l)}$:
\begin{equation}
h_{i}^{(l)}= t^{(l)}(b_{i}^{(l)}).
\end{equation}
This procedure iterates over all hidden layers, and in the end we obtain a single value for the output layer.
The activation functions used include the conventional choices of linear, and $\tanh$ types, as well
as a customized one of quadratic form, depending on the target functions and layers.
Note we constrain elements of the weight matrix of the output layer to be positive
for the NN associated with the $\chi^2$ since it is positive definite.
Elements in the weight matrix, $w^{(l)}_{i j}$, are trained to minimize the so-called loss
function, which is defined as
\begin{equation}
d_{loss} = \frac{1}{n}\sum_{k=1}^{n}\left(A^{k}_{{\rm NN}}(w_{i j})-A^{k}_{{\rm TR}}\right)^{2},
\label{loss}
\end{equation}
where $n$ is the total number of events in the training sample and $A^k_{\rm TR}$ and $A^{k}_{\rm NN}$
are the truth of the target function and the prediction from NNs for the $k_{th}$ event.

\subsection{PDF parametrization form}
The parametrization form of PDFs used at the initial scale $Q_0$ is
\begin{equation}\label{eq:para}
f_{i}\left(x, Q_{0}\right)=a_{0} x^{a_{1}-1}(1-x)^{a_{2}} P_{i}\left(y ; a_{3}, a_{4}, \ldots\right) ,
\end{equation}
where $\{a_{1}, a_{2}, \ldots\}$ are free parameters, and the behavior of $x^{a_1}$ at $x \to 0$ and $(1-x)^{a_2}$ at $x \to 1$ is guided by Regge theory and spectator counting rules respectively.
$P_{i}\left(y ; a_{3}, a_{4}, \ldots\right)$ is a polynomial dependent on $y \equiv \sqrt{x} ~(y \equiv 1-(1-\sqrt{x})^{a_3})$ for valence quark and gluon PDFs (light-quark sea PDFs).
Parametrization forms used here are the same as in the CT18 NNLO analysis~\cite{Hou:2019efy}.

For the valence-quark ($u_v$ and $d_v$) PDF,
\begin{equation}\begin{aligned}
f_{v}\left(x, Q_{0}\right) =&a_{0} x^{a_{1}-1}(1-x)^{a_{2}} P_{v}(y), \\
P_{v}(y) =&\sinh \left[a_{3}\right](1-y)^{4}+\sinh \left[a_{4}\right] 4 y(1-y)^{3}+\sinh \left[a_{5}\right] 6 y^{2}(1-y)^{2} \\
&+(1+\frac{1}{2}a_1) 4 y^{3}(1-y)+y^{4}.
\end{aligned}\label{eq_valence}\end{equation}

For the gluon PDF,
\begin{equation}\begin{aligned}
f_{g}\left(x, Q_{0}\right) =&a_{0} x^{a_{1}-1}(1-x)^{a_{2}} P_{g}(y), \\
P_{g}(y) =&\sinh \left[a_{3}\right](1-y)^{3}+\sinh \left[a_{4}\right] 3 y(1-y)^{2}+(3+2a_1) y^{2}(1-y)+y^{3}.
\end{aligned}\label{eq_gluon}\end{equation}

For the sea quark ($\bar{u}$, $\bar{d}$ and $s \equiv \bar{s}$) PDF,
\begin{equation}
\begin{aligned}
f_{\bar{q}}\left(x, Q_{0}\right) =&a_{0} x^{a_{1}-1}(1-x)^{a_{2}} P_{\bar{q}}(y), \\
P_{\bar{q}}(y) =&(1-y)^{5}+a_{4} 5 y(1-y)^{4}+a_{5} 10 y^{2}(1-y)^{3}+a_{6} 10 y^{3}(1-y)^{2} \\
&+a_{7} 5 y^{4}(1-y)+a_{8} y^{5}.
\end{aligned}
\label{eq_sea}
\end{equation}

In all, we have 8 free parameters for valence quarks after applying the valence sum rules
and letting $a_1$ be equal for $u_v$ and $d_v$.
We have 15 free parameters for sea quarks after fixing some of those $a_i$ or letting them be
equal for different flavors~\cite{Hou:2019efy}.
We are left with 5 free parameters for gluon after applying the momentum sum rule.
The total number of free PDF parameters is 28.

\subsection{Targets and samples}
\label{sec_data sets}
In this paper, we associated PDFs with $\chi^2$ and other physics quantities through our NNs.
Details of these target functions are described in the following:\\

$\bullet$~The individual $\chi^2$ of each data set in an NNLO global analysis of PDFs.
We use the same 39 experimental data sets as in CT18 NNLO global analysis.
These experimental data sets are summarized in Table~\ref{tab:data set}. 
The theoretical calculations used are explained in the CT18 paper~\cite{Hou:2019efy}.
We take those calculations from CT18 except for minor updates on NNLO K-factors of
several data sets.
The global $\chi^2$ is simply a sum of the 39 individual $\chi^2$.
$\bullet$~The cross sections of Higgs boson pair (top-quark pair with a Higgs boson) production in proton-proton collisions
at center of mass energy $\sqrt{s} = $ 13 TeV or 100 TeV. 
They are computed at leading (next-to-leading) order
in QCD using MG5\_aMC@NLO~\cite{Alwall:2014hca} 
and AMCfast~\cite{Bertone:2014zva} to provide an interface with APPLgrid~\cite{Carli:2010rw}.
We choose these two processes for demonstrations, and any scattering cross sections at hadron collisions
can be included in a similar way.

$\bullet$~The PDFs and PDF ratios at various $x$ and $Q$ values.
They are obtained using HOPPET~\cite{Salam:2008qg} with DGLAP evolutions at NNLO.\\

\begin{table}[htpb]
  \centering
  \resizebox{\textwidth}{54mm}{
  \begin{tabular}{|c|c|c||c|c|c|}
  \hline
  ID & Experimental data set & $N_{pt}$ &ID & Experimental data set & $N_{pt}$\\
  \hline
  160 & \makecell*[c]{HERA I+II 1 $fb^{-1}$, H1 and ZEUS\\NC and CC reduced cross sec. comb.}~\cite{H1:2015ubc}& 1120 &101& BCDMS $F_2^p$~\cite{BCDMS:1989qop}&337 \\
  \hline
  102 & BCDMS $F_2^d$~\cite{BCDMS:1989ggw}& 250 & 104 & NMC $F_2^d/F_2^p$~\cite{NewMuon:1996fwh}&123\\
  \hline
  108 & CDHSW $F_{2}^{p}$~\cite{Berge:1989hr} &85 & 109 & CDHSW $x_{B}F_{3}^{p}$~\cite{Berge:1989hr} & 96\\
  \hline
  110 & CCFR $F_{2}^{p}$~\cite{CCFRNuTeV:2000qwc} & 69 & 111 & CCFR $x_{B}F_{3}^{p}$~\cite{Seligman:1997mc} & 86\\
  \hline
  124 & NuTeV $\nu \mu \mu$ SIDIS~\cite{osti_879078} & 38 & 125 & NuTeV $\bar{\nu} \mu \mu$ SIDIS~\cite{osti_879078} & 33\\
  \hline
  126 & CCFR $\nu \mu \mu$ SIDIS~\cite{NuTeV:2001dfo} & 40 & 127 & CCFR $\bar{\nu} \mu \mu$ SIDIS~\cite{NuTeV:2001dfo} & 38\\
  \hline
  145 & H1 $\sigma_r^b$~\cite{H1:2004esl} & 10 & 147 & Combined HERA charm production~\cite{H1:2012xnw} & 47\\
  \hline
  169 & H1 $F_L$~\cite{H1:2010fzx} & 9 & 201 & E605 Drell-Yan process~\cite{Moreno:1990sf} & 119\\
  \hline
  203 & E866 Drell-Yan process $\sigma_{pd}/(2\sigma_{pp})$~\cite{NuSea:2001idv} & 15 & 204 & E866 Drell-Yan process $Q^{3}d^{2}\sigma_{pp}/(dQdx_{F})$~\cite{NuSea:2003qoe} & 184\\
  \hline
  225 & CDF Run-1 lepton $A_{ch}$, $p_{Tl}>$ 25 GeV~\cite{CDF:1998uzn} & 11 & 227 & CDF Run-2 electron $A_{ch}$, $p_{Tl}>$ 25 GeV~\cite{CDF:2005cgc} & 11\\
  \hline
  234 & D$\varnothing$ Run-2 muon $A_{ch}$, $p_{Tl}>$ 20 GeV~\cite{D0:2007pcy} & 9 & 260 & D$\varnothing$ Z rapidity~\cite{D0:2007djv} & 28\\
  \hline
  261 & CDF Run-2 Z rapidity~\cite{CDF:2010vek} & 29 & 266 & CMS 7 TeV 4.7 fb$^{-1}$, moun $A_{ch}$, $p_{Tl}>$ 35 GeV~\cite{CMS:2013pzl} & 11\\
  \hline
  267 & CMS 7 TeV 840 fb$^{-1}$, electron $A_{ch}$, $p_{Tl}>$ 35 GeV~\cite{CMS:2012ivw} & 11 & 268 & ATLAS 7 TeV 35 $pb^{-1}$ W/Z cross section, $A_{ch}$~\cite{ATLAS:2011qdp} & 41\\
  \hline
  281 & D$\varnothing$ Run-2 9.7 $fb^{-1}$ electron $A_{ch}$, $p_{Tl}>$ 25 GeV~\cite{D0:2014kma} & 13 & 504 & CDF Run-2 inclusive jet production~\cite{CDF:2008hmn} & 72\\
  \hline
  514 & D$\varnothing$ Run-2 inclusive jet production~\cite{D0:2008nou} & 110 & 245 & \makecell*[c]{LHCb 7 TeV 1.0 fb$^{-1}$ W/Z\\forward rapidity cross sec.}~\cite{LHCb:2015okr}& 33\\
  \hline
  246 & \makecell*[c]{LHCb 8 TeV 2.0 fb$^{-1}$ $Z\to e^- e^+$\\forward rapidity cross sec.}~\cite{LHCb:2015kwa}& 17 & 249 & \makecell*[c]{CMS 8 TeV 18.8 fb$^{-1}$ muon\\charge asymmetry $A_{ch}$}~\cite{CMS:2016qqr}& 11\\
  \hline
  250 & LHCb 8 TeV 2.0 fb$^{-1}$ W/Z cross sec.~\cite{LHCb:2015mad} & 34 & 253 & ATLAS 8 TeV 20.3 fb$^{-1}$, Z $p_T$ cross sec.~\cite{ATLAS:2015iiu} & 27\\
  \hline
   542 & \makecell*[c]{CMS 7 TeV 5 fb$^{-1}$, single incl.\\jet cross sec., R = 0.7 (extended in y)}~\cite{CMS:2014nvq} & 158 & 544 & \makecell*[c]{ATLAS 7 TeV 4.5 fb$^{-1}$, single incl.\\jet cross sec., R = 0.6}~\cite{ATLAS:2014riz}& 140\\
  \hline
   545 & \makecell*[c]{CMS 8 TeV 19.7 fb$^{-1}$, single incl.\\jet cross sec., R = 0.7, (extended in y)}~\cite{CMS:2016lna}& 185 & 573 & \makecell*[c]{CMS 8 TeV 19.7 fb$^{-1}$, $t\bar{t}$ norm.\\double-diff. top $p_T$ and y cross sec.}~\cite{CMS:2017iqf}& 16\\
  \hline
  580 & \makecell*[c]{ATLAS 8 TeV 20.3 fb$^{-1}$,\\$t\bar{t}$ $p_T^t$ and $m_{t\bar{t}}$ abs. spectrum}~\cite{ATLAS:2015lsn}& 15 & & & \\
  \hline
  \end{tabular}
 } 
  \caption{Experimental data sets involved in the global fit~\cite{Hou:2019efy}.} 
  \label{tab:data set}
\end{table}

We first generate randomly a training sample consisting of 6000 replicas
of PDFs and another test sample of 2000 replicas to prevent from over training.
Details about the generation of the replicas of PDFs can be found in Appendix~\ref{sec:para}.
We compute all the target functions ($\chi^2$ or physics quantities) for each of the replicas,
which can be time consuming depending on whether the fast interpolation approaches,
like APPLgrid or FastNLO, are used or not.
However, we only need to perform these heavy calculations once for all.
Afterwards we construct a NN for each of the target function considered
with the architectures shown in Table~\ref{tab:architecture}.
We train each NN for about 10 hours, depending slightly on the architecture,
on a single CPU-core (2.4 GHz) according to the loss function defined in Eq.~(\ref{loss}).
Thus for all $\chi^2$ of the 39 individual data sets that takes about 390 core-hours
in total for the training process.
We found a very good performance of the resulting NNs without much tuning on the
training process for all target functions studied, which will be reported in the next section.
In a later stage for the evaluation of the target functions with arbitrary PDF parameters,
we can simply use the optimized NNs rather than direct calculations. 
Comparison between computational cost of the NNs and the direct computations are
summarized in Appendix~\ref{sec:para} where substantial improvements in the
speed from the NNs are observed.

\section{Validation of NNs}\label{sec:validation}
In this section, we perform several comparisons between the truths and the predictions from
our NNs before we apply them to further phenomenological studies.
We emphasize that the entire NN approach we discussed so far and in the following is bound
to the CT18 parametrization form, especially with the CT18 PDF set.
All PDF replicas for training and testing are sampled from the CT18 PDFs.
A first attempt of generalization to other parametrization forms or even independent
of PDF parametrization shows promising results, and is detailed in Appendix A.
It should be noted that the NNs should be retrained in general if the underlying PDF
parametrization changes.

\subsection{$\chi^2$ of the global fit}
\begin{figure}[htbp]
  \centering
    \subcaptionbox{}[7.7cm] 
    {\includegraphics[width=7.7cm]{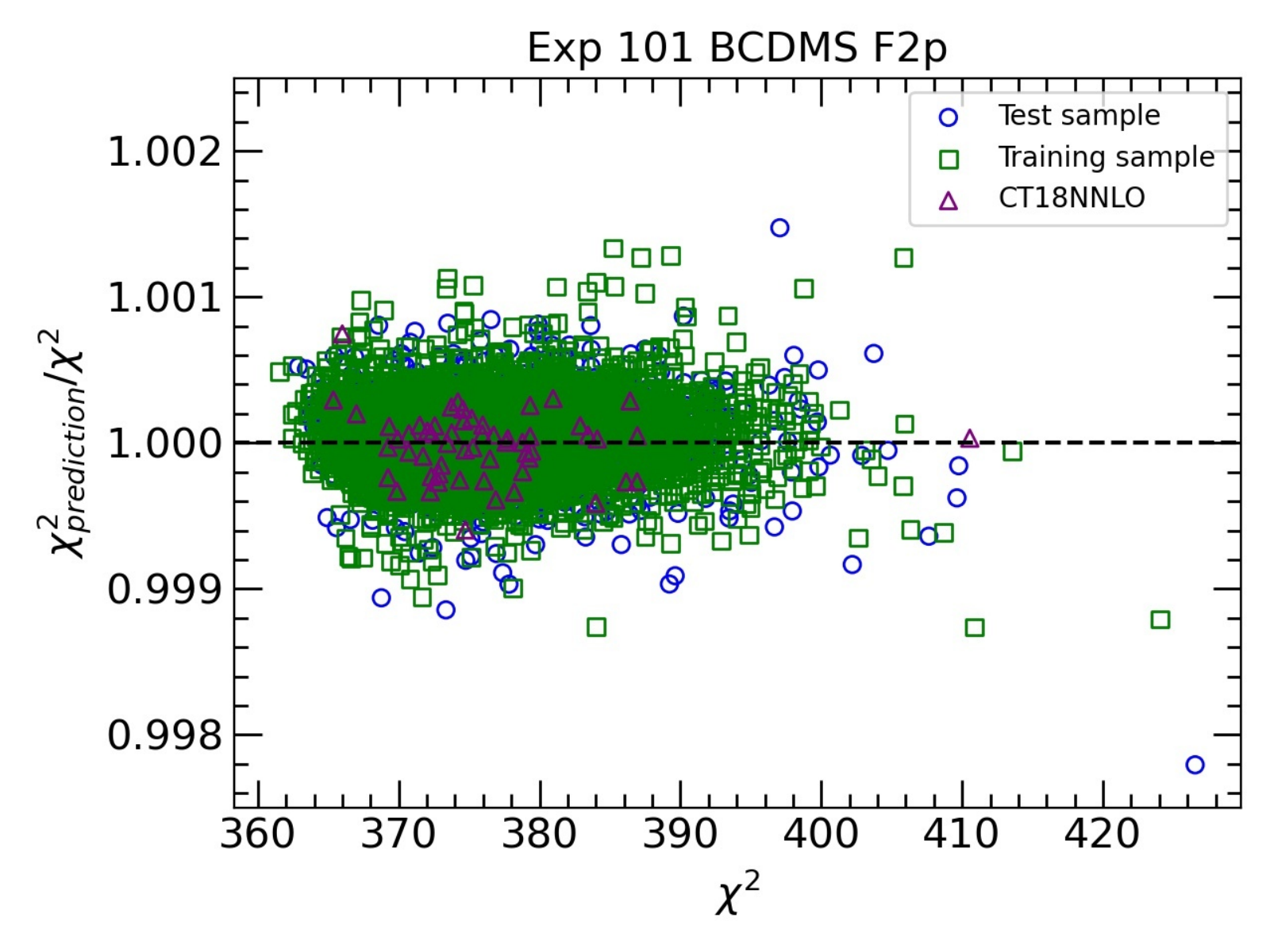}}
  \subcaptionbox{}[7.7cm] 
    {\includegraphics[width=7.7cm]{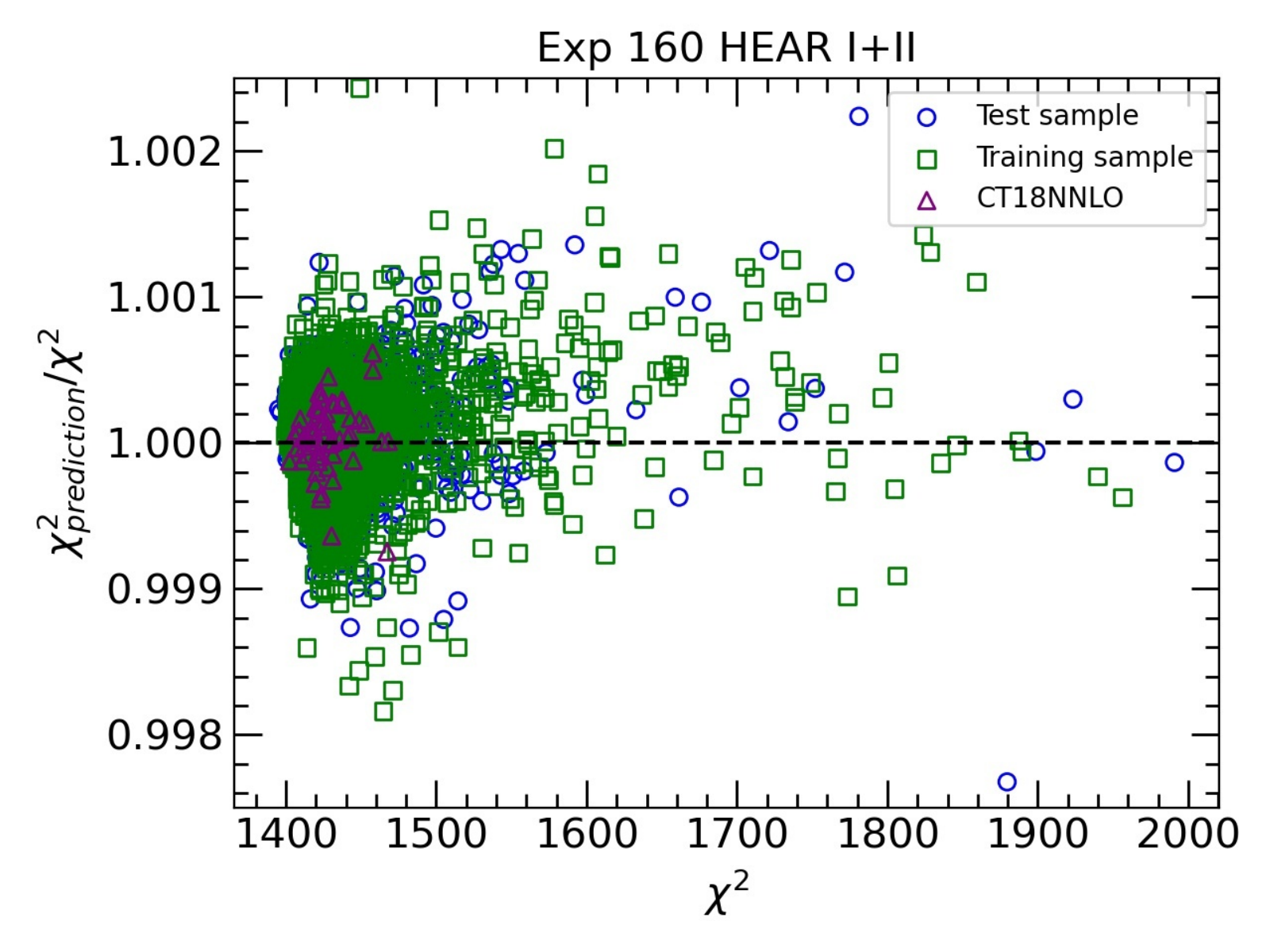}}
  \subcaptionbox{}[7.7cm]
    {\includegraphics[width=7.7cm]{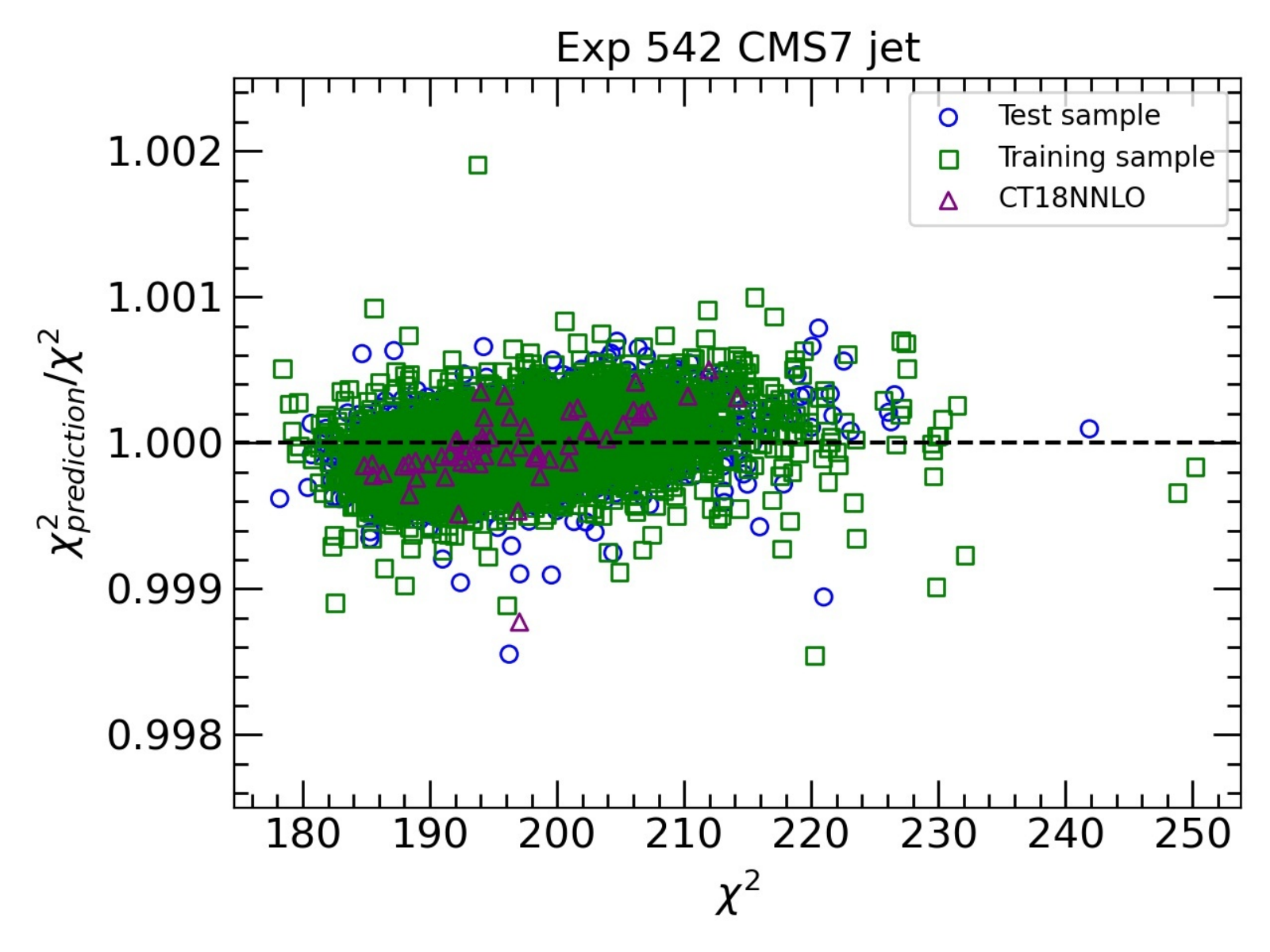}}
   \subcaptionbox{}[7.7cm]
     {\includegraphics[width=7.7cm]{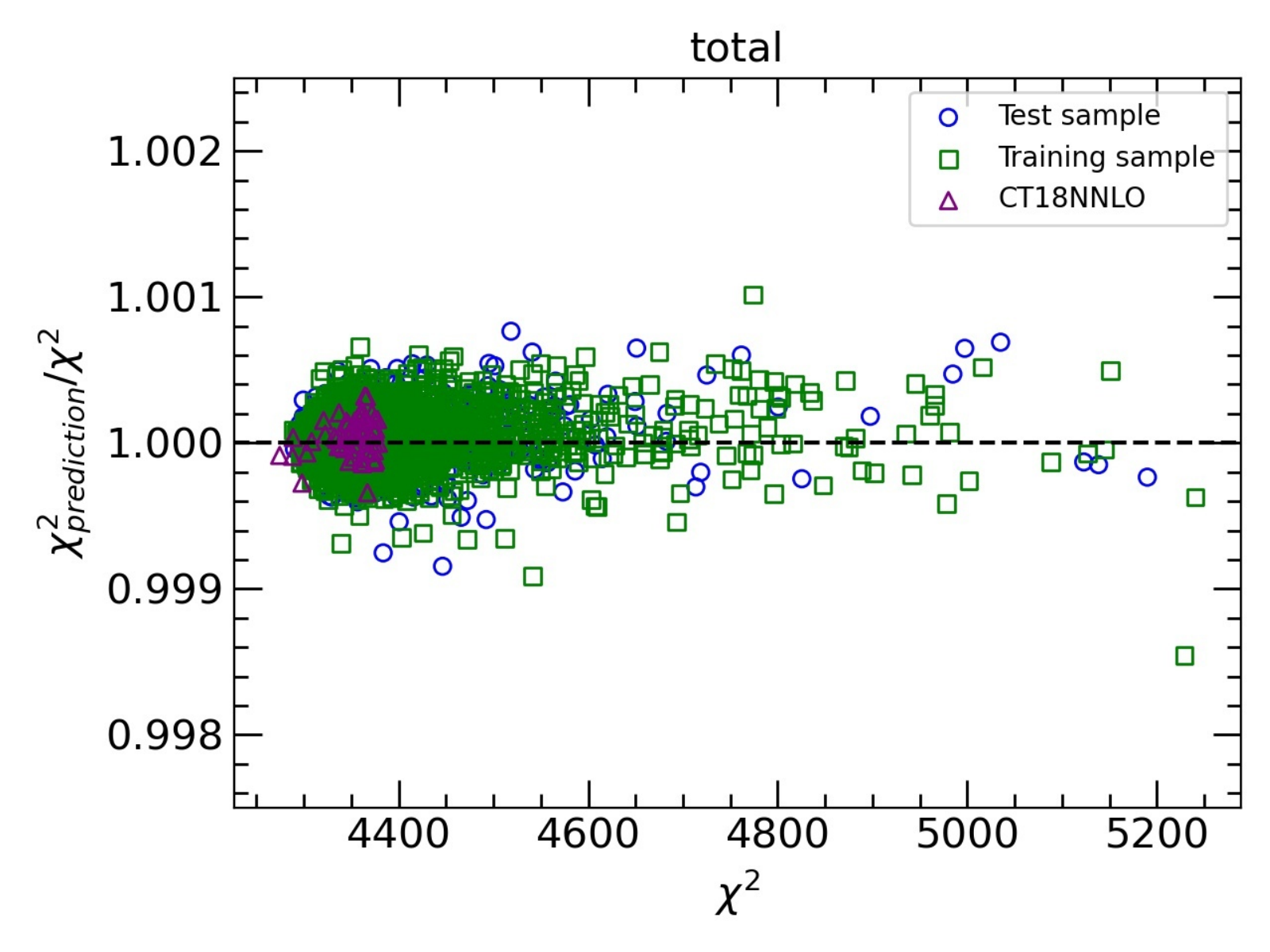}}
  \caption{The predictions to truths ratios of $\chi^2$ for experimental data sets 
  for measurements of the proton structure function by BCDMS, measurements of inclusive DIS reduced cross sections at HERA and 
  measurements of the inclusive jet cross sections at $\sqrt{s} =$ 7 TeV by CMS as well as for the full data set.}
  \label{Fig:NN_result_chi}
\end{figure}

In Fig.~\ref{Fig:NN_result_chi}, we show the predictions to truths ratios of $\chi^2$ for three experimental data sets: measurements of the proton structure function by BCDMS, measurements of inclusive DIS reduced cross sections at HERA and measurements of the inclusive jet cross sections at $\sqrt{s} =$ 7 TeV by CMS. The ratio of total $\chi^2$ for the full data set is also shown in the lower-right panel.
The ratios are calculated for the PDFs from the aforementioned training sample and test sample of NNs as well as the CT18 NNLO PDFs.
The CT18 NNLO PDFs consist of a central PDF set and 56 error PDFs in a total of 28 Hessian eigenvector directions.
The horizontal axis represents the truths of $\chi^2$.
Each mark corresponds to a PDF set from these three samples of PDF sets.
The green squares and the blue circles represent the ratios corresponding to the PDFs from the training sample and test sample respectively, and the purple triangles represent the ratios corresponding to the PDFs from CT18 NNLO.
We find good agreement between the training and test samples, although the NN produces greater deviation than the original CT18 NNLO PDFs.
We find that the predictions and the truths in general agree within 1 per mille for each data set.
For the total $\chi^2$, the deviation is within 0.6 per mille. 

\begin{figure}[htbp]
  \centering
  \includegraphics[width=1\textwidth,clip]{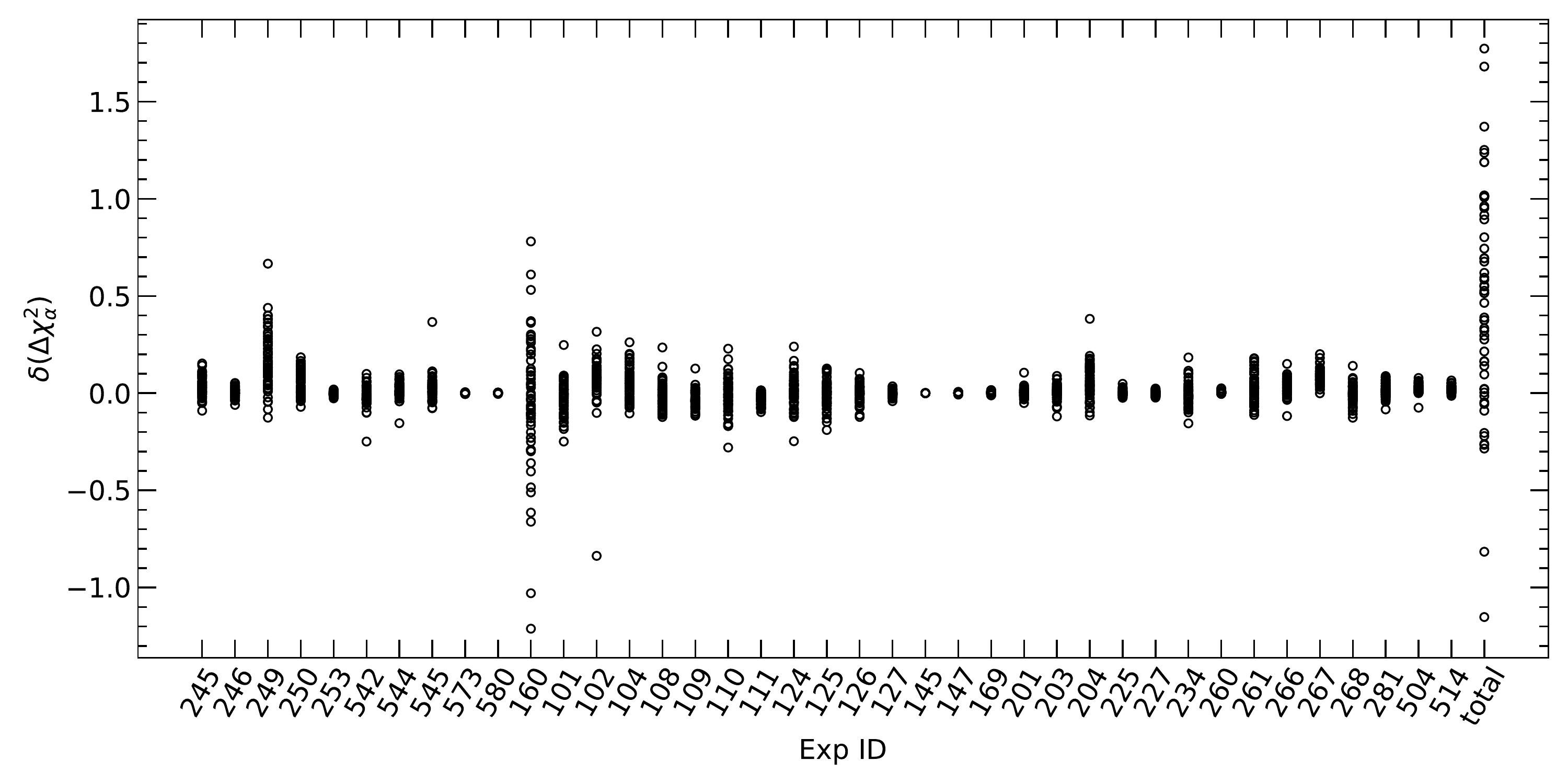}
  \hfill
  \caption{
 The $\delta(\Delta\chi^2_{\alpha})$ corresponding to the 56 error PDFs of CT18 NNLO are represented by the 56 marks distributed along the vertical direction for each individual data set and for the full data set.}
  \label{Fig:delta_chi2}
\end{figure}

We define $\Delta\chi^2$ as the difference between a certain $\chi^2$ value and its value at the best fit, which is conventionally used in the determination of PDF uncertainties.
In Fig.~\ref{Fig:delta_chi2}, differences between predicted and true $\Delta\chi^2$ denoted as $\delta(\Delta\chi^2_{\alpha}) \equiv \Delta\chi^2_{\alpha,pre}-\Delta\chi^2_{\alpha,tru}$ are demonstrated for each data set, where $\alpha$ represents the PDF set used in the comparison.
Here we choose a sample of PDF sets consisting of the 56 Hessian error PDFs in CT18 NNLO set, which are represented by the marks distributed along the vertical direction.
We also show similar results for the total $\chi^2$.
It can be seen that, for each individual data set the $\delta(\Delta\chi^2_{\alpha})$ is at most 1 unit,
and the $\delta(\Delta\chi^2_{\alpha})$ for the full data set are within 2 units.
The extent of $\delta(\Delta\chi^2_{\alpha})$ for HERA inclusive DIS data set and for total $\chi^2$ is slightly larger than other experimental data sets. 
It should be noticed that the number of data points of HERA inclusive DIS data set and the full data set is 1120 and 3671 respectively.
Besides, the $\Delta\chi^2_{\alpha}$ of the full data set for most of the 56 Hessian error PDFs is about 100 units. This indicates the relative deviation of NNs predictions is below 2\% for $\Delta\chi^2$.

\begin{figure}[htbp]
  \centering
    \subcaptionbox{}[7.5cm] 
    {\includegraphics[width=7.5cm]{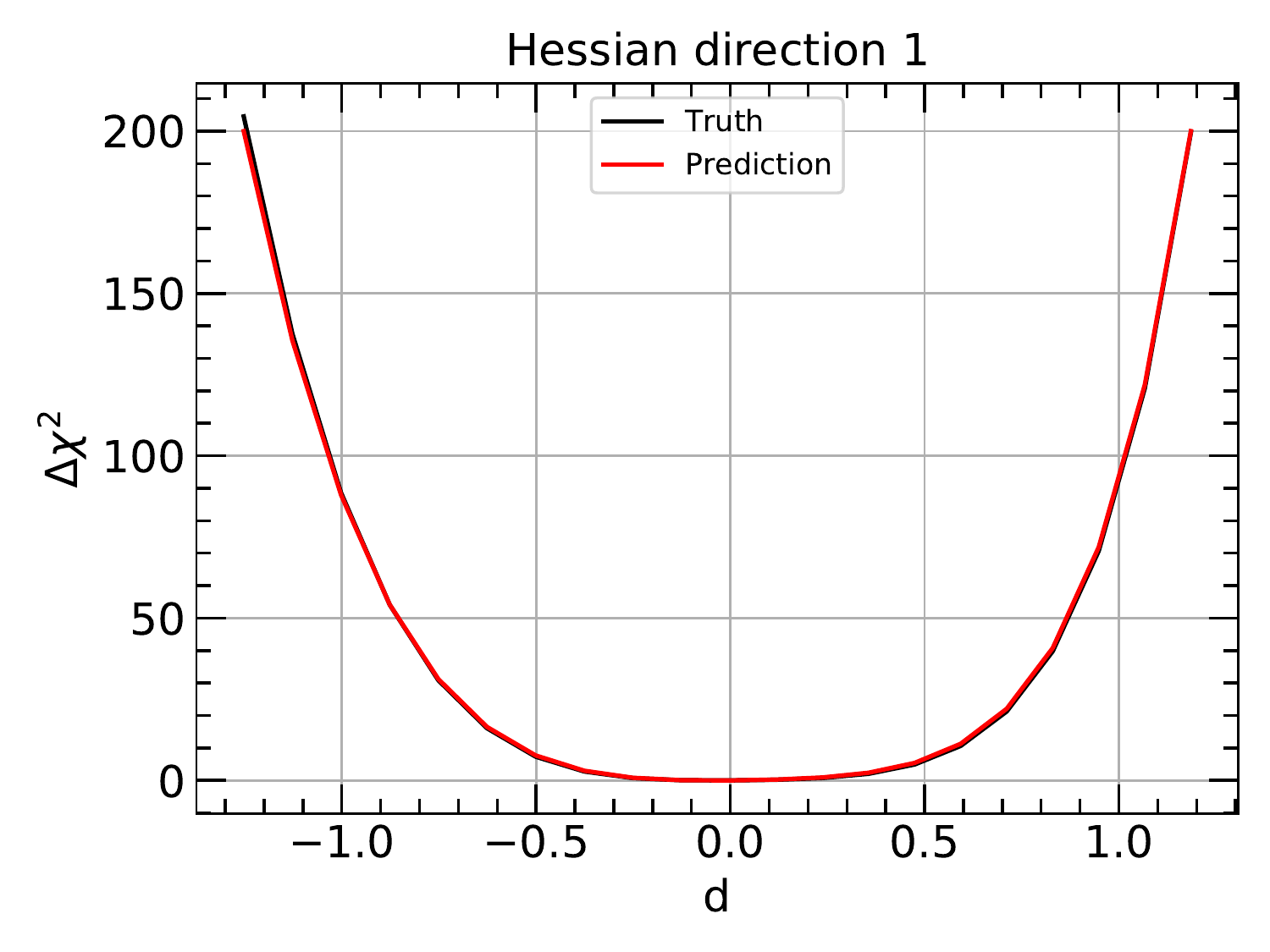}}
  \subcaptionbox{}[7.5cm] 
    {\includegraphics[width=7.5cm]{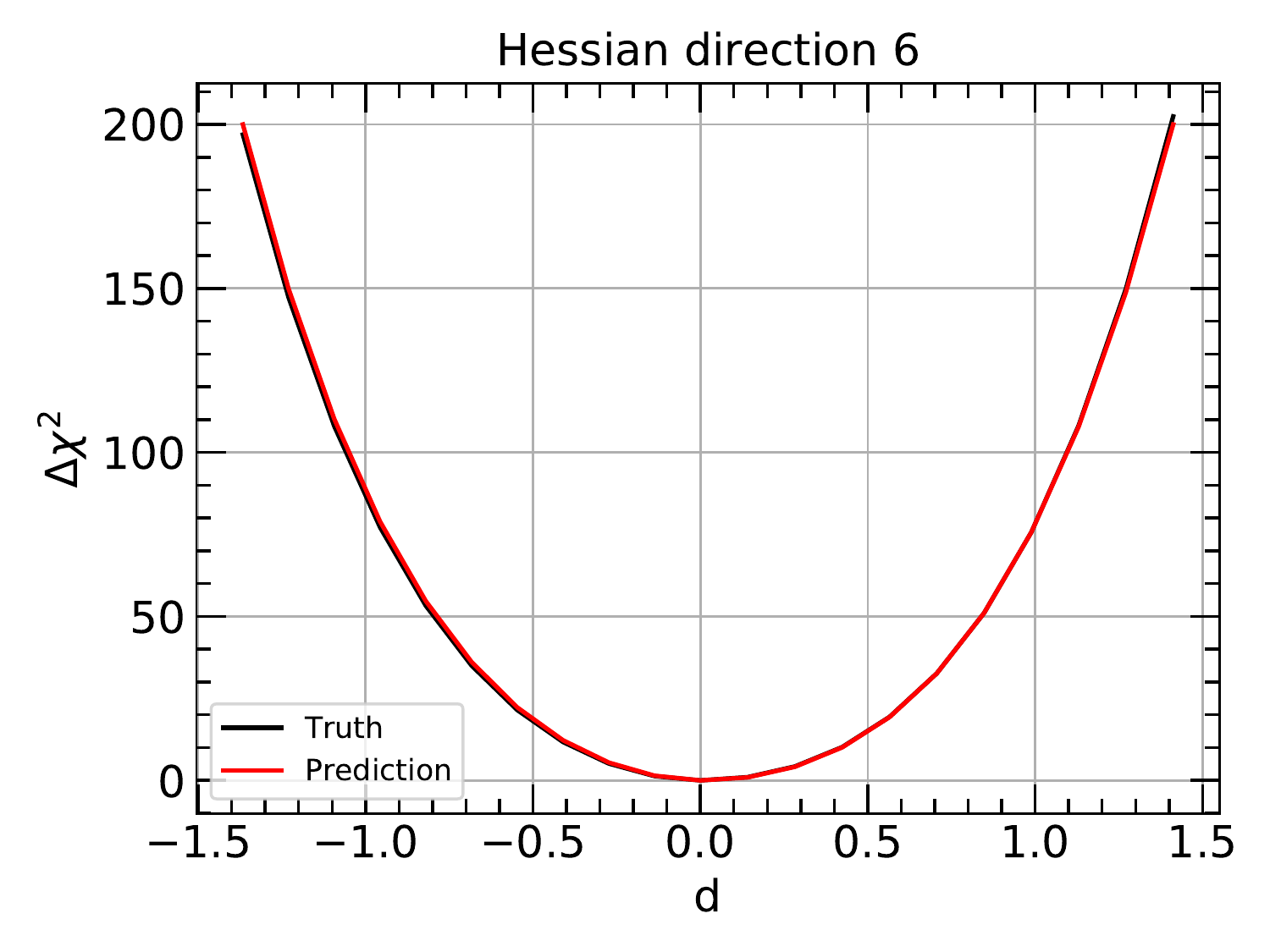}}
  \subcaptionbox{}[7.5cm] 
    {\includegraphics[width=7.5cm]{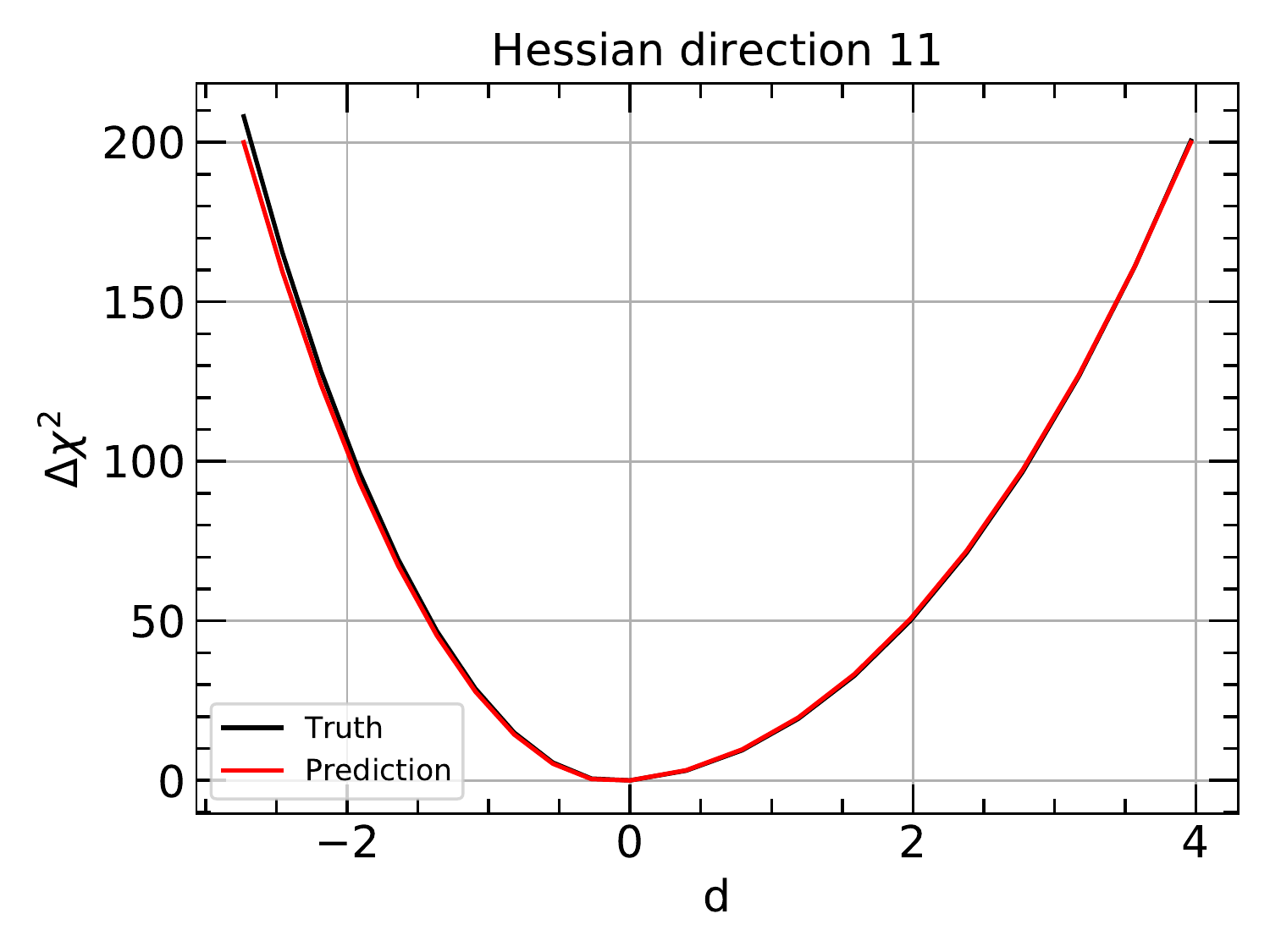}}
  \subcaptionbox{}[7.5cm] 
    {\includegraphics[width=7.5cm]{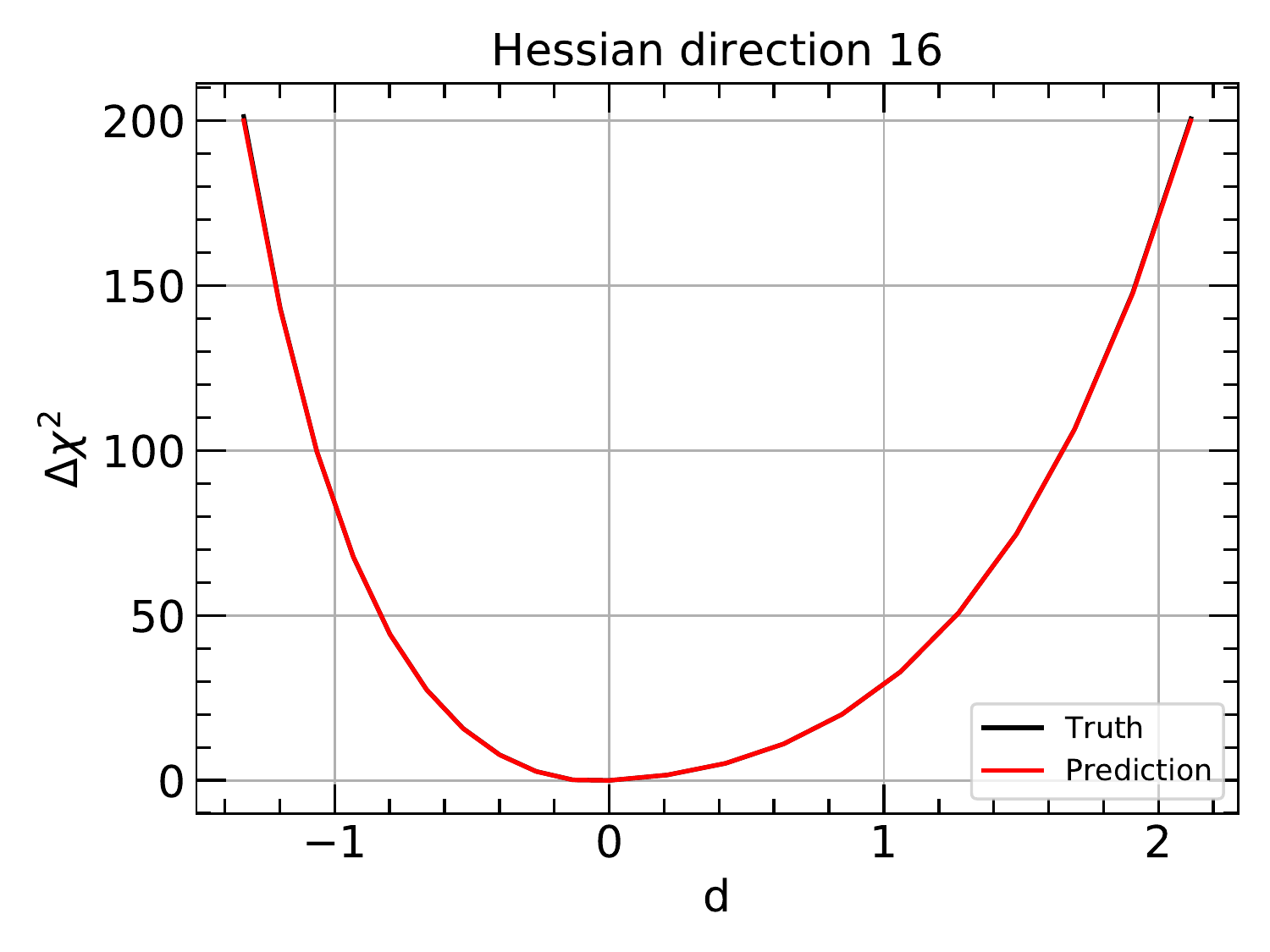}}
  \subcaptionbox{}[7.5cm] 
    {\includegraphics[width=7.5cm]{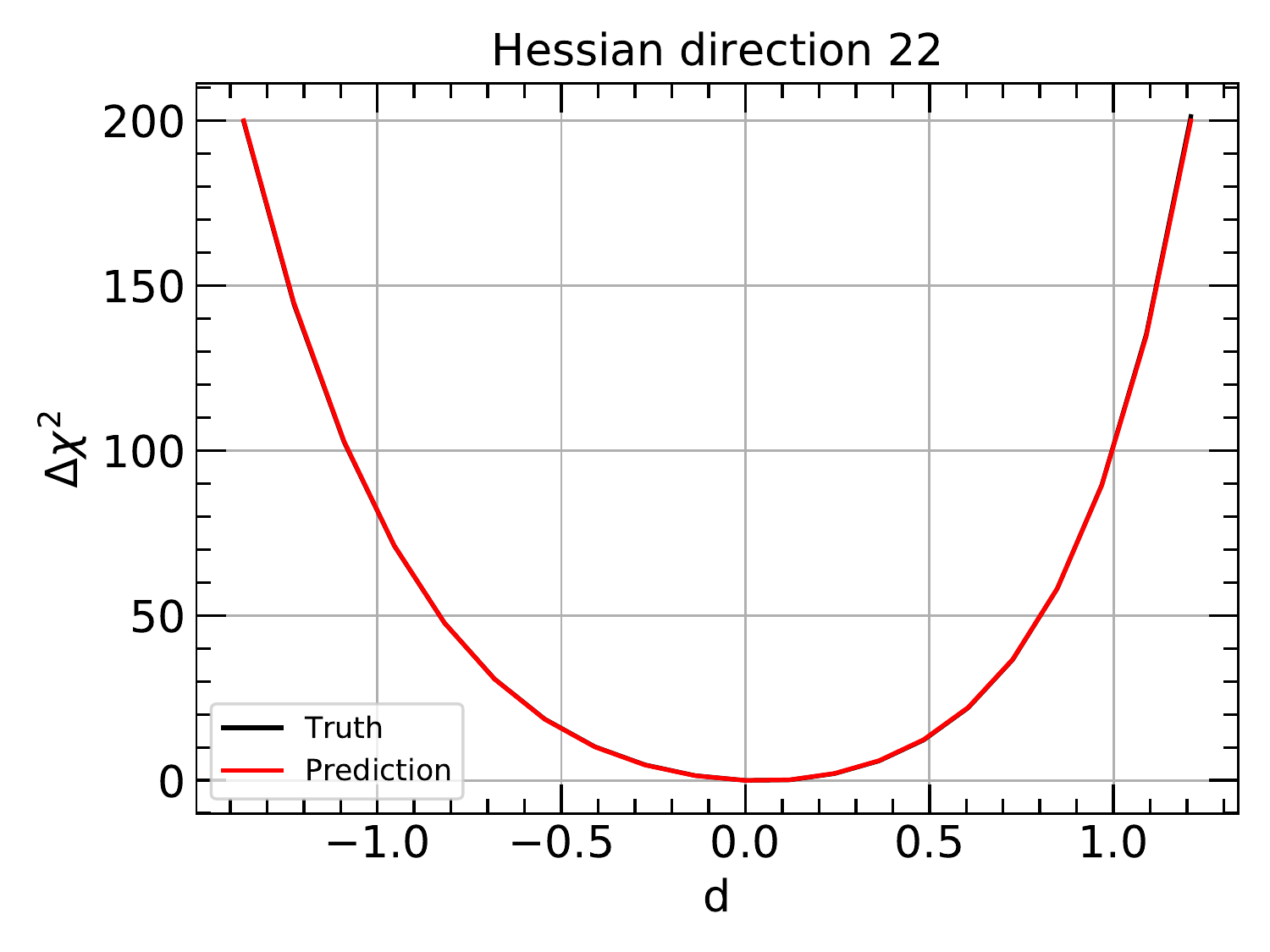}}
  \subcaptionbox{}[7.5cm] 
    {\includegraphics[width=7.5cm]{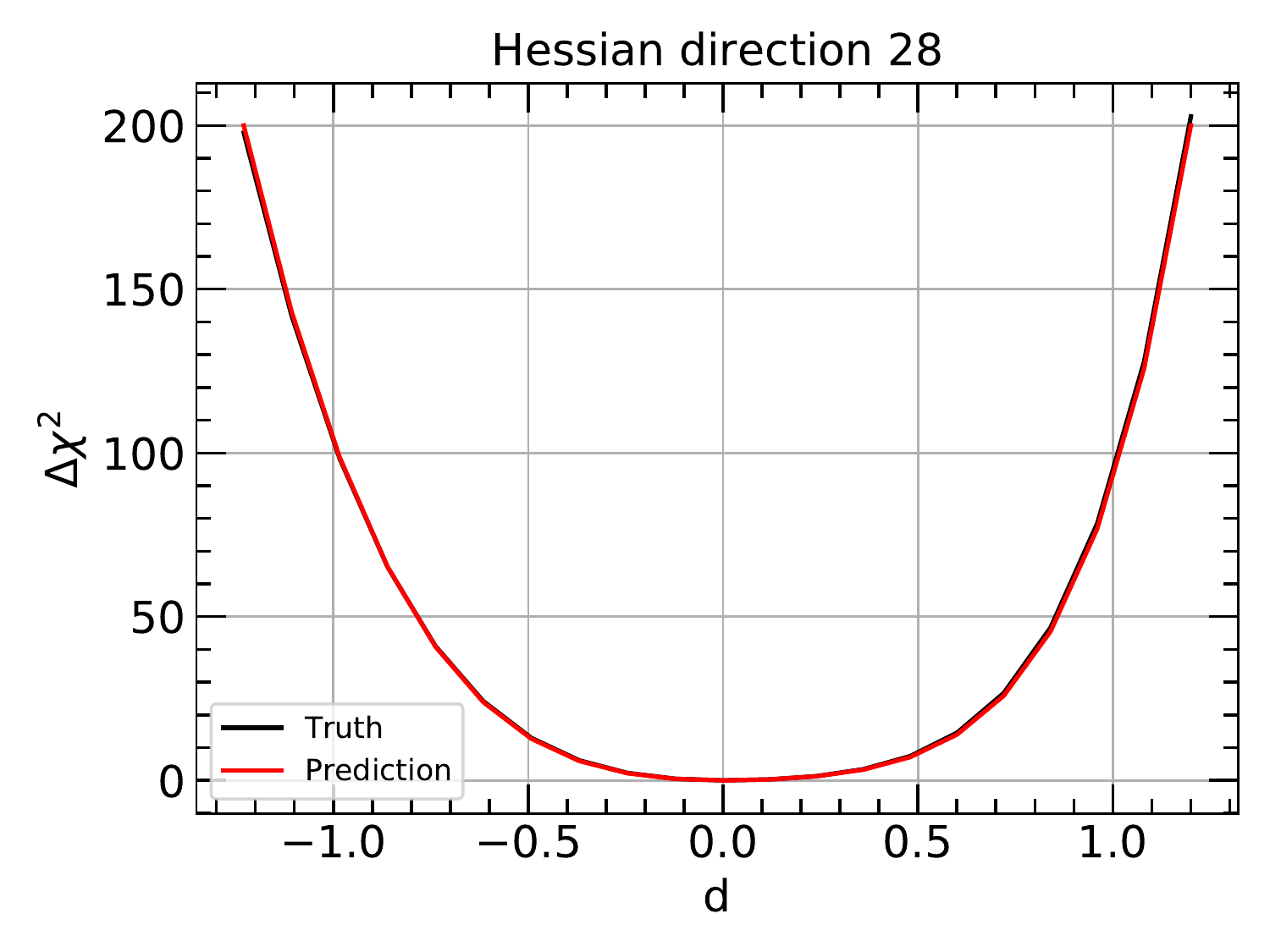}}
  \caption{The variation of $\Delta\chi^2$ with $d$ along the 1st, 6th, 11th, 16th, 22th and 28th CT18NNLO Hessian eigenvector directions.}
  \label{Fig:hes_direction}
\end{figure}

Furthermore, we compare the $\Delta\chi^2$ for the full data set along the 28 eigenvector directions of CT18 NNLO PDFs by scans of PDF parameters.
A variable $d$ is introduced to measure the distance that PDF parameters go along the direction of a certain eigenvector.
The variation of PDF parameters for the scan along the $j_{th}$ eigenvector direction can be written as:
\begin{equation}
    a_{i}^{j,scan}(d) = \left\{\begin{array}{ll}
    d(a_{i}^{2j-1} - a_{i}^{0})+a_{i}^{0}, & (d > 0) ,\\
    \\
    d(a_{i}^{0} - a_{i}^{2j})+a_{i}^{0}, & (d <0) ,
    \end{array}\right.
\end{equation}
where $i$ represents the index of the PDF parameters,
\{$a_{i}^{0}$\} represents PDF parameters for the central PDF of CT18 NNLO, \{$a_{i}^{2j-1}$\} and \{$a_{i}^{2j}$\} represent PDF parameters of the two error PDFs in the $j_{th}$ eigenvector direction of CT18 NNLO.
The total $\chi^2$ are computed using the new set of PDF parameters. We define $\Delta{\chi^2}\equiv\chi^2(d)-\chi^2(d=0)$, 
and compare the truths of $\Delta\chi^2$ and the predictions from NNs as a function of $d$ for a few selected eigenvector directions in Fig.~\ref{Fig:hes_direction}.

We find that the NNs can describe well the dependence of the $\Delta\chi^2$ on PDF parameters in all Hessian eigenvector directions.
The predictions and the truths in general agree within 2\% in all directions, which agrees with Fig.~\ref{Fig:delta_chi2}.
We also notice that $\Delta \chi^2$ has a sizable deviation from quadratic shape in some directions.
The NNs can reproduce well the asymmetric and non-quadratic behavior of $\Delta\chi^2$, which is one of the main advantages as comparing to the traditional Hessian method.
It is justified to say the deviation of $\chi^2$ due to the NNs approximation is negligible for the PDF parameter space of interest.

\subsection{Physics quantities}
\begin{figure}[htbp]
  \centering
  \subcaptionbox{}[7.7cm] 
    {\includegraphics[width=7.7cm]{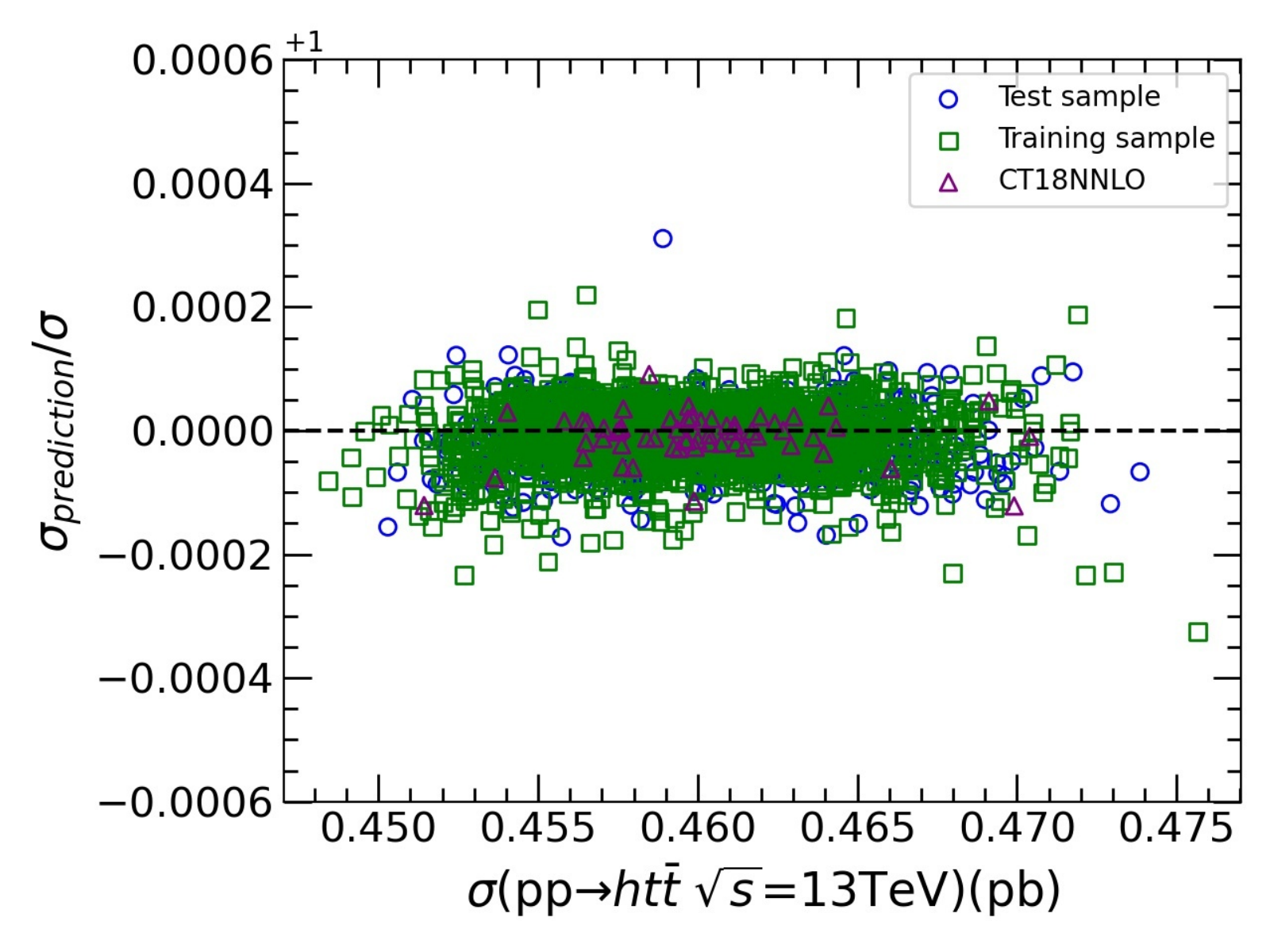}}
  \subcaptionbox{}[7.7cm] 
    {\includegraphics[width=7.7cm]{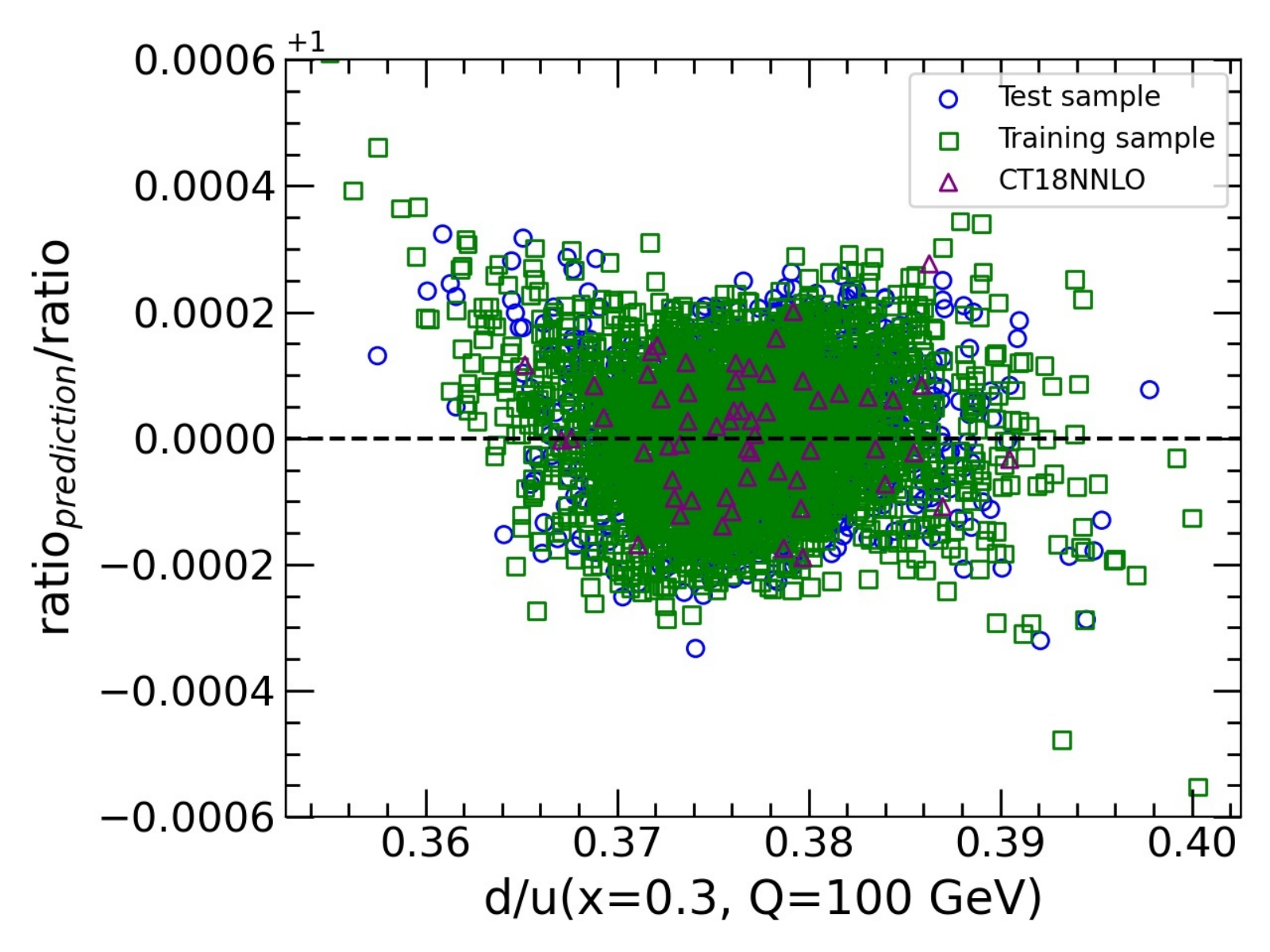}}
  \caption{The ratios of the predictions from NNs to the truths for the cross section of top-quark pair with a Higgs boson production in proton-proton collisions at a center of mass energy $\sqrt{s}$ = 13 TeV and the PDF ratio $d/u$($x = 0.3$, $Q = 100$ GeV).}
  \label{Fig:NN_result_phy_quan}
\end{figure}

In Fig.~\ref{Fig:NN_result_phy_quan}, we show the ratios of the predictions to the truths for the cross section of top-quark pair with a Higgs boson production in proton-proton collisions at a center of mass energy $\sqrt{s}$ = 13 TeV, and for the PDF ratio $d/u$($x = 0.3$, $Q = 100$ GeV).
The ratios are calculated for the PDFs from the training sample and test sample of NNs as well as the CT18 NNLO PDFs.
We find good agreements between the distributions of marks for training sample and for test sample.
Deviations for these two physics quantities are in general within 0.15 per mille and 0.2 per mille, respectively. 
The performance of the NNs for these two physics quantities is better than that for $\chi^2$, which is because the dependence of $\chi^2$ on PDFs is more complex than the cases for cross sections or PDF ratios.
The dependence of these physics quantities on PDFs is even close to linear.

\begin{figure}[htbp]
  \centering
  \includegraphics[width=1\textwidth,clip]{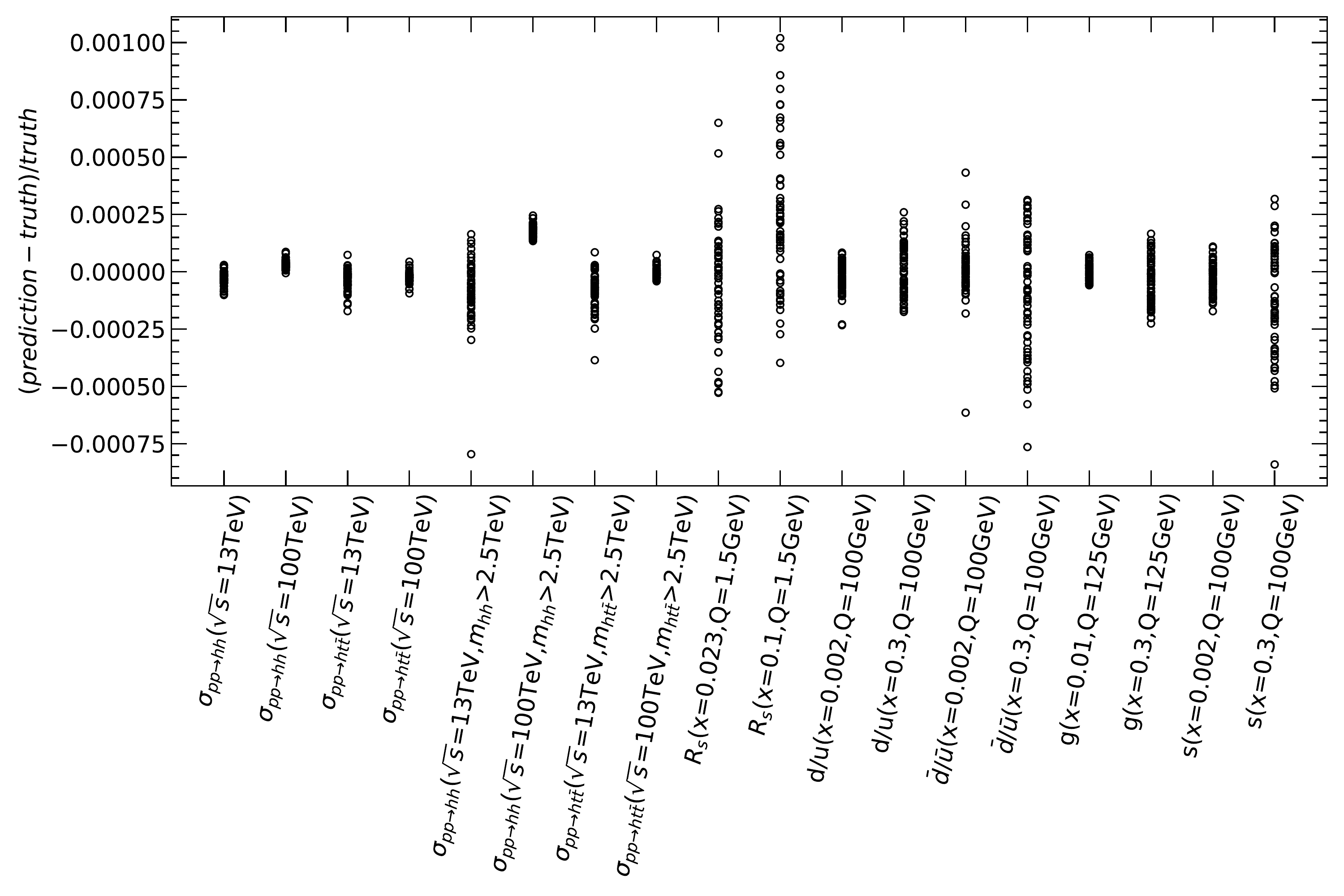}
  \hfill
  \caption{The relative difference between the predictions from NNs and the truths for various physics quantities. For each physics quantity, 57 marks distributed along the vertical direction correspond to the results from 57 PDFs of CT18 NNLO.}
  \label{Fig:compare_phy_quan}
\end{figure}

We further summarize the relative difference between the predictions from NNs and the truths for various physics quantities in Fig.~\ref{Fig:compare_phy_quan}.
For each physics quantity, 57 marks distributed along the vertical direction correspond to the results from the 57 CT18 NNLO PDFs.
The results for the cross sections of Higgs boson pair production and top-quark pair with a Higgs boson production in proton-proton collisions at center of mass energy $\sqrt{s} = $ 13 TeV or 100 TeV are shown in this figure.
We find the predictions from NNs and the truths for these cross sections agree within 0.2 per mille.
In addition, the results for cross sections of $pp \to hh$ and $pp \to ht\bar{t}$ with high invariant mass $m_{hh} > $ 2.5 TeV or $m_{ht\bar{t}} > $ 2.5 TeV are also shown in Fig.~\ref{Fig:compare_phy_quan}, and the relative difference between the predictions and truths in general agree within 0.3 per mille.
Comparisons for strangeness ratio $R_s \equiv \dfrac{s(x, Q)+\bar{s}(x, Q)}{\bar{u}(x, Q)+\bar{d}(x, Q)}$ and PDF ratios $d/u$ and $\bar{d}/\bar{u}$ at various $x$ and $Q$ are also shown in this figure, and the predictions and the truths agree within 1 per mille.
We also show the results for PDF values for $g$ and $s$-quark at various $x$ and $Q$, and the deviations between the predictions and the truths are within 0.75 per mille.

\section{Lagrange Multiplier scans}\label{sec:LM}
LM scan is a robust method to estimate PDF uncertainties, which was originally developed in Refs.~\cite{Pumplin:2000vx,Stump:2001gu}. 
In this method a physics quantity $X(\{a_{i}\})$ is introduced to the global fit as a Lagrange multiplier.
Then the new function that needs to be minimized in the global fit becomes
\begin{equation}
\Psi\left(\lambda,\left\{a_{i}\right\}\right) \equiv \chi^{2}\left(\left\{a_{i}\right\}\right)+\lambda {X}\left(\left\{a_{i}\right\}\right),
\label{eq:LM}
\end{equation}
where $\lambda$ is a specified constant.
For each value of $\lambda$, one can determine a set of \{$a_i$\}, $X(\{a_{i}\})$ and $\chi^2$ by minimizing $\Psi$.
Here the $\chi^2$ corresponds to the minimum of a constrained fit with $X\{a_i\}$ fixed to the corresponding value.
Specially, the central value of $X(\{a_{i}\})$ and the global minimum of $\chi^2$, $\chi^2_{min}$, can be determined by setting $\lambda = 0$.
A parametrically defined curve ($X$, $\chi^2$) can be determined by repeating the minimization for many values of $\lambda$.
This means the $\chi^2$ of the global fit depends on the value of $X(\{a_{i}\})$ and can be represented as $\chi^2=\chi^2_{min}+\Delta\chi^2$.
The PDF uncertainty of $X(\{a_{i}\})$ can be determined by requiring $\Delta\chi^2 + P = T$, here $T$ is the so-called ``tolerance factor".
We assume that 90\% CL region corresponds to $T = 100$.
The penalty term $P$, called Tier-2 penalty, is introduced to ensure the tolerance will be reached as soon as any data set shows disagreement at 90\% CL.
The detailed definition of the penalty term can be found in Refs.~\cite{Gao:2017yyd,Dulat:2015mca}. 

In comparison, we briefly describe the calculation of PDF uncertainties in the framework of the Hessian method.
Given the physics quantity $X(\{a_{i}\})$, the asymmetric PDF uncertainties can be calculated as~\cite{Lai:2010vv}
\begin{equation}
\begin{array}{l}
\delta^{+} X=\sqrt{\sum_{i=1}^{N_d}\left[\max \left(X_{2i-1}-X_{0}, X_{2i}-X_{0}, 0\right)\right]^{2}}, \\
\delta^{-} X=\sqrt{\sum_{i=1}^{N_d}\left[\max \left(X_{0}-X_{2i-1}, X_{0}-X_{2i}, 0\right)\right]^{2}},
\end{array}
\end{equation}
where $X_0$ represents the value of the physics quantity with the central PDF of the Hessian set, $X_{2i-1}$ ($X_{2i}$) represents the value of the physics quantity with the error PDF of the Hessian set in the positive (negative) direction of the $i_{th}$ eigenvector in the $N_d$-dimensional PDF parameter space.

\subsection{LM scans on PDFs}
\begin{figure}[htbp]
  \centering
  \subcaptionbox{}[7.7cm] 
    {\includegraphics[width=7.7cm]{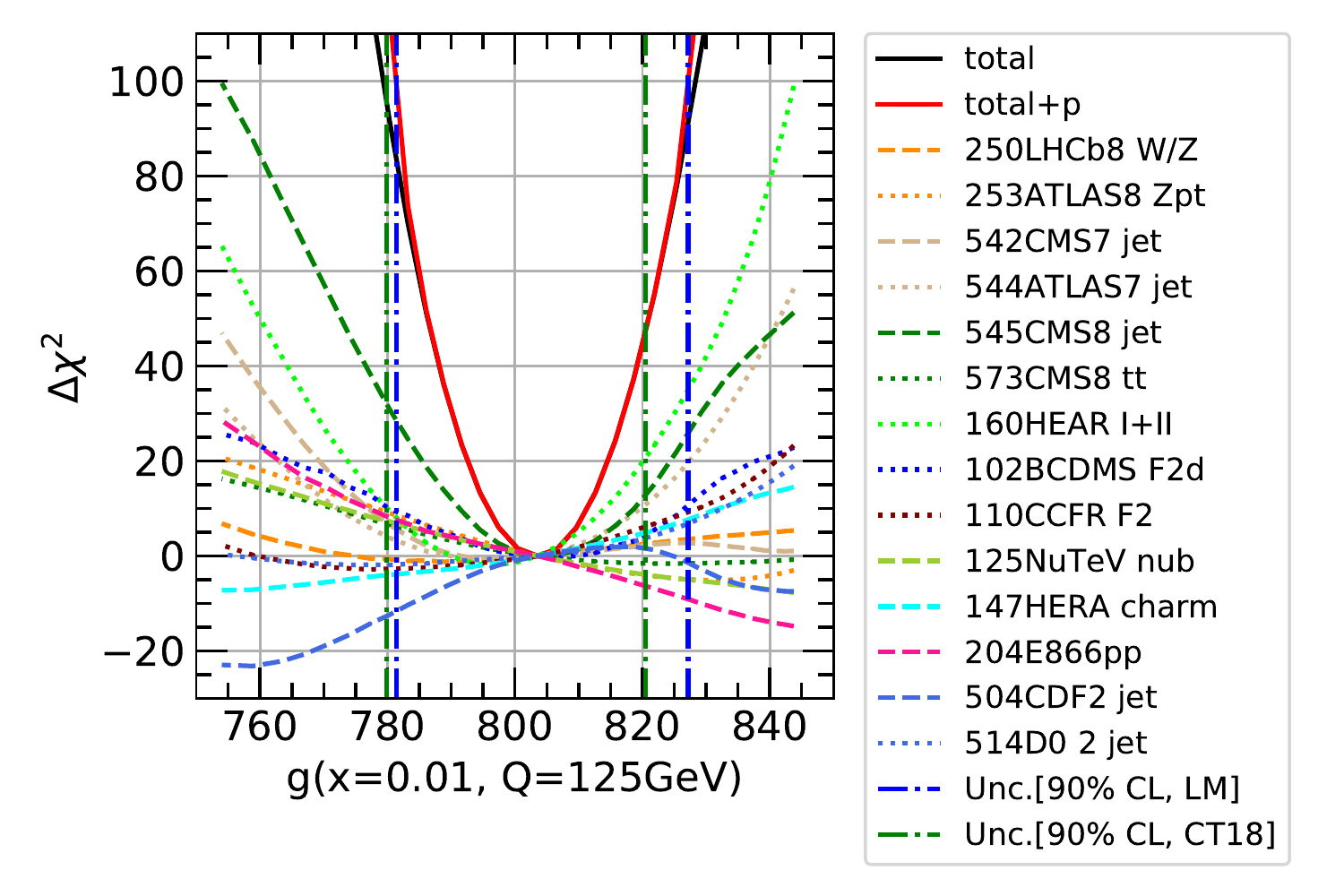}}
  \subcaptionbox{}[7.7cm]
    {\includegraphics[width=7.7cm]{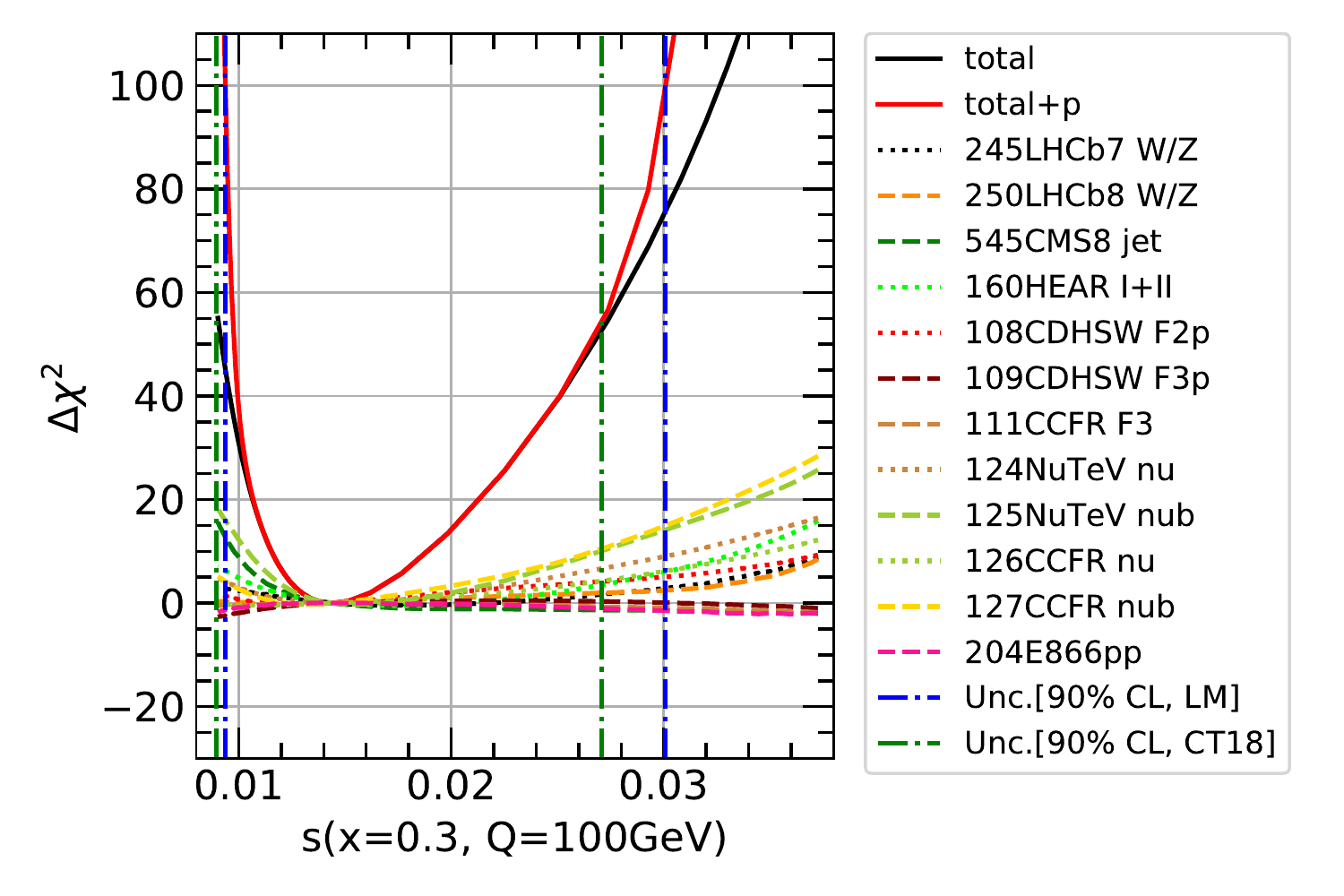}}
  \subcaptionbox{}[7.7cm]
      {\includegraphics[width=7.7cm]{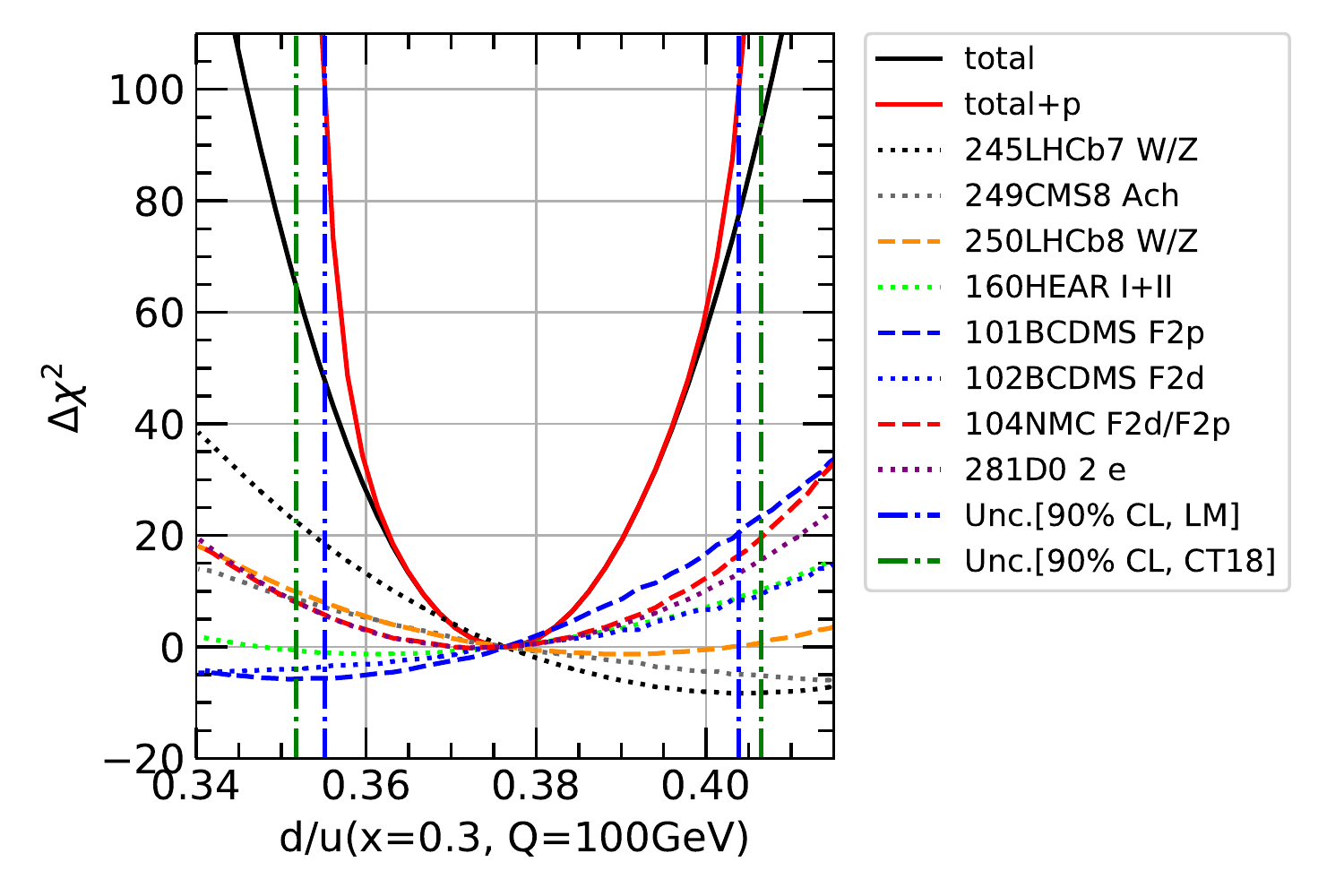}}
  \subcaptionbox{}[7.7cm]
    {\includegraphics[width=7.7cm]{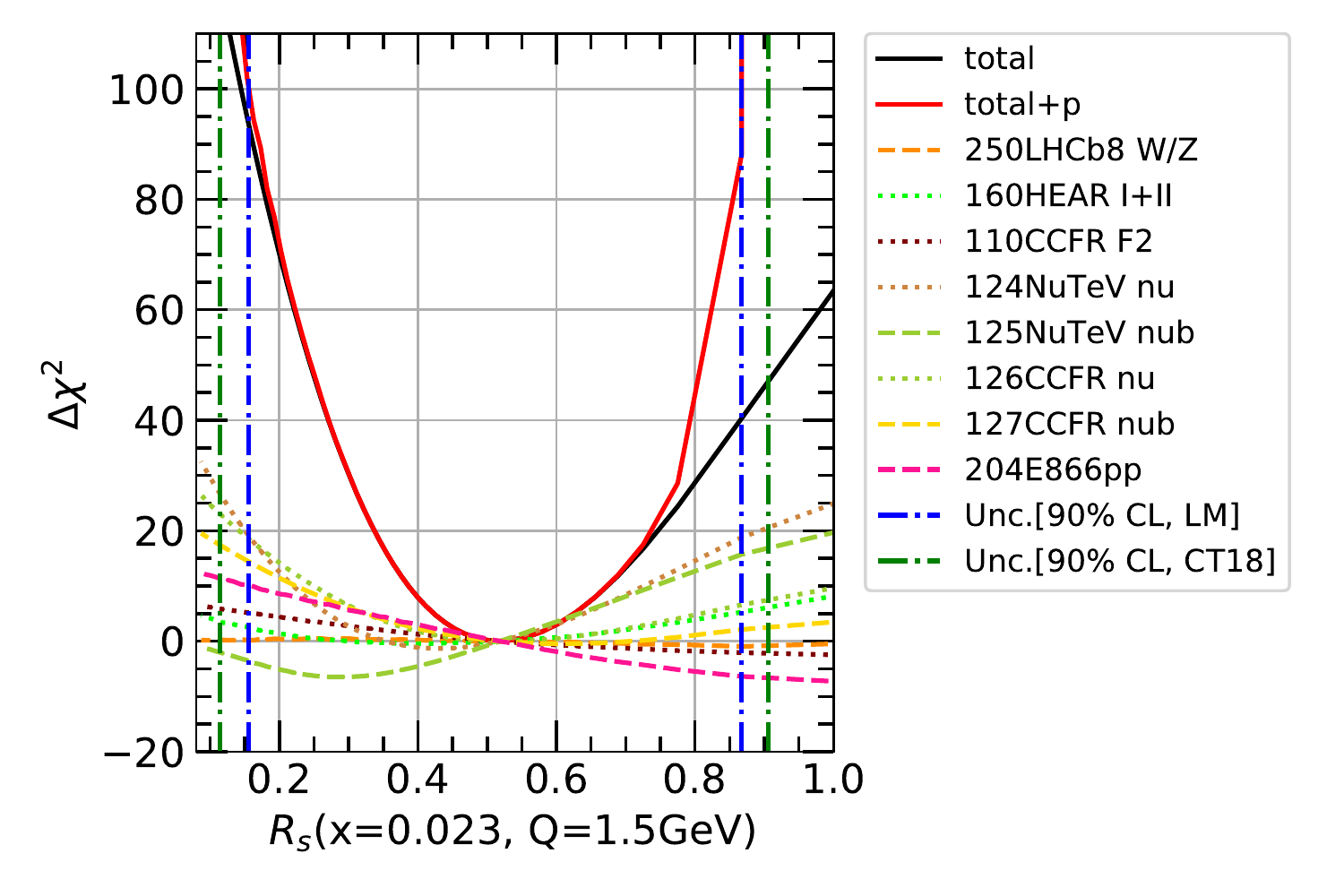}}
  \caption{LM scans on the $g$ ($x = 0.01$ GeV, $Q = 125$ GeV), $s$ ($x = 0.3$, $Q = 100$ GeV), $d/u$ ($x = 0.3$, $Q = 100$ GeV) and $R_s$ ($x = 0.023$, $Q = 1.5$ GeV).
The black and the red solid lines represent $\Delta\chi^2$ and $\Delta\chi^2 + P$ respectively.
The dot and the dash lines indicate the contributions to $\Delta\chi^2$ from individual data sets.
The blue and the green vertical dot-dash lines indicate the uncertainties at 90\% CL determined with the LM method by requiring $\Delta\chi^2 + P$ = 100 and with the Hessian method from the published CT18 NNLO PDFs, respectively.
  }
  \label{Fig:LM_g}
\end{figure}

We first study PDF values and ratios with LM scans based on the aforementioned NNs approximation of $\chi^2$ and physics quantities.
The results are shown in Fig.~\ref{Fig:LM_g}.
The black and the red solid lines represent $\Delta\chi^2$ and $\Delta\chi^2 + P$ respectively.
The dot and the dash lines indicate the contributions to $\Delta\chi^2$ from individual data sets.
The blue and the green vertical dot-dash lines indicate the uncertainties at 90\% CL determined with the LM method by requiring $\Delta\chi^2 + P$ = 100 and with the Hessian method from the published CT18 NNLO PDFs, respectively.
Among the generic features of the scans, it can be seen that the profile of the total $\Delta\chi^2$ and individual $\Delta\chi^2$ show almost a quadratic dependence on the variable at the neighborhood of the global minimum, which is a requirement of the Hessian method.
Some individual data sets prefer PDF values or ratios that differ significantly from those at the global minimum.
Besides, the HERA inclusive DIS data play important roles in all cases, which can be understood as due to the high experimental precision and the large number of data points.
The penalty term also gives strong constraints on some PDF values or ratios.
In addition, there are some slight differences between the uncertainties determined with the Hessian method and the LM scans, which is expected.

In the upper-left panel of Fig.~\ref{Fig:LM_g}, we show the results of LM scans on the gluon PDF at $Q = 125$ GeV and $x = 0.01$.
We find that the HERA inclusive DIS data and the LHC jet data give the leading constraints.
In addition, the CDF inclusive jet data (Exp. ID = 504) prefers a smaller value of the gluon PDF.
At the global minimum, the $\chi^2$ for the CDF inclusive jet data is elevated by about 20 units.
The Hessian method gives a smaller PDF uncertainty than the estimation based on the LM scans.

In the upper-right panel, we show the results of LM scans on the $s$-quark PDF at $Q = 100$ GeV and $x = 0.3$. 
In this panel, the NuTeV and the CCFR dimuon data together with the HERA inclusive DIS data give the dominant constraints.
These experimental data are consistent with the global fit.
A marked deviation from the quadratic shape can be observed in the profile of the $\Delta\chi^2$. 
In this case, a notable difference in uncertainties manifests between the LM method and the Hessian method, and the LM method should give the more reliable result.

In the bottom-left panel of Fig.~\ref{Fig:LM_g}, we show the results of LM scans on the PDF ratio $d/u$ at $Q = 100$ GeV and $x = 0.3$. 
The $d/u$ ratio is dominantly constrained by the LHCb W and Z boson production and the fixed target experiments BCDMS and NMC.
Contrasted with previous situations, the LM method gives a smaller PDF uncertainty for the $d/u$ ratio.

In the bottom-right panel, we show the results of LM scans on the strangeness ratio $R_s$ at $Q = 1.5$ GeV and $x = 0.023$.
We find that the NuTeV and the CCFR dimuon data and HERA inclusive DIS data give the dominant constraints.
It is also worthy noting that the NuTeV dimuon data with anti-neutrinos (Exp. ID = 125) prefers $R_s \approx 0.25$ which is smaller than the best fit value from the global fit $R_s$ = 0.52 and results in a large penalty term.
At the global minimum, the $\chi^2$ for the NuTeV dimuon data with anti-neutrinos is elevated by about 7 units.
The LM method predicts $R_s = 0.52^{+0.35}_{-0.36}$ at 90\% CL that has smaller uncertainties than $R_s = 0.52^{+0.39}_{-0.41}$ from the Hessian method.

Above scans have also been performed in the CT18 analysis~\cite{Hou:2019efy}, and our results are consistent with those, which further proves the validity of our approach.
After demonstrating the great efficiency and validity of our approach on LM scans, 
we are now ready to perform a systematic study on the PDF values and ratios for all flavors at
a series of $x$ values spreading over a wide range.

\begin{figure}[htbp!]
  \centering
  \subcaptionbox{}[6.9cm]
    {\includegraphics[width=6.9cm]{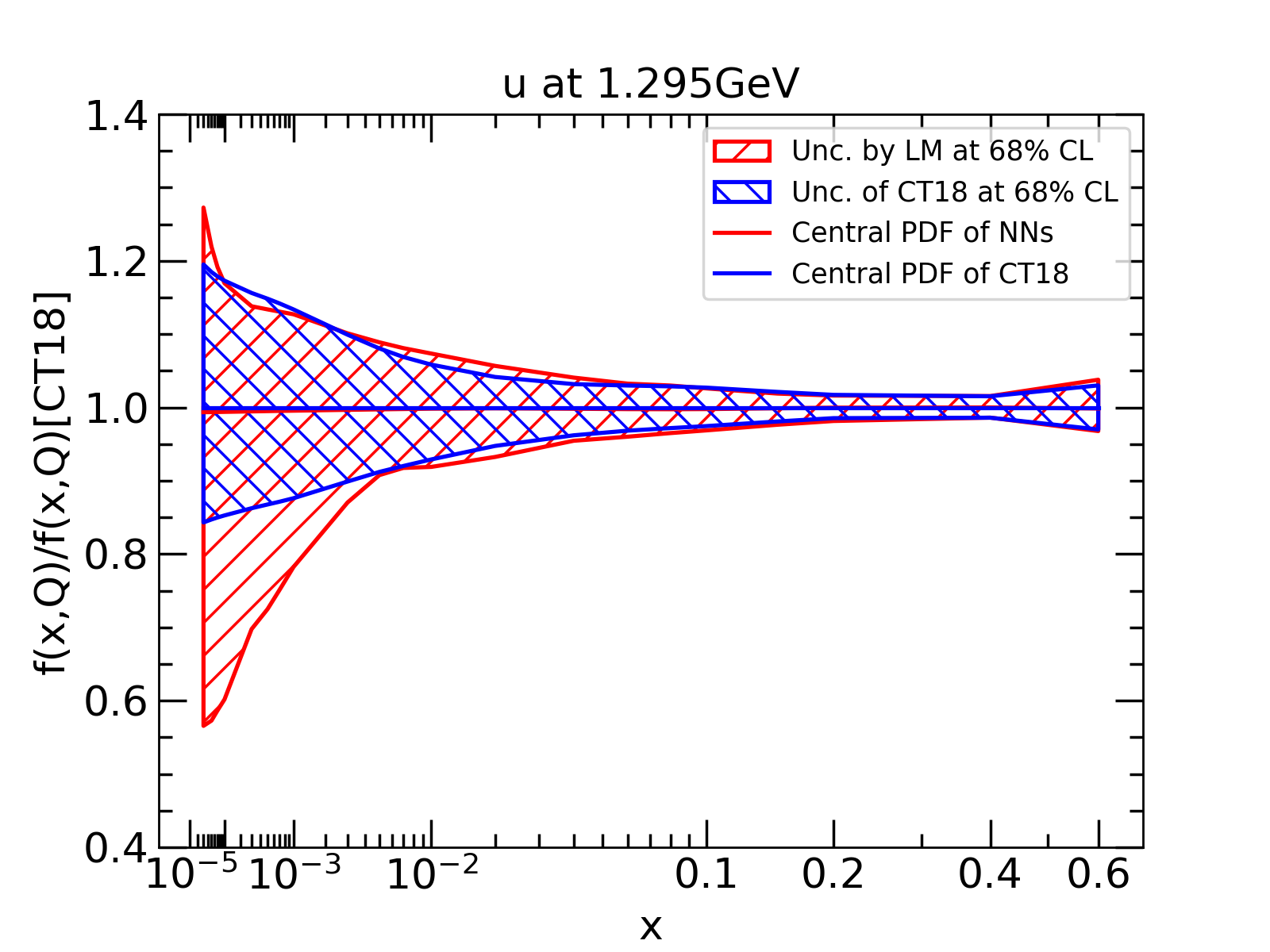}}
  \subcaptionbox{}[6.9cm]
    {\includegraphics[width=6.9cm]{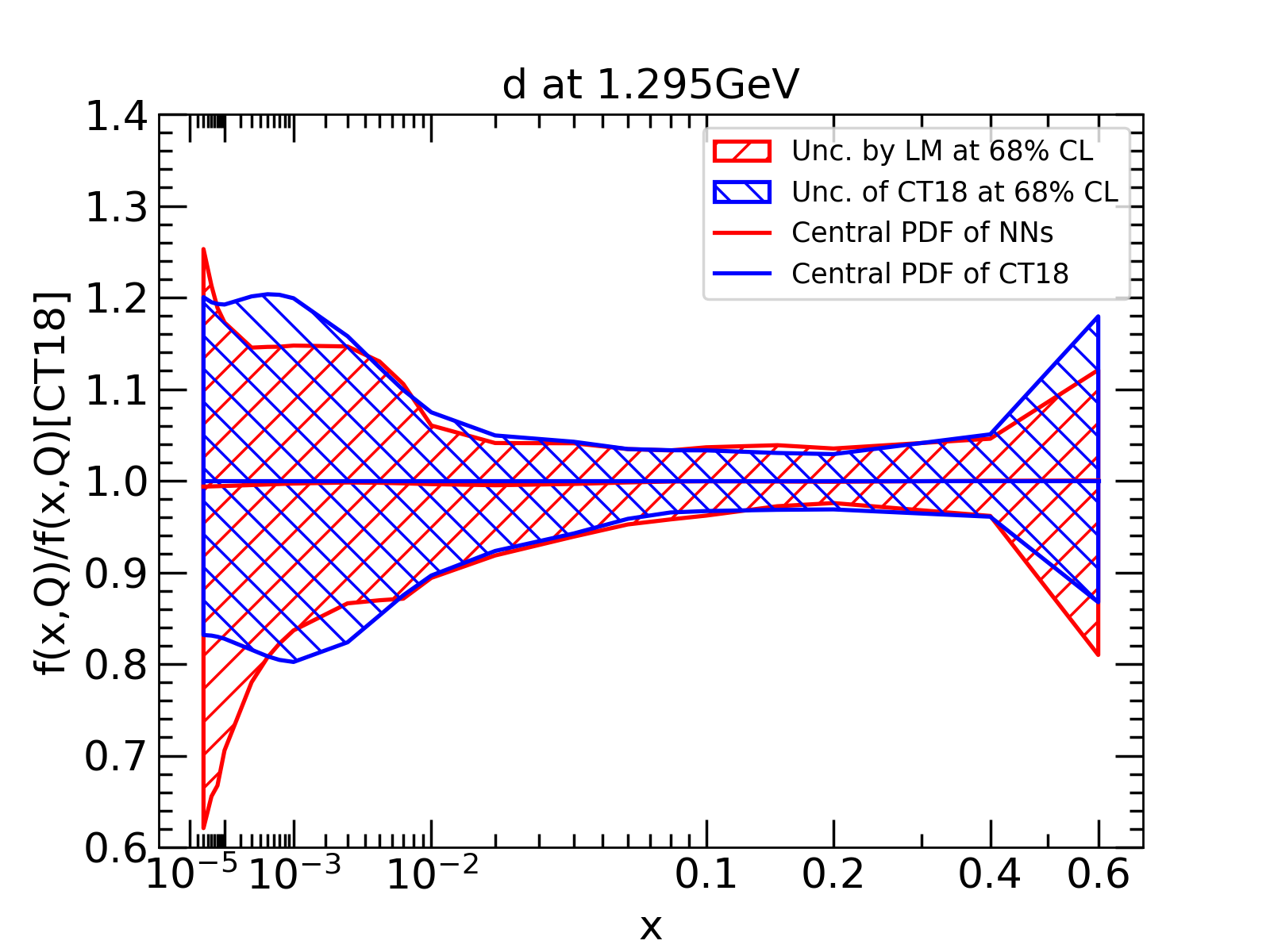}}
  \subcaptionbox{}[6.9cm]
    {\includegraphics[width=6.9cm]{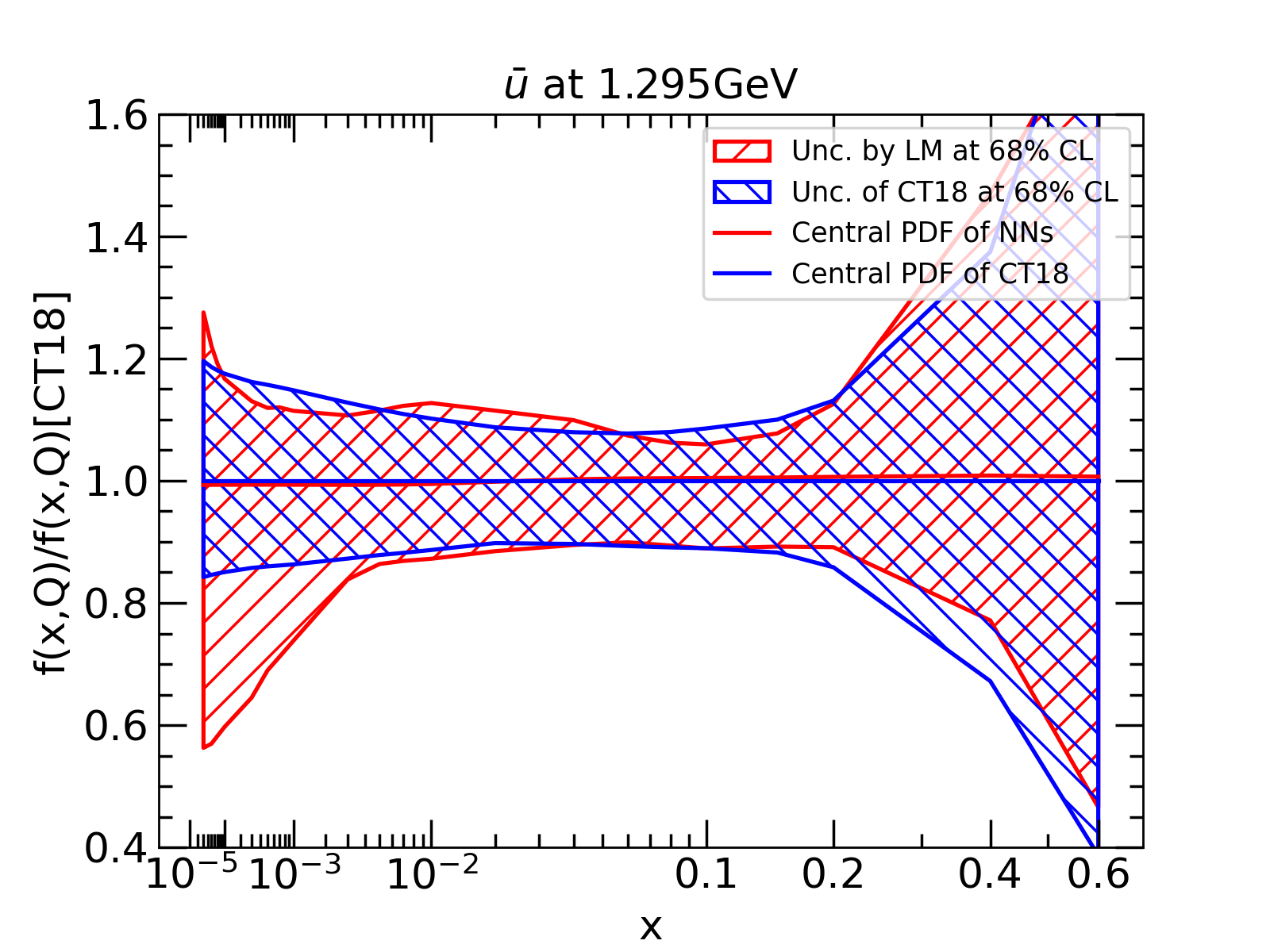}}
  \subcaptionbox{}[6.9cm]
    {\includegraphics[width=6.9cm]{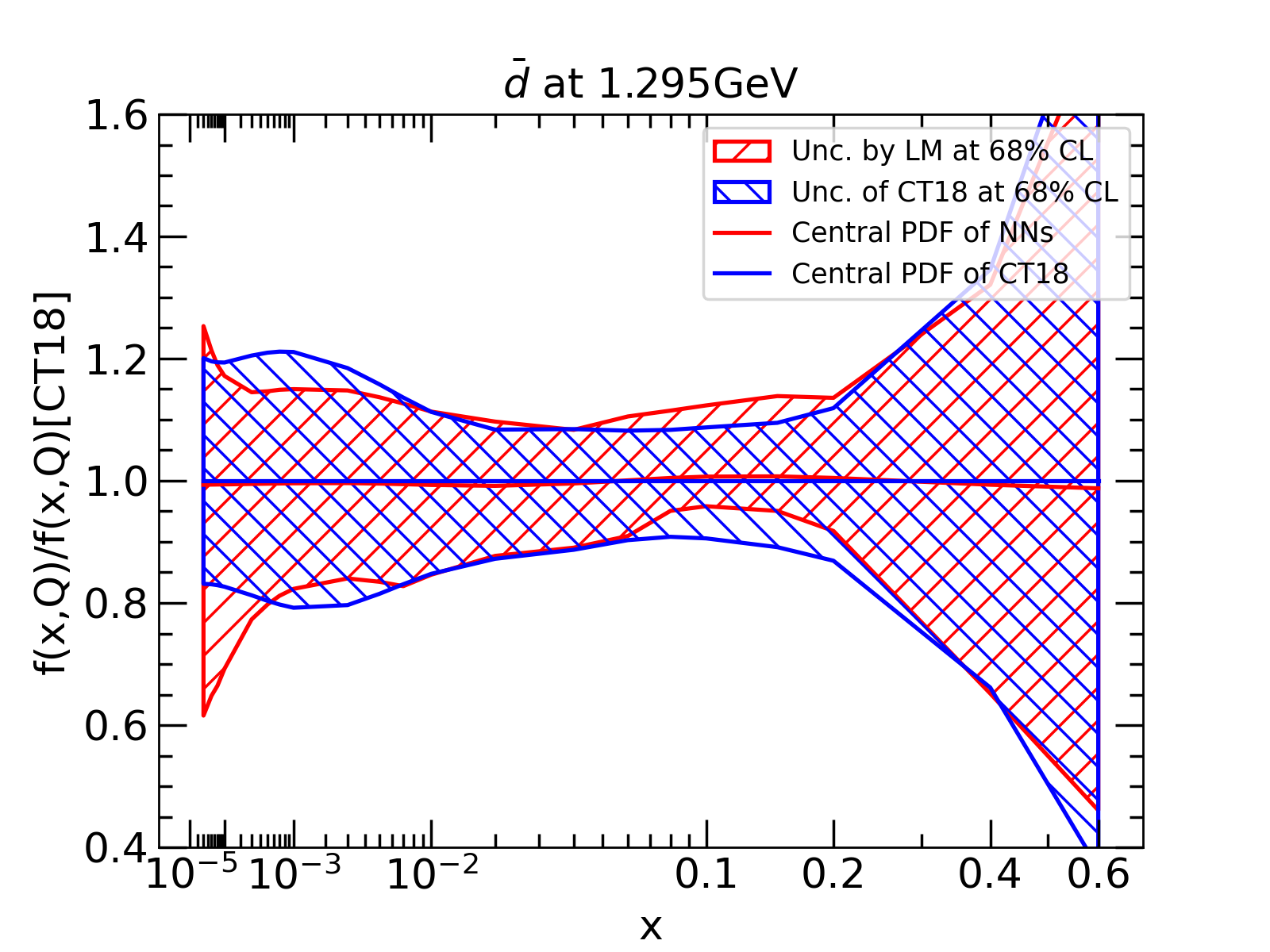}}
  \subcaptionbox{}[6.9cm]
    {\includegraphics[width=6.9cm]{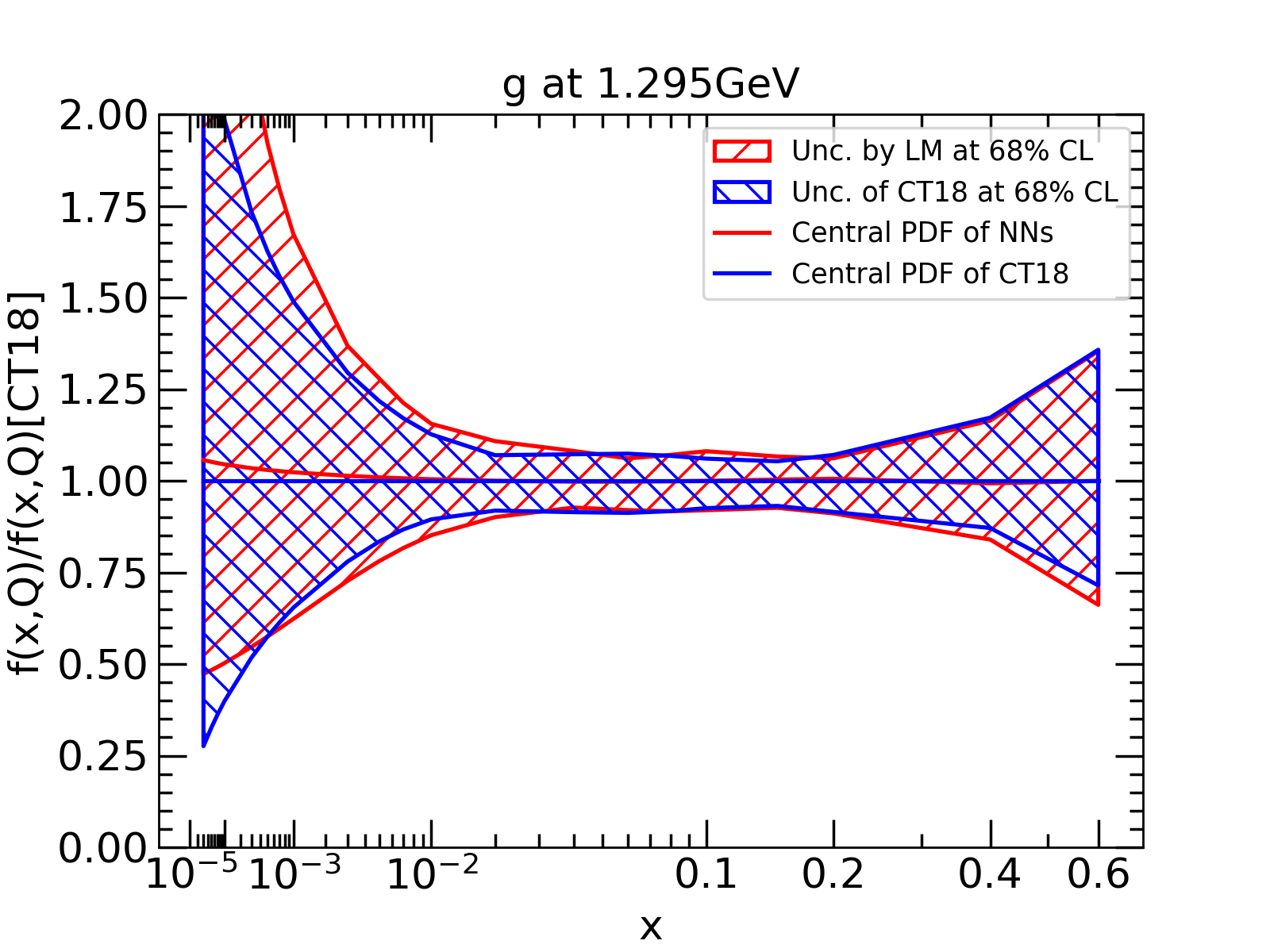}}
  \subcaptionbox{}[6.9cm]
    {\includegraphics[width=6.9cm]{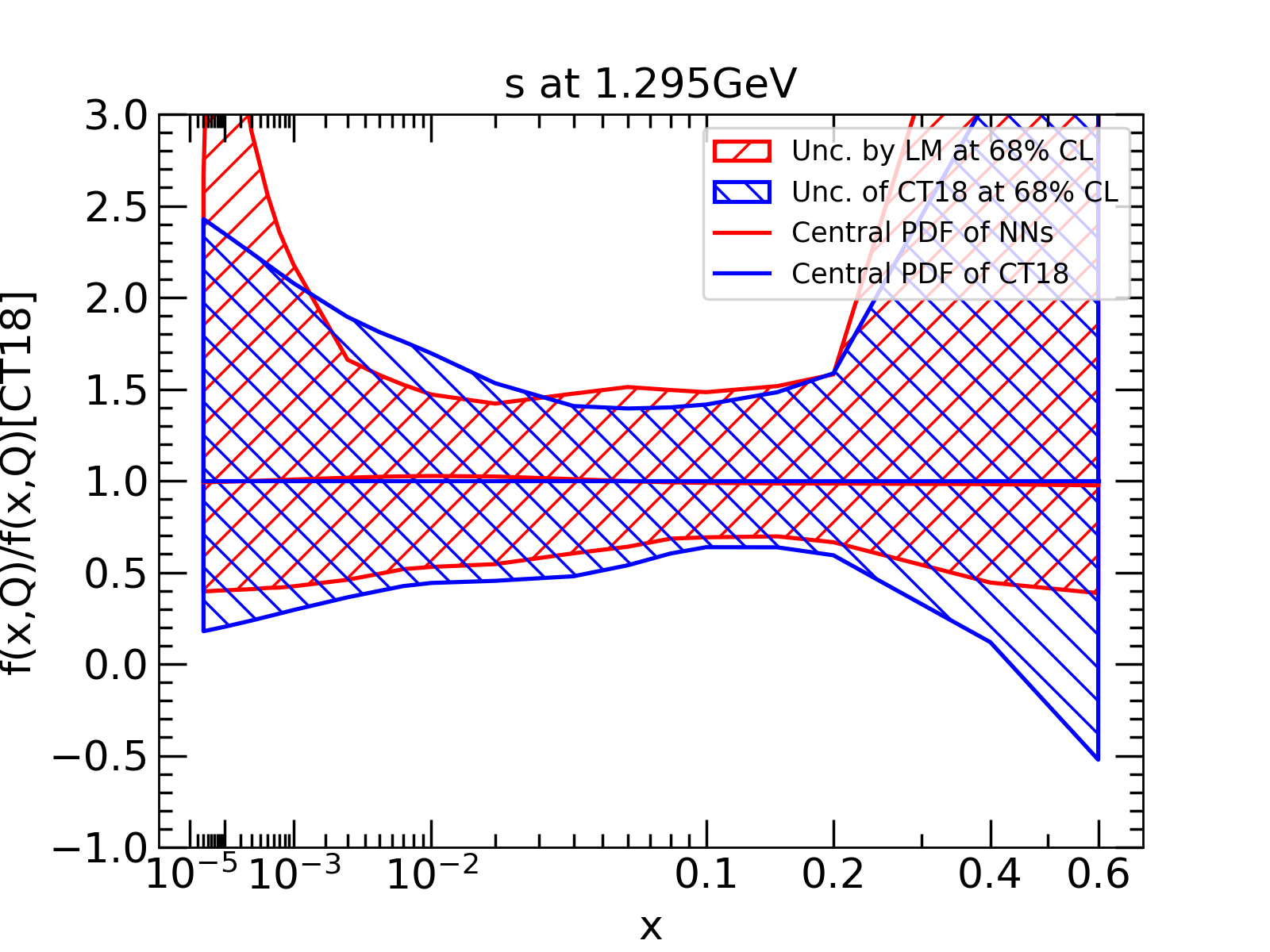}}
  \caption{The parton distribution functions at $Q = 1.295$ GeV for $u$, $d$, $\bar{u}$, $\bar{d}$, $g$ and $s$. 
  The blue and the red solid lines represent the central values of the CT18 NNLO and the PDFs determined with the aforementioned NNs approximation of $\chi^2$ respectively.
The blue and the red hatched areas represent the uncertainties at 68\% CL determined with the Hessian method and the LM scans respectively.
}
\label{Fig:LM_total}
\end{figure}

In Fig.~\ref{Fig:LM_total}, we compare the PDF uncertainties at 68\% CL at $Q$ = 1.295 GeV between the LM method and the Hessian uncertainties from the CT18 NNLO PDFs.
The blue and the red solid lines represent the central values of the CT18 NNLO and the PDFs determined with the aforementioned NNs approximation of $\chi^2$ respectively.
The blue and the red hatched areas represent the uncertainties determined with the Hessian method and the LM method respectively.
The results are normalized to the central value of CT18 NNLO PDFs. 
We find good agreements of both the uncertainties and the central values between the two methods.
A notable difference, however, can be seen for $u$, $d$, $\bar{u}$, $\bar{d}$ and $s$-quark at small-$x$ ($\lesssim 10^{-4}$), as well as for $d$, $\bar{u}$ and $s$-quark at large-$x$ ($\gtrsim 0.4$).
This indicates a failure of the quadratic approximation in these regions.
The uncertainties from the LM method can be either larger or smaller than the uncertainties from the Hessian method depending on the flavor and the $x$ value.

In the lower-right panel, we find that $s$-quark PDFs have large uncertainties for both $x \lesssim 0.001$ and $x \gtrsim 0.4$.
This is because the large-$x$ and small-$x$ behavior of the $s$-quark are mostly constrained by the extrapolation of the PDF parametrization.
The error band of $s$-quark of CT18 NNLO PDFs covers negative PDF values at $x \gtrsim 0.4$. This unphysical behavior implies a limitation of the Hessian method.
On the contrary, the error band determined with the LM method is bounded above zero in all regions.

\begin{figure}[htbp]
  \centering
  \subcaptionbox{Hessian}[7.7cm]
    {\includegraphics[width=7.7cm]{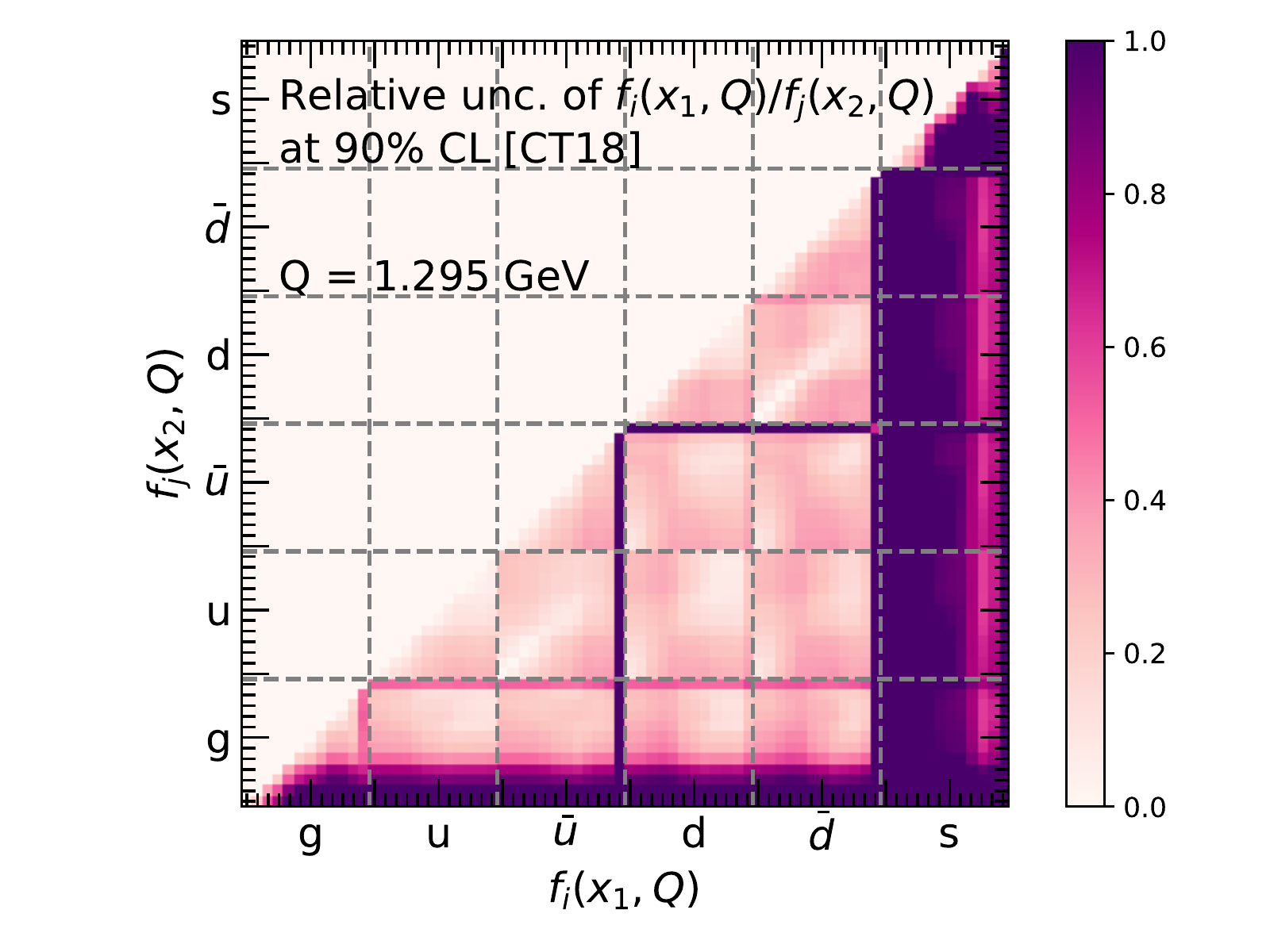}}
  \subcaptionbox{LM}[7.7cm]
    {\includegraphics[width=7.7cm]{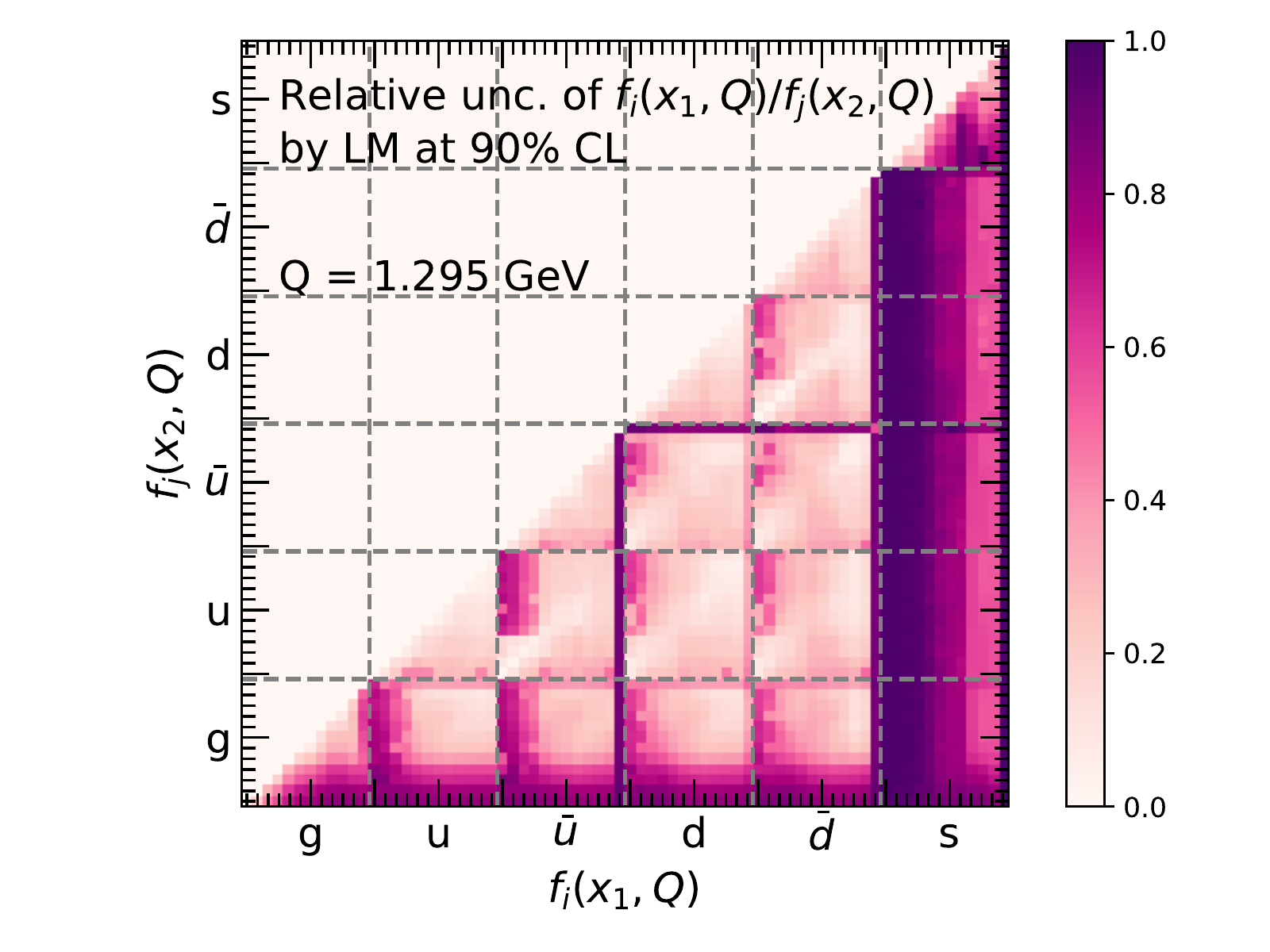}}
  \caption{The relative uncertainties of $R_f = f_{i}(x_1,Q)/f_{j}(x_2,Q)$ determined with the Hessian method and the LM method at 90\% CL are shown in panel (a) and (b) respectively.
The color code represents the relative uncertainties of the ratio $R_{f}$.
The relative uncertainties are calculated at $Q = 1.295$ GeV with $x_1$ and $x_2$ selected among 12 values from 3$\times 10^{-5}$ to 0.6 listed in Table~\ref{tab:x_value_ratio}, and $i, j\in \{g, u, \bar{u}, d, \bar{d}, s\}$ runs over all parton flavors.
}
  \label{Fig:delta_r}
\end{figure}

We also perform the LM scans on the general PDF ratios that is defined as
\begin{equation}
R_{f} \equiv \frac{f_{i}(x_1,Q)}{f_{j}(x_2,Q)}.
\label{eq:r}
\end{equation}
The relative uncertainties of $R_f$ are calculated at $Q = 1.295$ GeV with $x_1$ and $x_2$ selected among 12 values from 3$\times 10^{-5}$ to 0.6 listed in Table~\ref{tab:x_value_ratio}, and $i, j\in \{g, u, \bar{u}, d, \bar{d}, s\}$ runs over all parton flavors.
The results at 90\% CL are shown in Fig.~\ref{Fig:delta_r} (a) and (b) for the Hessian and LM method respectively.
The x and the y axis indicate the numerator and the denominator, and color code represents the relative uncertainties of
the ratio $R_{f}$.
By comparison of the two panels, we find good agreements between the uncertainties determined with the LM method and
the Hessian method in most regions.
Similar to Fig.~\ref{Fig:LM_total}, there are notable differences at small-$x_1$, especially for those with sea quarks
in the numerator.

\begin{table}[htpb]
  \centering
  \begin{tabular}{c|cccccc}
  \hline
   & 1 & 2 & 3 & 4 & 5 & 6\\
  \hline
  $x$ &  $3 \times 10^{-5}$ & $7 \times 10^{-5}$ & $3 \times 10^{-4}$ & $7 \times 10^{-4}$ & $ 3\times 10^{-3}$ & $7 \times 10^{-3}$ \\
  \hline
  \hline
   & 7 & 8 & 9 & 10 & 11 & 12 \\
  \hline 
  $x$ & 0.01 & 0.02 & 0.06 & 0.1 & 0.2 & 0.6 \\
  \hline
  \end{tabular}
  \caption{The $x$ values selected for the calculation of the uncertainties of $R_f$.} 
  \label{tab:x_value_ratio}
\end{table}

\subsection{LM scans on cross sections}
Higgs bosons are produced dominantly through gluon fusions at the LHC.
The inclusive gluon-fusion cross-section has been calculated to next-to-next-to-next-to-leading order in QCD~\cite{Anastasiou:2015vya},
which further reduces the scale variations and makes the PDF uncertainties even more important.
Besides, the production of Higgs boson pair and top-quark pair associated with a Higgs boson are of equal
importance for studies of the Higgs boson self-coupling and the top-quark Yukawa coupling. 

\begin{figure}[htbp]
  \centering
  \subcaptionbox{}[7.7cm] 
    {\includegraphics[width=7.7cm]{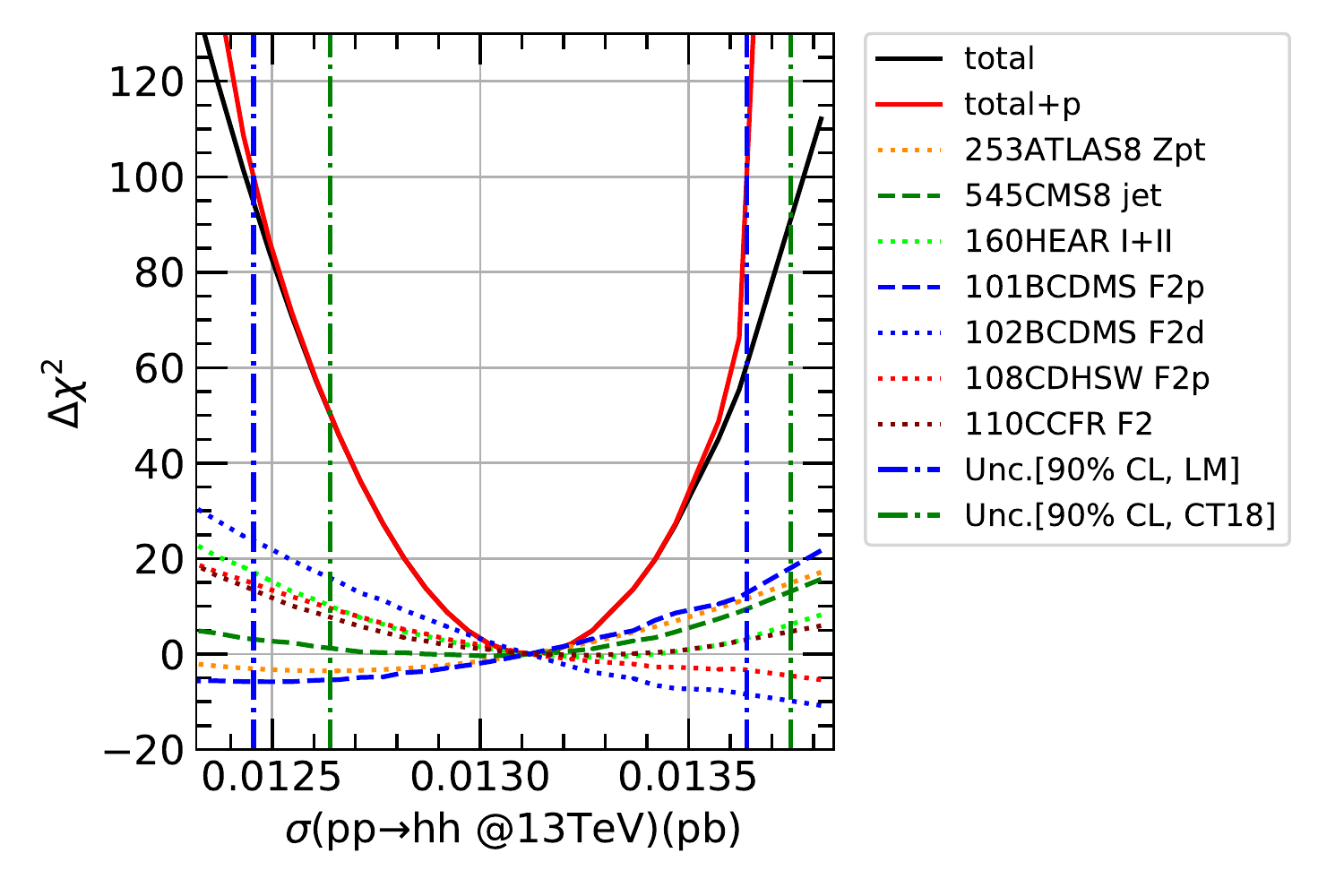}}
  \subcaptionbox{}[7.7cm]
    {\includegraphics[width=7.7cm]{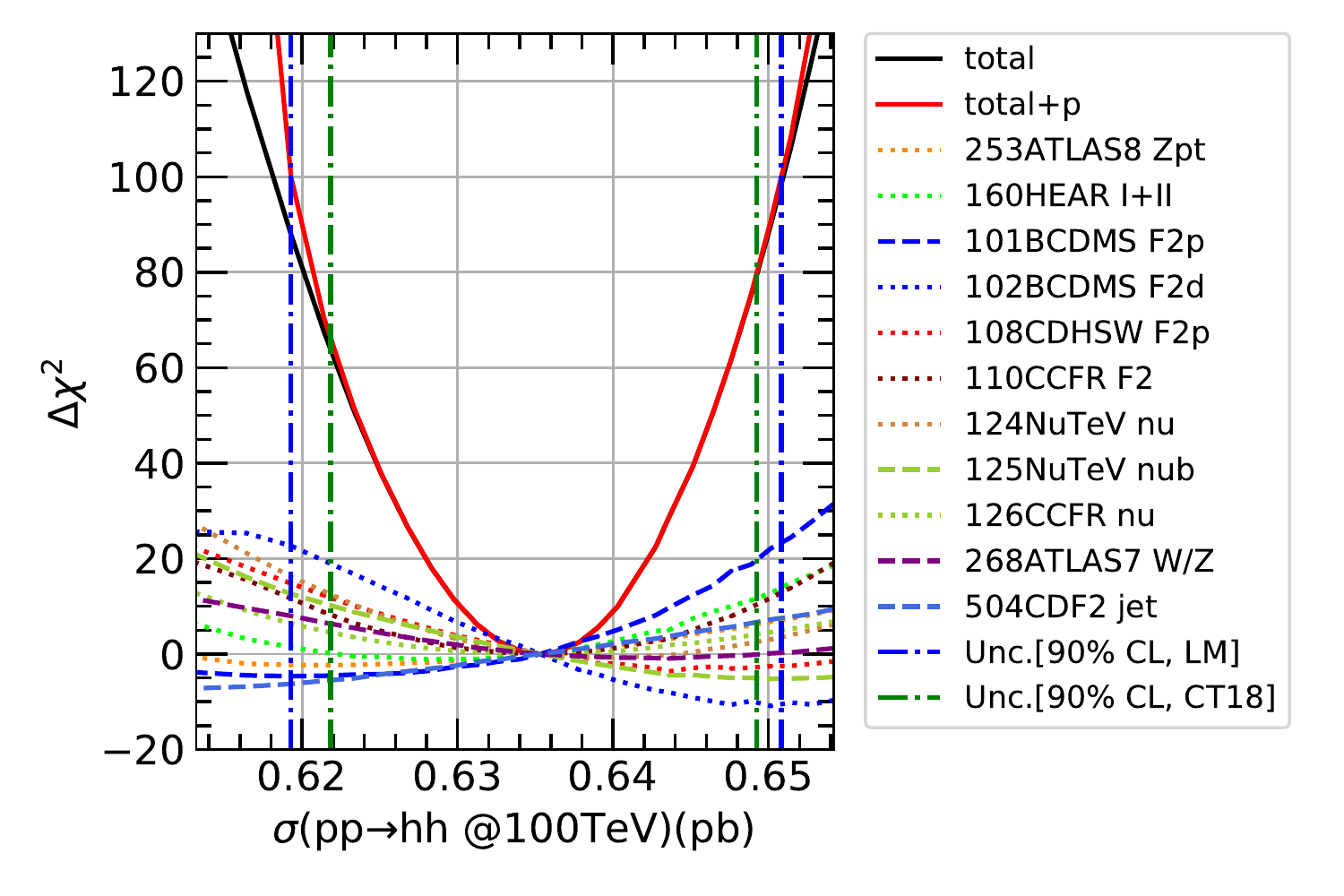}}    
  \subcaptionbox{}[7.7cm]
    {\includegraphics[width=7.7cm]{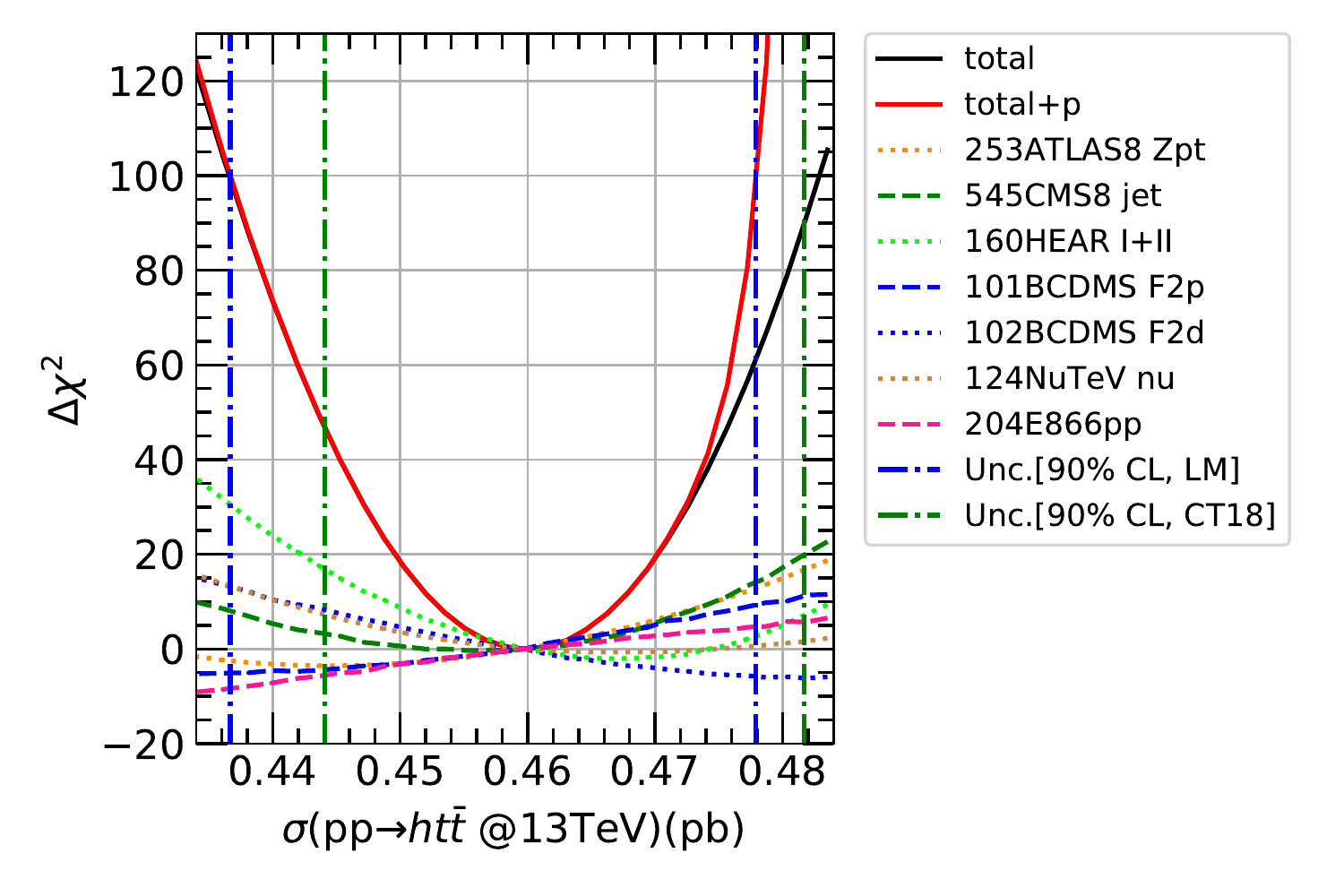}}
  \subcaptionbox{}[7.7cm]
    {\includegraphics[width=7.7cm]{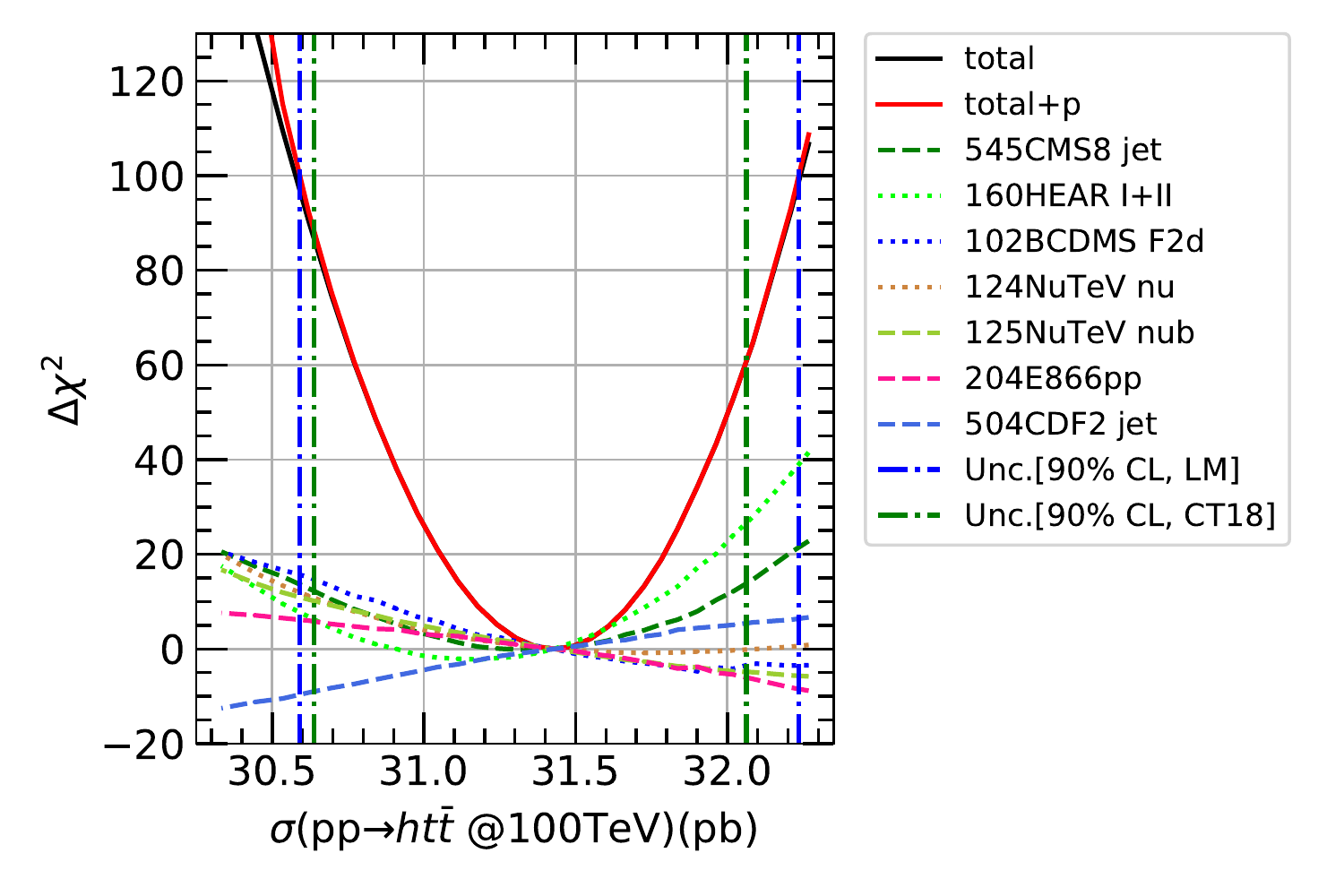}}    
  \caption{LM scans on the $\sigma_{pp\to hh}$ and $\sigma_{pp\to ht\bar{t}}$ at $\sqrt{s}=$ 13 TeV or 100 TeV.}
  \label{Fig:LM_cs_hh}
\end{figure}

In Fig.~\ref{Fig:LM_cs_hh} we show the results of LM scans on $\sigma_{pp \to hh}$ and $\sigma_{pp \to ht\bar{t}}$ at $\sqrt{s}$ = 13 TeV or 100 TeV. 
For $\sigma_{pp \to hh}$ at $\sqrt{s}$ = 13 TeV, in the upper-left panel, the behaviors of $\chi^2$ are very much similar to that shown in
the upper-left panel of Fig.~\ref{Fig:LM_g} for the gluon PDF.
That is because the cross section of $pp \to hh$ at 13 TeV is strongly correlated with the gluon PDF at $x\sim 0.02$.
Constraints from HERA inclusive DIS data, BCDMS proton and deuterium data, CMS 8 TeV jet data and ATLAS 8 TeV Z $p_T$ data stand out as expected.
In addition, the BCDMS proton data and ATLAS 8 TeV Z $p_T$ data both prefer a larger cross section contrasted with the BCDMS deuterium
data which prefers a smaller value.
For $\sigma_{pp \to hh}$ at $\sqrt{s}$ = 100 TeV, in the upper-right panel, the constraints are distributed among more data sets
and are related to PDFs at small-$x$.

The cross section of $pp \to ht\bar{t}$ mainly depends on the gluon, $u$-quark and $d$-quark PDFs.
For $\sigma_{pp \to ht\bar{t}}$ at $\sqrt{s}$ = 13 TeV, in the lower-left panel of Fig.~\ref{Fig:LM_cs_hh}, similar behaviors of the $\chi^2$ as
$\sigma_{pp \to hh}$ are observed. 
At $\sqrt{s}$ = 100 TeV, the constraints from HERA inclusive DIS data predominate.
In addition, constraints from NuTeV dimuon data, CMS jet data and BCDMS proton data also
play important roles.

\subsection{Study on impact of individual data sets}

In order to assess the contribution from an individual experimental data set, we remove one data set at a time, and repeat the LM scans
on physics quantities with the rest of the data sets.
Difference between the fit with and without the data set can be an assessment of its contribution.

\begin{figure}[htbp]
  \centering
  \includegraphics[width=0.95\textwidth,clip]{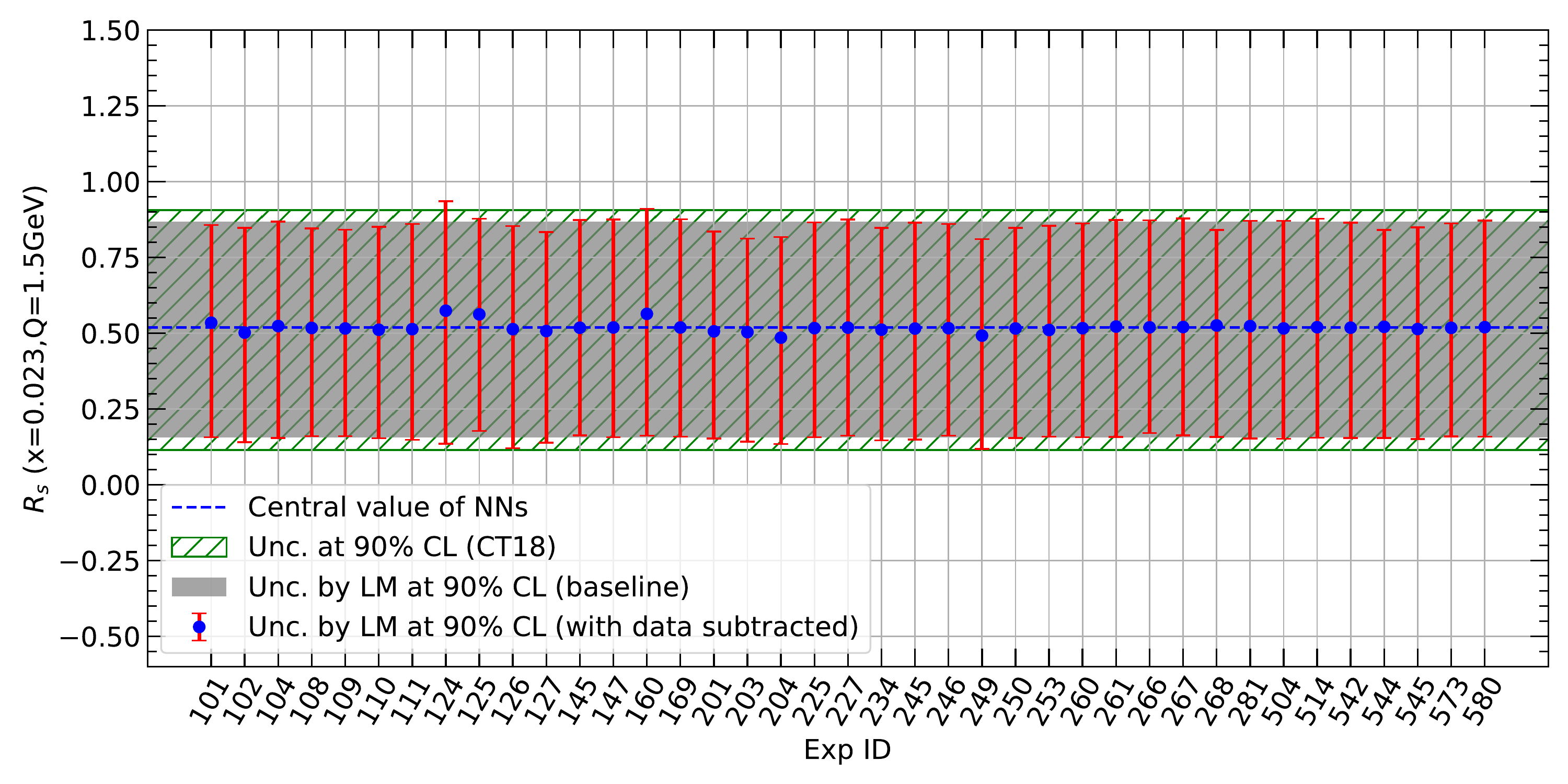}
  \hfill
  \caption{The results of LM scans on the $R_s$ ($x = 0.023$, $Q = 1.5$ GeV) with data subtracted.
  The horizontal axis represents the experimental data set removed from the LM scans.
  The blue mark and the red error bar respectively indicate the central value and uncertainties at 90\% CL determined with the LM method with the rest of the data sets.
  The green hatched area and the gray band represent the uncertainties at 90\% CL determined with the Hessian method and the LM method with the full data set respectively.
  }
  \label{Fig:indi_rs1}
\end{figure}

The results for $R_s$ at $x = 0.023$ and $Q = 1.5$ GeV are shown in Fig.~\ref{Fig:indi_rs1}.
After the removal of each data set, we find that $R_s$ value and its uncertainty are only changed slightly,
as represented by those error bars comparing to the uncertainty from LM scans with the full data set represented by
the gray band.
The subtraction of a single data set shows largest effects for NuTeV dimuon production data (Exp. ID = 124, 125), CCFR dimuon production data (Exp. ID = 126, 127), E866 Drell-Yan data (Exp. ID = 204) and HERA inclusive DIS data (Exp. ID = 160).
In addition, the NuTeV dimuon production data and HERA inclusive DIS data prefer a smaller $R_s$ contrasted with E866 Drell-Yan data which prefers a larger value, that is consistent with the bottom-right panel of Fig.~\ref{Fig:LM_g}.

\begin{figure}[htbp]
  \centering
  \includegraphics[width=0.95\textwidth,clip]{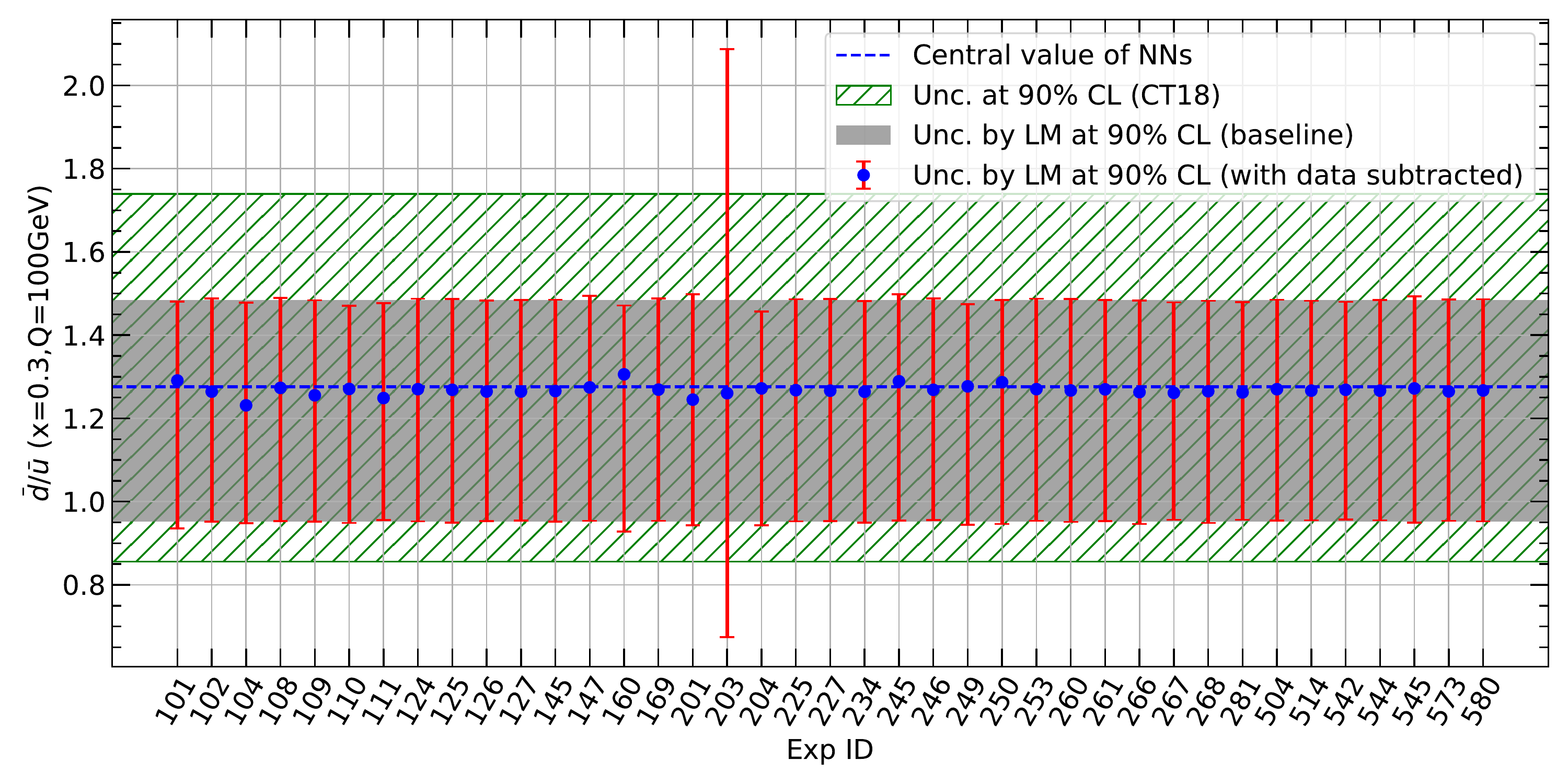}
  \hfill
  \caption{The same as Fig.\ref{Fig:indi_rs1}, but for the results of LM scans on the $\bar{d}/\bar{u}$ ( $x = 0.3$ and $Q = 100$ GeV).}
  \label{Fig:indi_udbar2}
\end{figure}

In Fig.~\ref{Fig:indi_udbar2} we show the results for $\bar{d}/\bar{u}$ at $x = 0.3$ and $Q = 100$ GeV.
The E866 Drell-Yan ratio data (Exp. ID = 203) gives the dominant constraints.
The fit without E866 Drell-Yan ratio data predicts a result of $\bar{d}/\bar{u} = 1.26^{+0.82}_{-0.59}$, while the fit with the full data set expects $\bar{d}/\bar{u} = 1.28^{+0.20}_{-0.33}$.
After the inclusion of E866 Drell-Yan ratio data, the uncertainties of $\bar{d}/\bar{u}$ are reduced by almost 60\%.
That is because the penalty term of E866 Drell-Yan ratio data provides a strong constraint on $\bar{d}/\bar{u}$.
In addition, constraints from NMC deuteron data (Exp. ID = 104) and HERA inclusive DIS data also play important roles.

\begin{figure}[htbp]
  \centering
  \includegraphics[width=0.95\textwidth,clip]{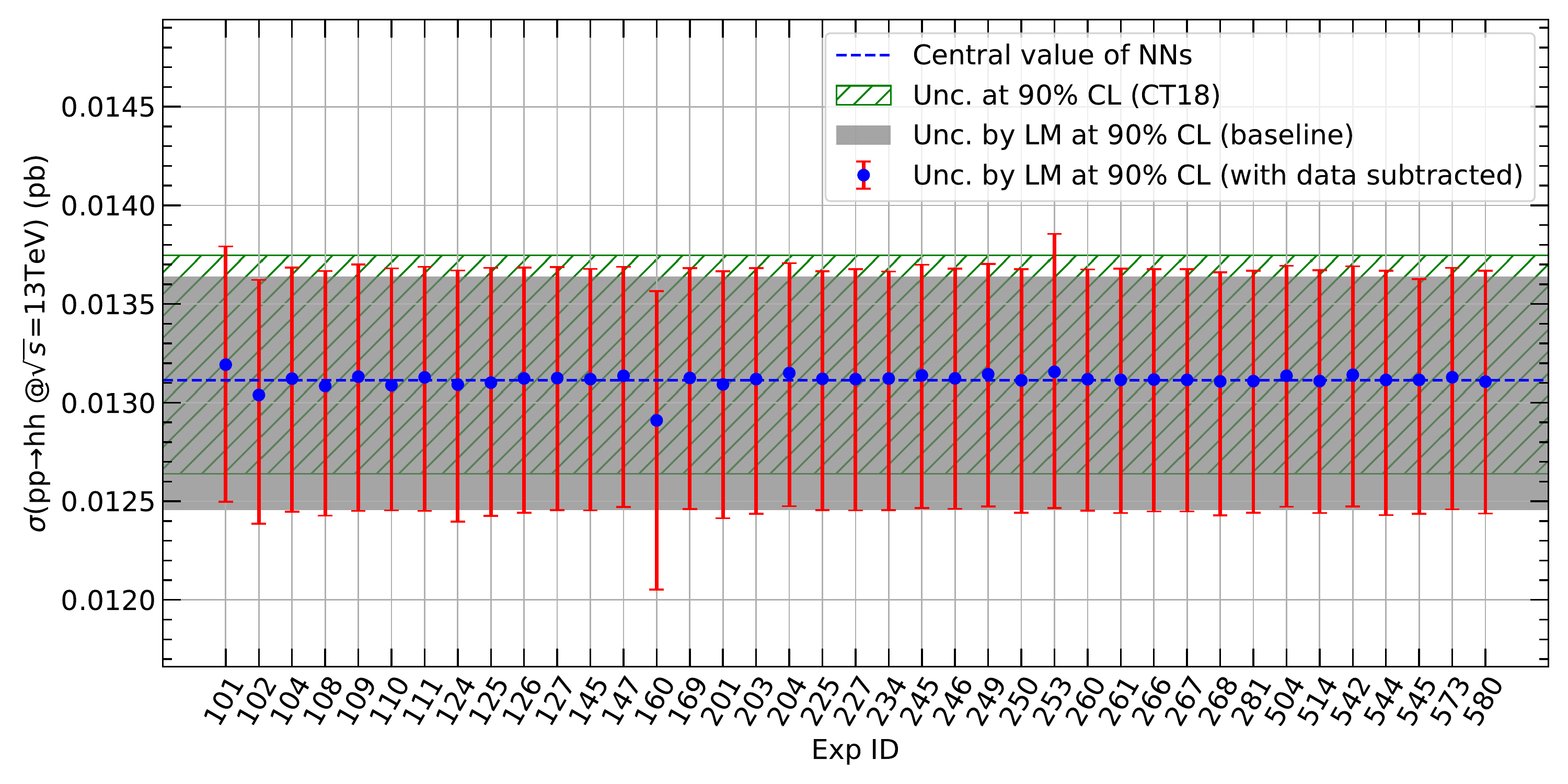}
  \hfill
  \caption{The same as Fig.\ref{Fig:indi_rs1}, but for the results of LM scans on the $\sigma_{pp\to hh}$ at $\sqrt{s}$ = 13 TeV.}
  \label{Fig:indi_hh13}
\end{figure}

In Fig.~\ref{Fig:indi_hh13} we show the results for $\sigma_{pp\to hh}$ at $\sqrt{s}$ = 13 TeV.
The constraints from HERA inclusive DIS data predominate as expected.
In addition to that, constraints from BCDMS proton and deuterium data (Exp. ID = 101, 102) and ATLAS 8 TeV Z $p_T$ data (Exp. ID = 253) also play important roles.
The fit without HERA inclusive DIS data expects $\sigma_{pp\to hh} = 0.0129^{+0.0007}_{-0.0009}$ pb, while the fit with the full data set gives $\sigma_{pp\to hh} = 0.0131^{+0.0005}_{-0.0007}$ pb.
An upward shift of about $2\times10^{-4}$ pb is observed when we incorporate HERA inclusive DIS data, and the uncertainties of $\sigma_{pp\to hh}$ are reduced by almost 20\%.
In addition, the HERA inclusive DIS data and BCDMS deuterium data both prefer a larger $\sigma_{pp\to hh}$ contrasted with BCDMS proton data 
and ATLAS Z $p_T$ data which prefer a smaller value, that is consistent with the upper-left panel of Fig.~\ref{Fig:LM_cs_hh}.

\subsection{Two-Dimensional LM scans}

Besides the PDF uncertainties, it is possible to quantify other statistical estimators such as the correlation
between two physics quantities with two-dimensional LM (2-D LM) scans.
That can be achieved by adding a second physics quantity into Eq.~(\ref{eq:LM}).
The new function that needs to minimized in the global fit becomes
\begin{equation}
\Psi\left(\lambda_1,\lambda_2,\left\{a_{i}\right\}\right) \equiv \chi^{2}\left(\left\{a_{i}\right\}\right)+\lambda_{1} {X_{1}}\left(\left\{a_{i}\right\}\right)+\lambda_{2} {X_{2}}\left(\left\{a_{i}\right\}\right) ,
\label{eq:LM_2d}
\end{equation}
where $\lambda_1$ and $\lambda_2$ are specified constants, and $X_{1}(\{a_i\})$ and $X_{2}(\{a_i\})$ represent the two physics quantities of interest.
Similar to Eq.~(\ref{eq:LM}), the constrained minimum of $\chi^2$ from the global fit depends on
$X_1$ and $X_2$, and can be written as $\chi^2=\chi^2_{min}+\Delta\chi^2$, where $\chi^2_{min} = \chi^2 (\lambda_1=0, \lambda_2=0)$.
The contour of $\Delta\chi^2 + P$ in the plane of $X_{1}$ vs. $X_{2}$ can be an assessment of the correlation between $X_{1}$ and $X_{2}$.

\begin{figure}[htbp]
  \centering
  \subcaptionbox{}[7.7cm]
    {\includegraphics[width=7.7cm]{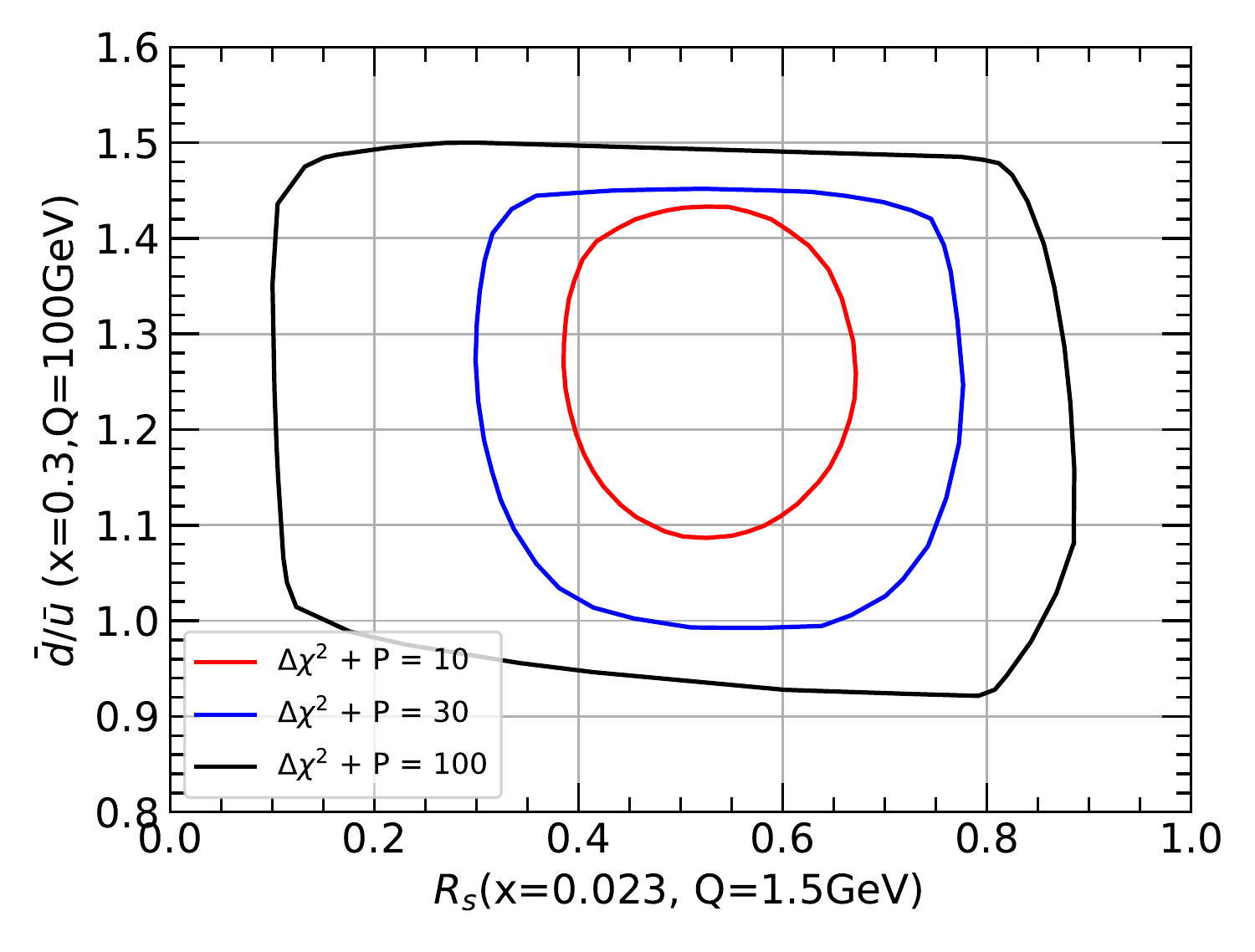}}
  \subcaptionbox{}[7.7cm]
    {\includegraphics[width=7.7cm]{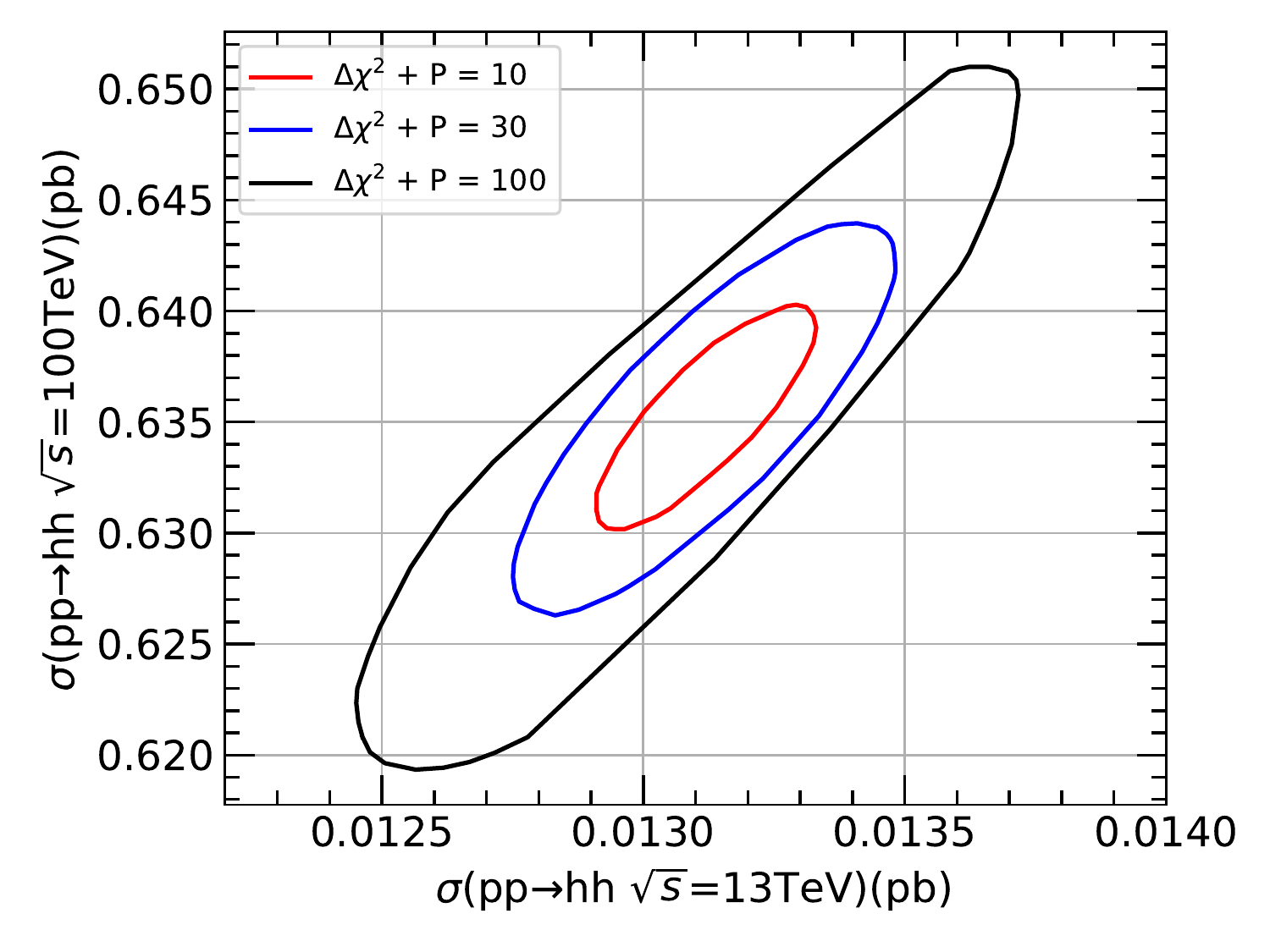}}
  \caption{Contour plot of $\Delta \chi^2$ plus Tier-2 penalty term on the plane of $\bar{d}/\bar{u}$ ($x = 0.3$, $Q = 100$ GeV) vs. $R_s$ ($x = 0.023$, $Q = 1.5$ GeV) 
  and $\sigma_{pp \to hh}$ ($\sqrt{s}$ = 13 TeV) vs. $\sigma_{pp \to hh}$ ($\sqrt{s}$ = 100 TeV).}
  \label{Fig:LM_2d}
\end{figure}

As examples in Fig.~\ref{Fig:LM_2d} we show contours for $\bar{d}/\bar{u}$ ($x = 0.3$, $Q = 100$ GeV) vs. $R_s$ ($x = 0.023$, $Q = 1.5$ GeV) and $\sigma_{pp \to hh}$ ($\sqrt{s}$ = 13 TeV) vs. $\sigma_{pp \to hh}$ ($\sqrt{s}$ = 100 TeV) determined with the 2-D LM scans.
In the left panel, a weak correlation between strangeness ratio $R_s$ and $\bar{u}/\bar{d}$ ratio is observed. 
That is because the two quantities are dominantly constrained by different experimental data sets.
At small $\Delta\chi^2 + P$ the contour shows an elliptic shape. 
When $\Delta\chi^2 + P$ gets larger, the shape of the contour becomes irregular due to the increase
of penalty term contributions.
On the contrary, the right panel demonstrates a strong correlation between $\sigma_{pp \to hh}$ ($\sqrt{s}$ = 13 TeV) and $\sigma_{pp \to hh}$ ($\sqrt{s}$ = 100 TeV) since both processes are sensitive to gluon PDFs and constrained by the relevant experimental data sets.

\section{Applications}\label{sec:application}
In this section, we evaluate the impact of the NOMAD measurements and of two pseudo-data sets of HL-LHC on PDFs based on the new approach.
In addition, we study constraints on the new physics with a joint fit of both
PDFs and the Wilson coefficient of lepton-quark contact interactions in the
framework of the SMEFT.

\subsection{Constraint from NOMAD data}\label{subsec:nomad}

The charm-quark production in CCDIS process provides a unique sensitivity to the strange-quark
distribution in the nucleon, with a clean signal of two muons with opposite charges in the final state.
Recently, NOMAD collaboration reported a measurement of dimuon production in the neutrino-iron scattering experiment~\cite{NOMAD:2013hbk}.
A sample of about $9\times10^6$ inclusive CCDIS events, including 15344 dimuon events, is collected, providing a reduced statistical uncertainty.
Observables are taken to be the ratios of dimuon to inclusive cross-sections,
which provides a large cancellation of the common systematic uncertainties presented in
both the numerator and the denominator.
Final results are distributed among three differential variables: the reconstructed
neutrino energy $E_\nu$, the Bjorken $x$ and the partonic center of mass energy $\sqrt{\hat{s}}$. 
By the supplement of data from NOMAD, the improvement in the constraint on $s$-quark PDFs
are studied in this section using the same NNs approach on $\chi^2$ mentioned in previous sections.
On the theoretical side, structure functions in S-ACOT-$\chi$ general mass scheme up to NNLO are constructed,
so that a full consideration of the charm-quark mass is included~\cite{Berger:2016inr,Gao:2017kkx,Gao:2021fle}.
Predictions of inclusive CCDIS and open charm production cross-sections are made from these constructions,
and dimuon cross sections are derived by further applying the inclusive decay branching ratio of charm quark to
muon.
The significant uncertainties of the decay branching ratio contribute as one of the dominant
systematic errors on the dimuon cross sections, which are summarized in Appendix~\ref{sec:vano}.

In Fig.~\ref{Fig:nomad_exp_vs_theo} we show comparison of NOMAD data and our predictions at
both NLO and NNLO, as well as the Hessian PDF uncertainties at 68\% CL for distributions over
$E_{\nu}$ or $x$.
The PDF uncertainties can be as large as 10\% in most regions.
This directly comes from the large uncertainties of the predictions
of dimuon cross-sections, and can be further traced back to the poor knowledge about $s$-quark PDFs.
In both distributions, most of the data points are consistent with our predictions,
while a significant deviation can be found in the last two points of the distribution over Bjorken $x$.
That can be due to the modeling of heavy nuclear corrections used in the experimental analysis.
We will discard those two data points when including NOMAD data in our later global fit.
It is also noted that the inclusion of NOMAD data to the global fit can improve the consistency with almost no cost of tension with the other data sets~\cite{Alekhin:2014sya,Faura:2020oom}.
Most of the data lie above our NLO predictions of central values.
Given this fact, an increased $s$-quark PDF is expected after the inclusion of NOMAD data, and this increase gets larger due to the negative corrections from NNLO.
\begin{figure}[htbp]
	\centering
	\subcaptionbox{}[7.7cm]
	{\includegraphics[width=7.7cm]{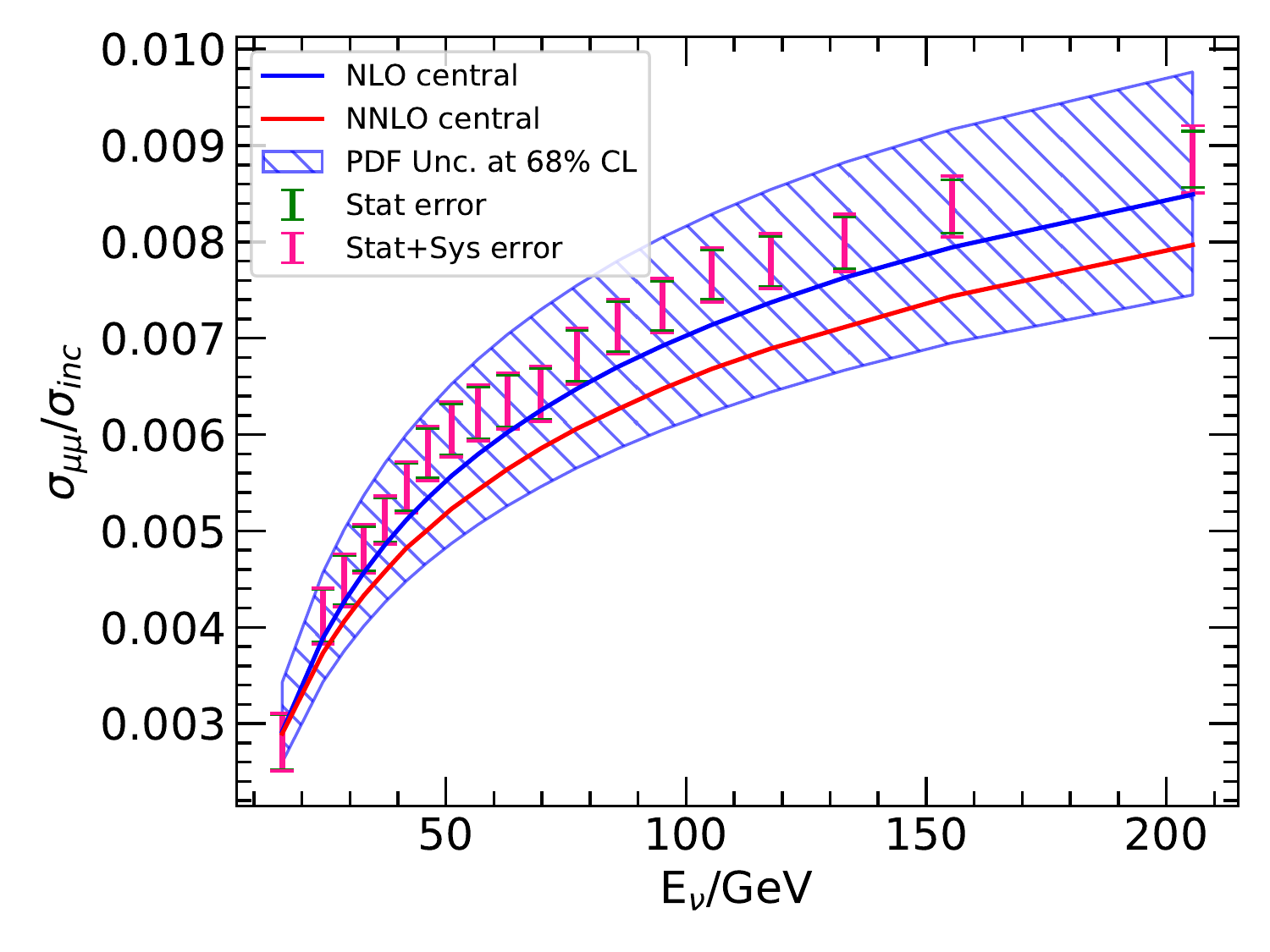}}
	\subcaptionbox{}[7.7cm]
	{\includegraphics[width=7.7cm]{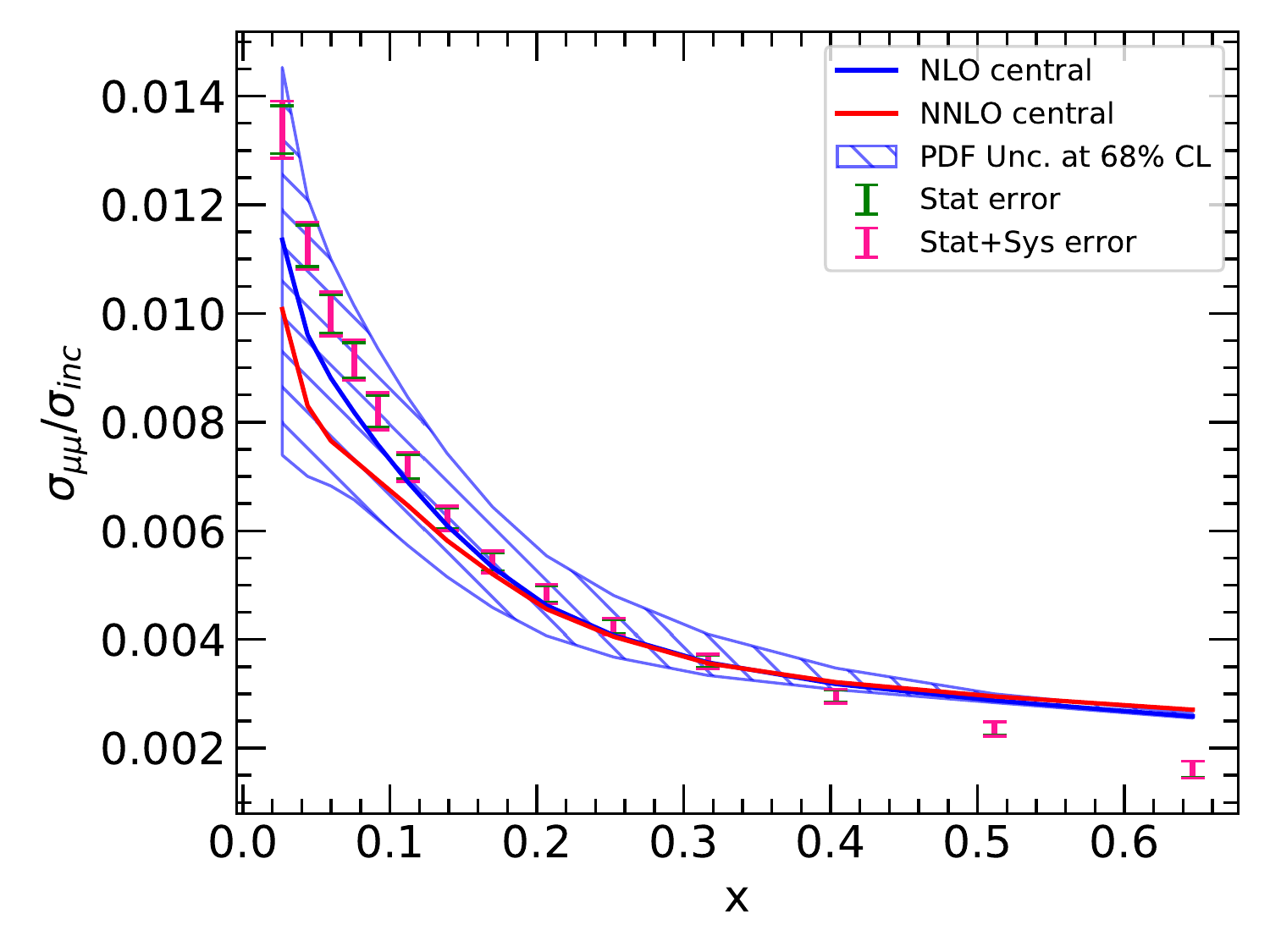}}
	\caption{
	NLO (blue line) and NNLO (red line) predictions for ratios of dimuon to CCDIS inclusive differential cross-sections with respect to neutrino energy (panel~a) and Bjorken $x$ (panel.~b).
	The blue hatched areas represent the Hessian PDF uncertainties of the NLO predictions at 68\% CL.
	NOMAD data are also shown with statistical uncertainties and the combination of both statistical and systematic uncertainties. }
	\label{Fig:nomad_exp_vs_theo}
\end{figure}
\begin{figure}[htbp]
	\centering
	\subcaptionbox{}[7.7cm]
	{\includegraphics[width=7.4cm]{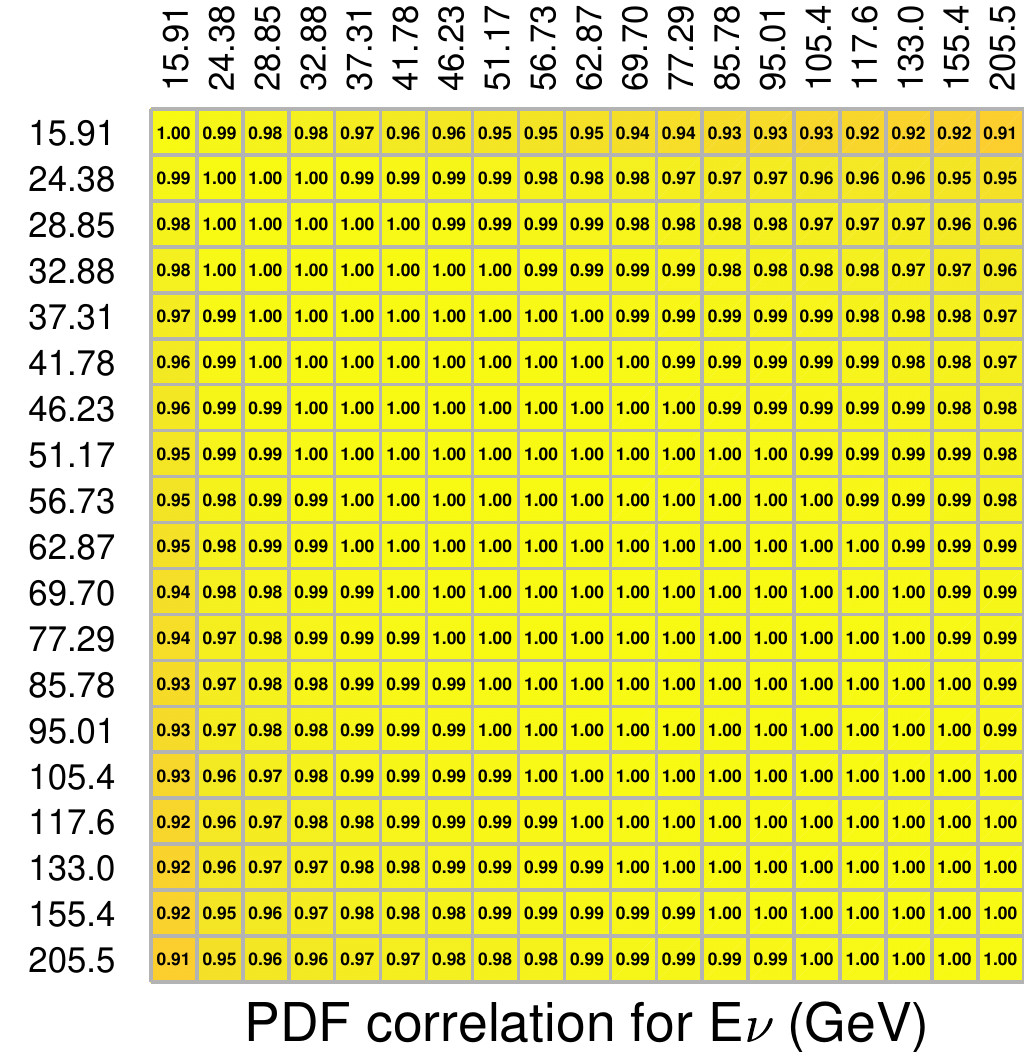}}
	\subcaptionbox{}[7.7cm]
	{\includegraphics[width=7.4cm]{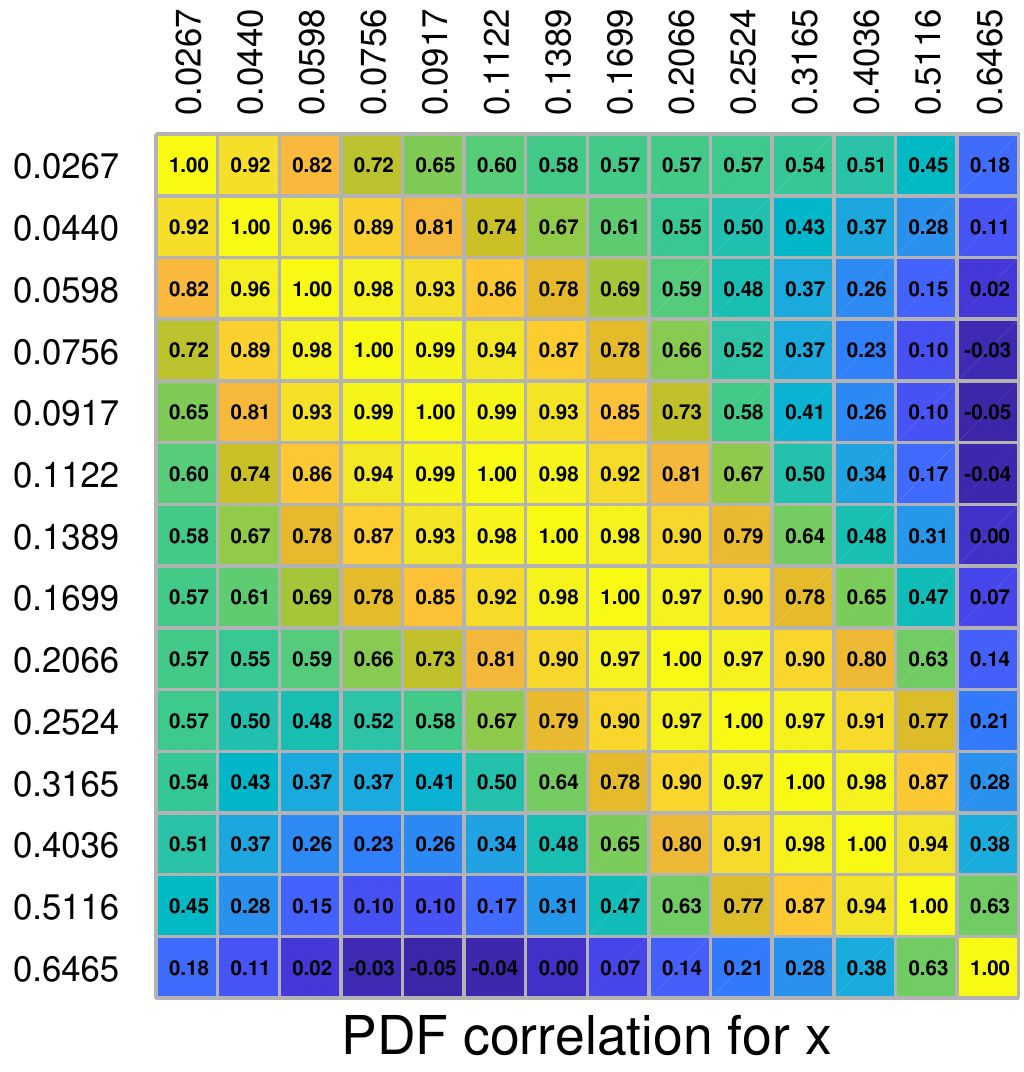}}
	\caption{PDF induced correlations between theory predictions for different experimental
	bins, for NOMAD distribution in neutrino energy (a) and in Bjorken-$x$ (b), calculated with
	CT18 NNLO Hessian PDF set. Numbers in the axis represent the center of each bin, and
    numbers	in the table represent the correlation cosine for each pair of bins.}
	\label{Fig:nomad_56PDF_correlation}
\end{figure}
As mentioned earlier, NOMAD presents measurements on three distributions.
The different sensitivities of distributions over $E_{\nu}$ or $x$ are illustrated in Fig.~\ref{Fig:nomad_56PDF_correlation}.
It shows the PDF induced correlations among bins of each distributions calculated with CT18 NNLO Hessian PDFs.
We find that for $E_{\nu}$ distribution, all data points are strongly correlated, and similar results are found for $\sqrt{\hat s}$ distribution which is not shown here.
Both of them only impose constraints on the overall normalization of the $s$-quark distribution.
Thus their constraints are diluted due to the systematic errors on the inclusive branching ratio of
charm quark to muon (0.094$\pm$0.01).
On the other hand, the correlation pattern is nontrivial for $x$ distribution which imposes
further constraints on the shape of $s$-quark PDFs.
We can not simply combine all these distributions from NOMAD data due to the lack of public statistical correlation
between these distributions.
Hence in the following, only the $x$ distribution is included in our global analysis.

\begin{figure}[htbp]
  \centering
  \subcaptionbox{}[7.7cm]
    {\includegraphics[width=7.7cm]{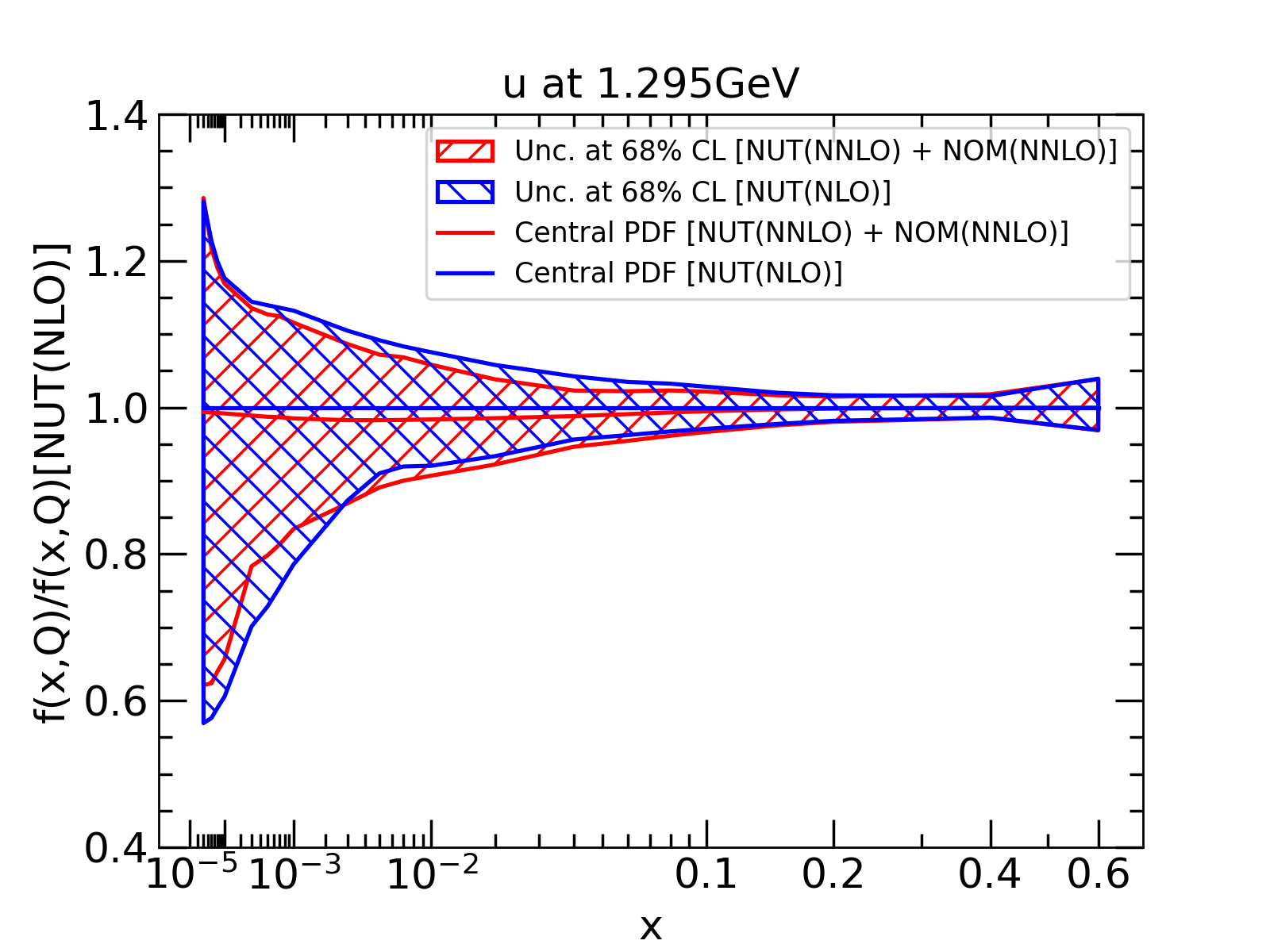}}
  \subcaptionbox{}[7.7cm]
    {\includegraphics[width=7.7cm]{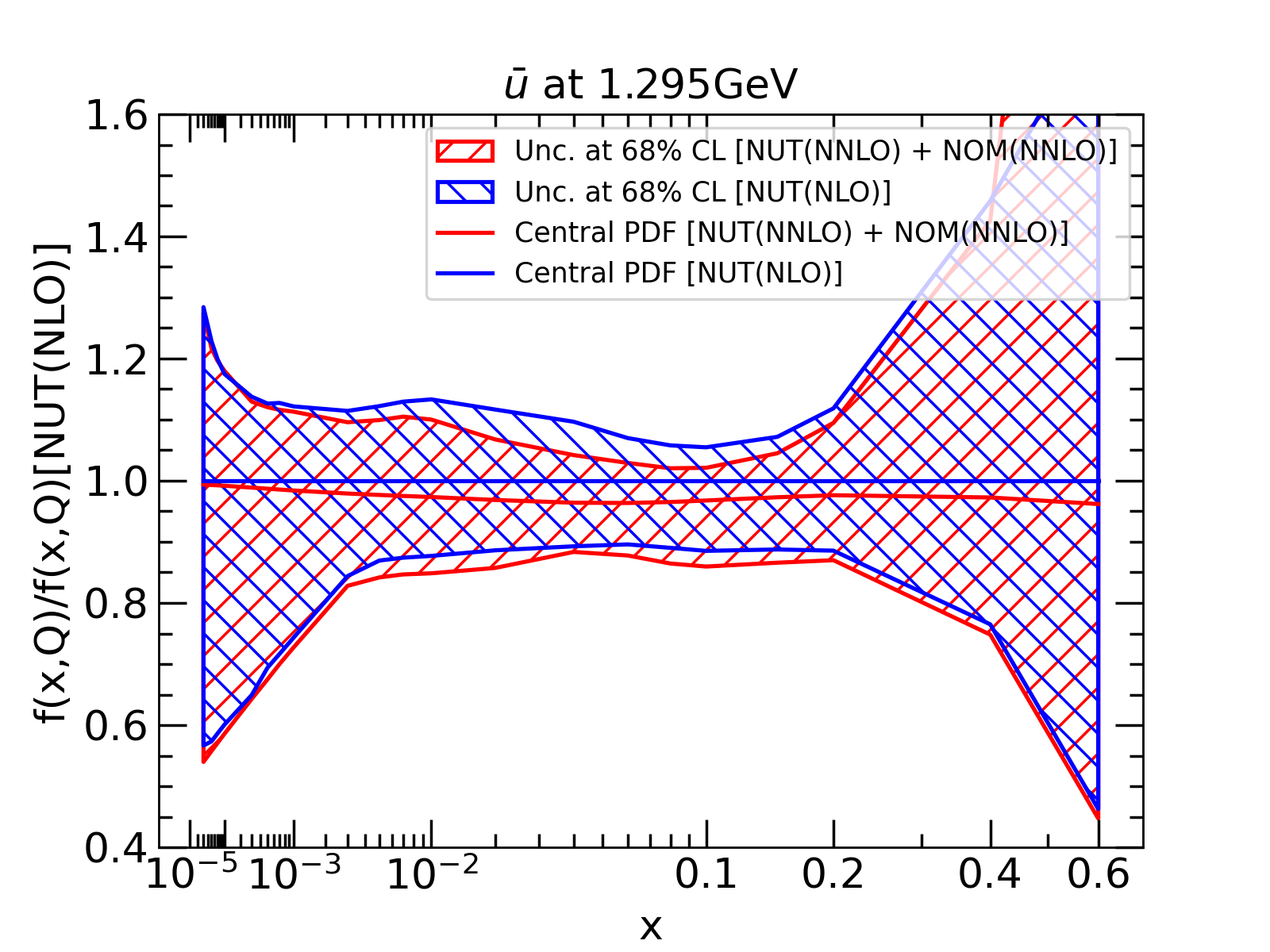}}
  \subcaptionbox{}[7.7cm]
    {\includegraphics[width=7.7cm]{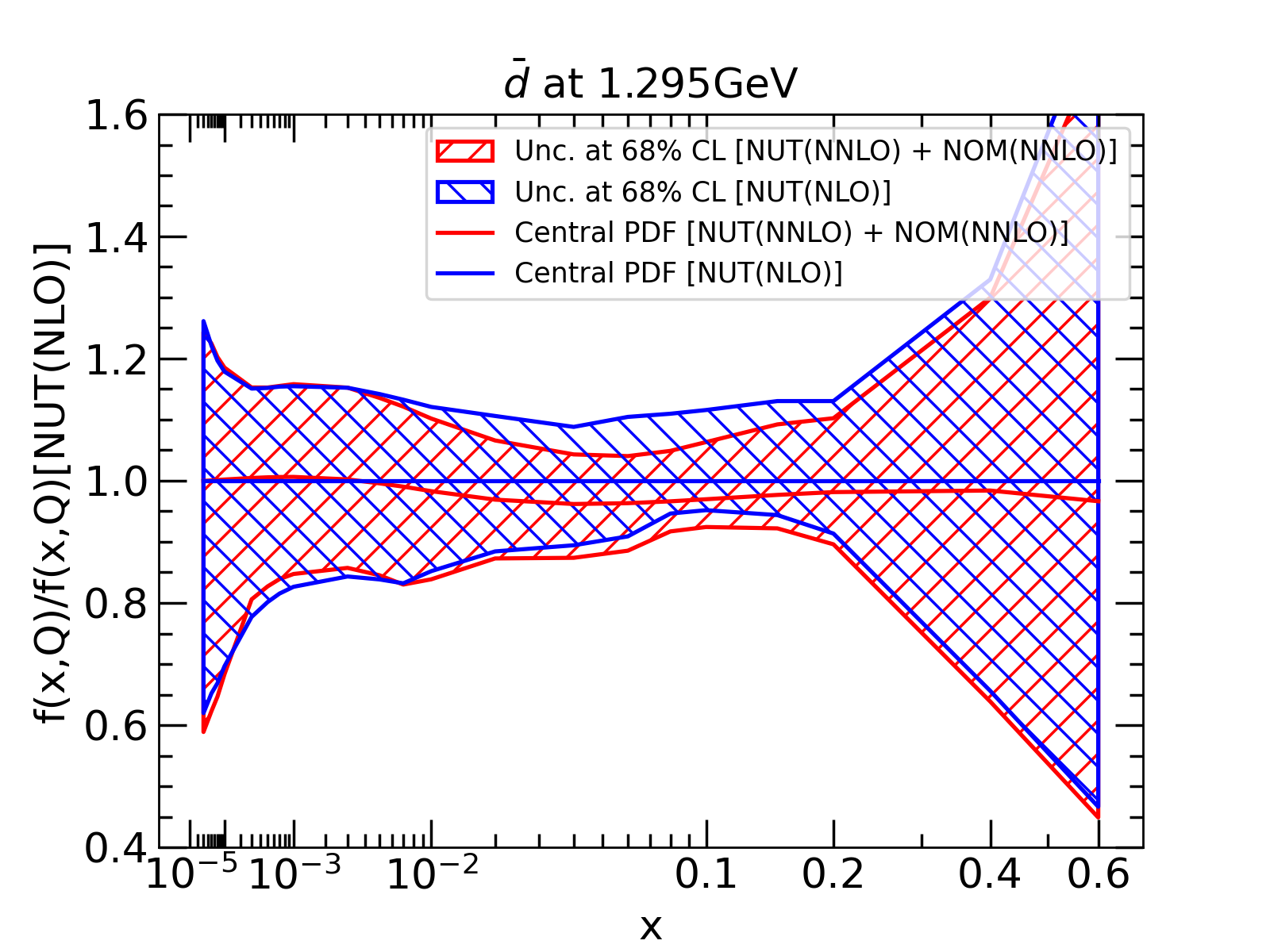}}
  \subcaptionbox{}[7.7cm]
    {\includegraphics[width=7.7cm]{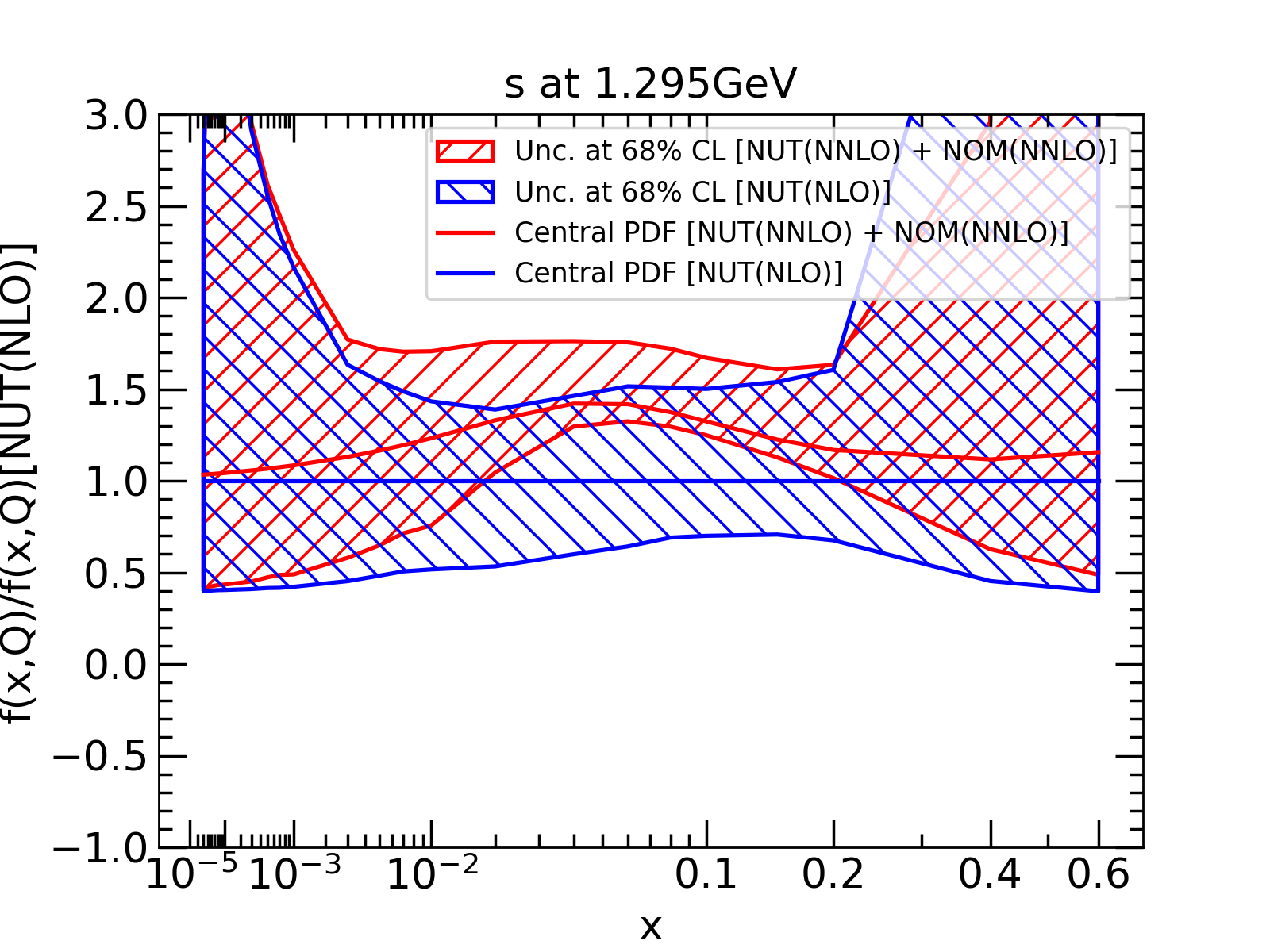}}
  \caption{
  The parton distribution functions at $Q = 1.295$ GeV for $u$, $\bar{u}$, $\bar{d}$, and $s$. 
  The red and the blue solid lines represent the central values with and without NNLO NOMAD data respectively, and the red and the blue hatched areas represent the respect uncertainties at 68\% CL.
  When NOMAD (NOM) data are incorporated, 124-127 data sets (NUT) are replaced with their NNLO version.
  Central values are normalized to the NLO version.
  }
\label{Fig:LM_total_xnnlo}
\end{figure}

\begin{figure}[htbp]
  \centering
  \subcaptionbox{}[7.7cm]
    {\includegraphics[width=7.7cm]{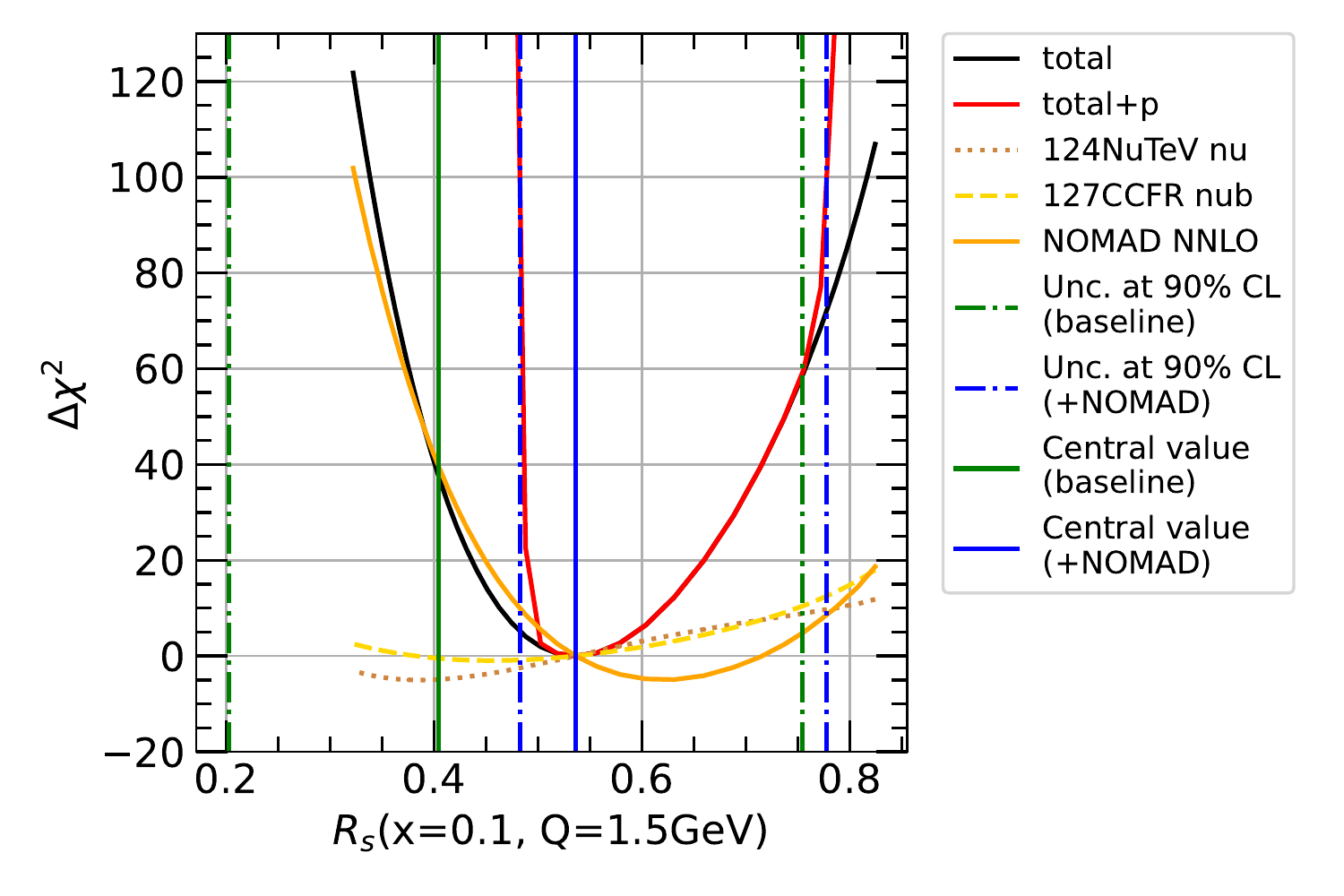}}
  \subcaptionbox{}[7.7cm]
    {\includegraphics[width=7.7cm]{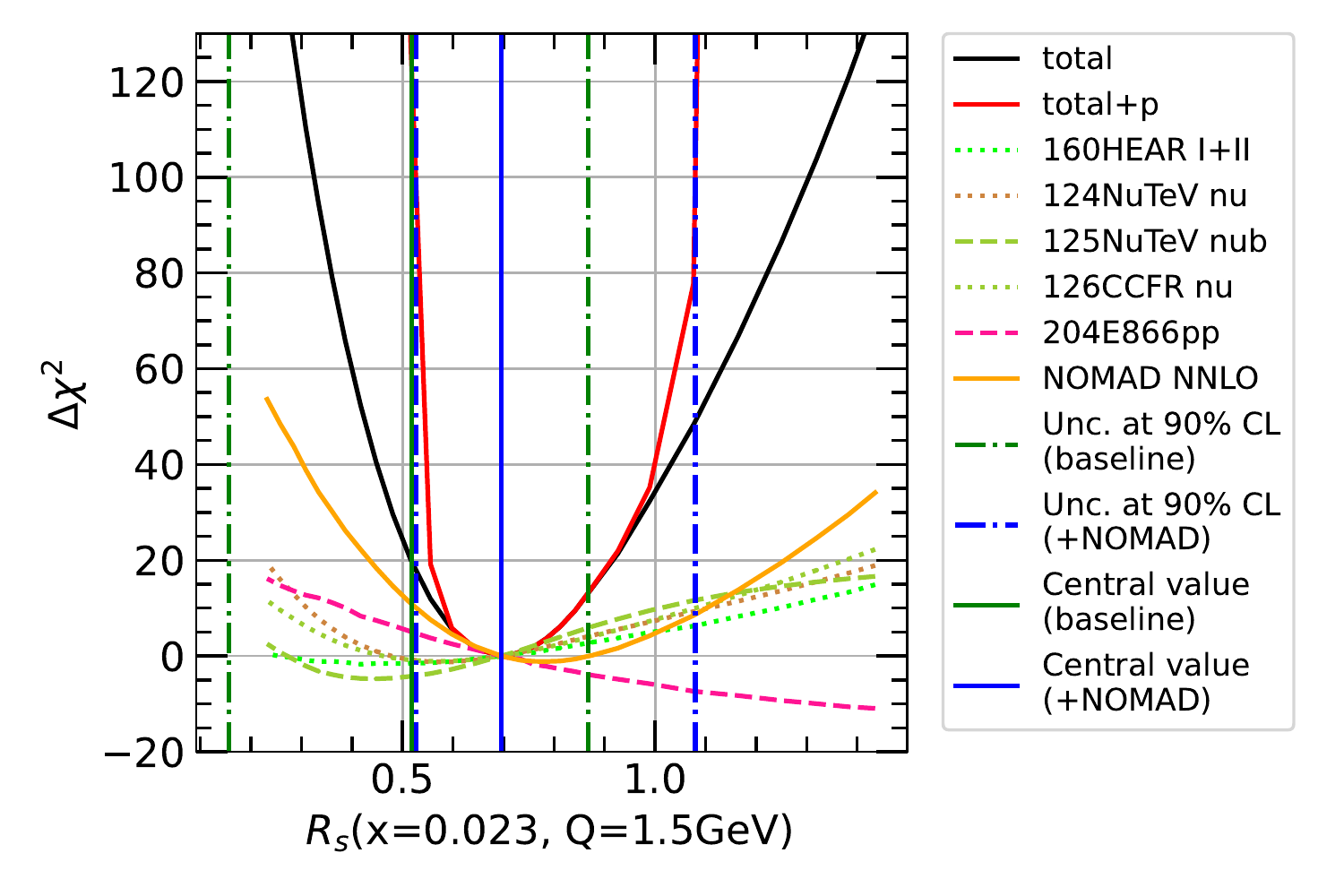}}
  \subcaptionbox{}[7.7cm]
    {\includegraphics[width=7.7cm]{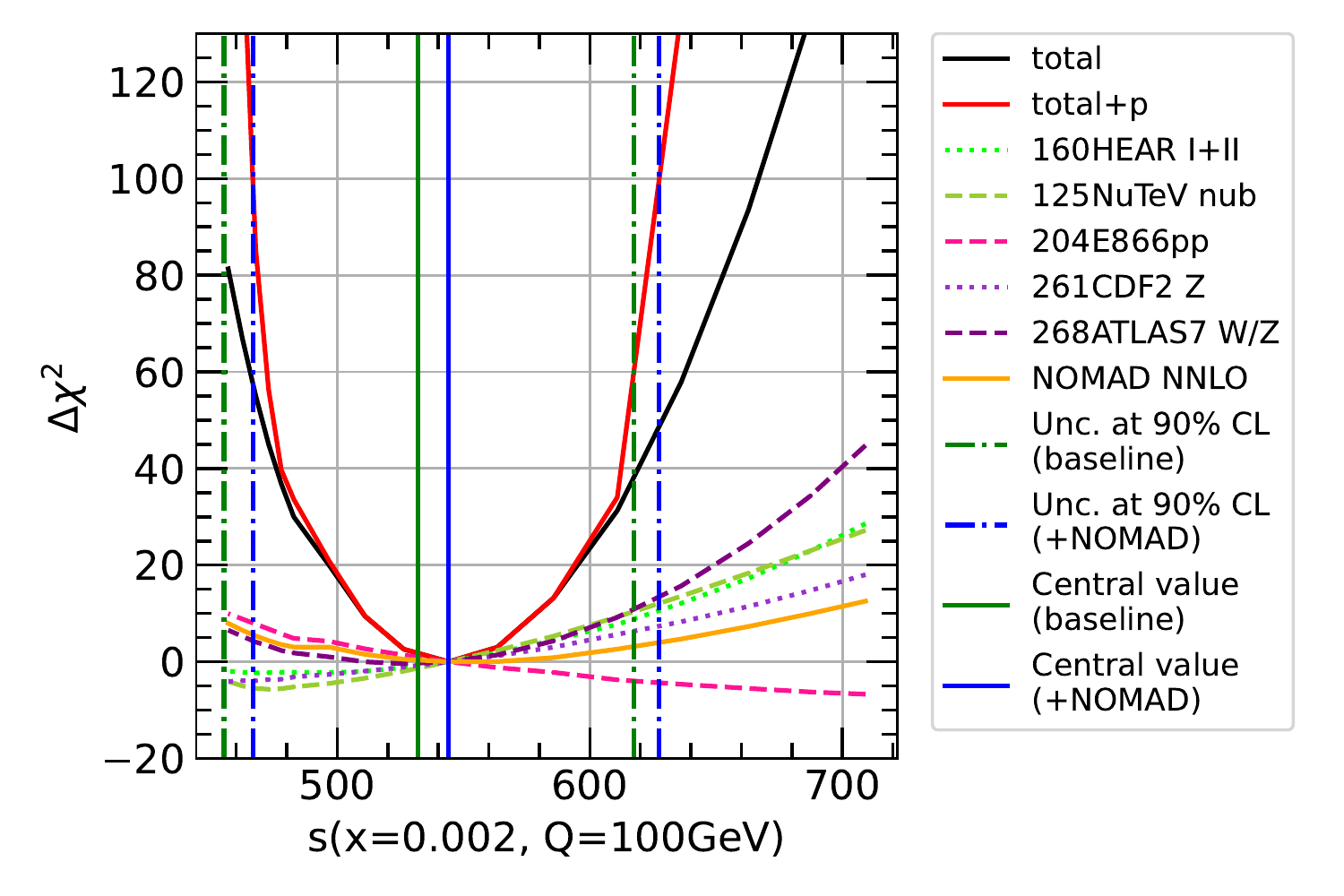}}
  \subcaptionbox{}[7.7cm]
    {\includegraphics[width=7.7cm]{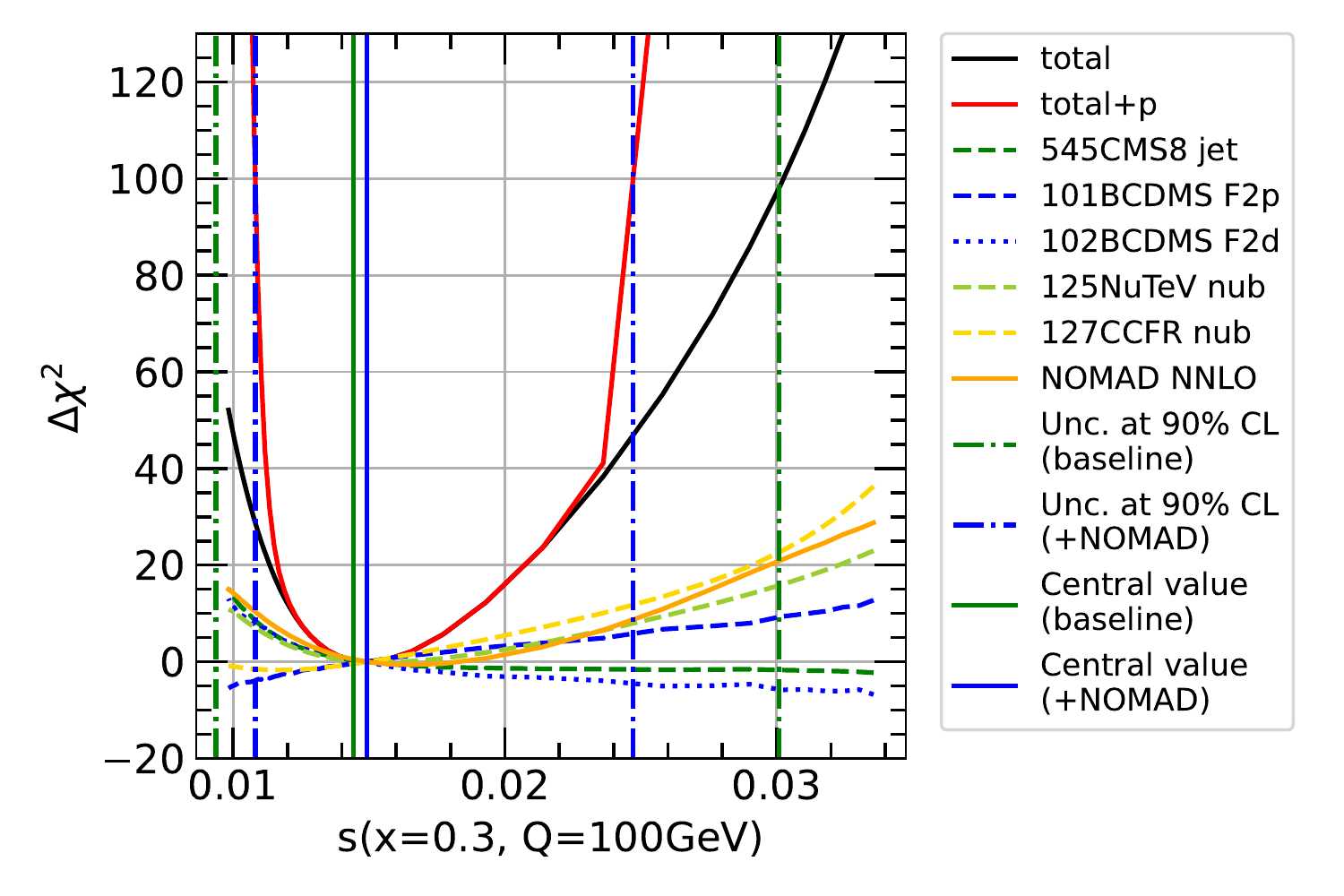}}
  \caption{
  LM scans on the $R_s$ at $Q = 1.5$ GeV and $x = 0.1$ or 0.023 (upper panels), and LM scans on the $s$-quark at $Q = 100$ GeV and $x = 0.002$ or 0.3 (lower panels).
  The blue and the green vertical solid (dot-dash) lines represent the central values (uncertainties) with and without NOMAD data respectively.
  }
  \label{Fig:LM_rs_s_xnnlo}
\end{figure}

In Fig.~\ref{Fig:LM_total_xnnlo}, we compare $u$, $\bar{u}$, $\bar{d}$ and $s$-quark PDFs at $Q = 1.295$ GeV from fits with and without the inclusion of NOMAD data. 
The PDF uncertainties are shown through hatched areas with relevant colors.
NOMAD data are taken from the distribution over Bjorken $x$ excluding the last two points, with
predictions calculated up to NNLO in QCD.
Predictions for data sets 124-127 (dimuon measurements from NuTeV and CCFR) in the global fit are replaced with their NNLO versions when including
NOMAD data, in order to match on the theoretical precision.
Note in the fit without NOMAD data the predictions for data sets 124-127 are evaluated at NLO similar to those in CT18.
All PDFs are normalized to the central value without NOMAD data in Fig.~\ref{Fig:LM_total_xnnlo}.
In the upper-left panel, almost no change occurs in the region $x \gtrsim 0.1$ of $u$-quark PDF, and a negligible downward shift smaller than 2\% can be seen for $x \lesssim 0.05$. 
Slight downward shifts on both central value and uncertainty region can also be observed in the $\bar{u}$ (upper-right panel) and the $\bar{d}$-quark (lower-left panel) PDFs.
The downward shifts observed are required to stabilize the $W$ and $Z$ production
cross sections at collider experiments.
The improvement in the constraints on $u$, $\bar{u}$ and $\bar{d}$-quark PDFs are smaller than about 3\%.
This insensitivity of $u$, $\bar{u}$ and $d$-quark PDFs to NOMAD data is an indication of the CKM suppression in the charm-quark production.
The constraint on $s$-quark PDF is, however, markedly improved around $x = 0.05$. In the region of $x \sim 0.05$, the $s$-quark PDF achieves a factor of two better precision when NOMAD data are incorporated.
This is because NOMAD data peak at neutrino energy $E_\nu \approx 30$ GeV, which implies a sensitivity to kinematic region with Bjorken $x \sim 1/(1+2M_{nucleon}E_{\nu}/Q^2) \sim 0.03$ at $Q=1.295$ GeV.
An upward shift of more than 15\% is also observed in most regions. 
It is indeed a manifestation of the trend of prediction-data comparison shown in Fig.~\ref{Fig:nomad_exp_vs_theo}.
Both ABM and NNPDF groups considered the impact of NOMAD data~\cite{Alekhin:2014sya,Faura:2020oom}.
As to the analysis of ABM group, an at most 5\% downward shift is reported near region $x \approx 0.05$ at scale $Q=3$ GeV when NOMAD data are incorporated into the fit with only NuTeV/CCFR data (data sets 124-127 in this paper)~\cite{Alekhin:2014sya}.
More data sets are considered in the work of NNPDF group~\cite{Faura:2020oom}.
With the analysis performed there, NOMAD data together with ATLAS W/Z data sets~\cite{ATLAS:2012sjl,ATLAS:2016nqi} contribute to a marked enhancement of $s$-quark PDF at $Q=10$ GeV compared with CT18 data sets.
It is noted that, between these two kinds of data sets, ATLAS W/Z data sets are already reported to give a larger $s$-quark PDF compared with CT18 data sets~\cite{Hou:2019efy},
and the work of NNPDF group further demonstrated that ATLAS W/Z data sets prefer a larger $s$-quark PDF compared with NOMAD data. 
Finally, both the two groups and our analysis indicate strong constraints on $s$-quark PDF in the region near $x \approx 0.05$, given the incorporation of NOMAD data.
We also compare the sensitivities to $s$-quark PDF between NOMAD data and the other data sets in Fig.~\ref{Fig:LM_rs_s_xnnlo}, in which we show LM scans on $s$-quark PDF and $R_s$.
This comparison is set at a scale of $Q=1.5$ GeV and $x = 0.1$ in panel (a). In this panel, NOMAD data predominate over the other experimental data sets.
When $x$ gets smaller to be 0.023 as in panel (b), NuTeV and CCFR neutrino DIS experiments become more important but still the NOMAD data show the most prominence. 
In panel (a), the fit without NOMAD data predicts $R_s(x=0.1, Q=1.5\rm{GeV})=0.40^{+0.35}_{-0.20}$ at 90\% CL,
while fit including NOMAD data expects $R_s(x=0.1,Q=1.5\rm{GeV})=0.54^{+0.24}_{-0.06}$, giving improved constraints by a factor of two.
In panel (b), the corresponding values are $R_s(x=0.023,Q=1.5\rm{GeV})=0.53^{+0.33}_{-0.38}$ and $R_s(x=0.023,Q=1.5\rm{GeV})=0.70^{+0.40}_{-0.17}$, respectively.
NOMAD data hence give about 20\% reduction on PDF uncertainties.
It is also noted that a slight tension exists between NOMAD data and data from the other two neutrino DIS experiments, i.e., NuTeV and CCFR, in both panels.
The latter two experiments both prefer smaller $R_s$s contrasted with NOMAD data which prefer a larger one.
Further investigations on the interplay of the three experiments and of different theories are included in Appendix~\ref{sec:vano}.
Moreover, the ATLAS W/Z data~\cite{ATLAS:2012sjl}, which prefer an especially larger $R_s(x=0.023,Q=1.38\rm{GeV}) \sim 1$, show an even stronger tension with these two neutrino DIS experiments.
NOMAD data, however, compromise between these two extremes.
This conclusion is also observed in the analysis of~\cite{Faura:2020oom}. Meanwhile, a similar result of $R_s(x=0.023, Q=1.6\rm{GeV})=0.71\pm0.1$ is obtained in that work once NOMAD data are included.
On the other hand, we let the scale increase to be $Q = 100$ GeV in panel (c) and panel (d).
The case with $x=0.3$ shows more sensitive than that with $x=0.002$.
In panel (d), it can be seen that NuTeV and CCFR data become comparable with NOMAD data.
No significant shift in the central value is found when we incorporate NOMAD data, but an almost 30\% better constraints on $s$-quark PDF is achieved.   
In panel (c), the sensitivity of NOMAD data becomes worse due to the favor of large-$x$ at this scale,
and collider data now play important roles. 
Only improvement of a few percent in the constraint on $s$-quark PDF can be obtained.

\subsection{Impact of High-luminosity LHC}\label{subsec:HL}

LHC data play important roles on constraining PDFs as shown in Table~\ref{tab:data set}.
And the upgrade of the LHC, the HL-LHC, is expected to accumulate a total integrated luminosity of 
$\mathcal{L} = 3000$ fb$^{-1}$ for ATLAS and CMS and $300$ fb$^{-1}$ for LHCb.
In this section, we take two of those HL-LHC pseudo-data sets constructed in Ref.~\cite{AbdulKhalek:2018rok}, and evaluate their impact on PDFs within the framework of CT18 based on our new approach.

The HL-LHC pseudo-data are generated for processes of Drell-Yan production with high dilepton invariant mass and W and Z boson production in the forward region.
Details of these pseudo-data are described as follows:

$\bullet$~The distribution of dilepton invariant mass $d\sigma (pp\to l^+l^-)/dm_{ll}$ of high-mass Drell-Yan process at $\sqrt{s} = $14 TeV, covered by the ATLAS experiment, is generated according to the following requirements:
$p_T^{l_{1(2)}} \ge$ 40 (30) GeV, $|\eta^{l}|\leq$ 2.5, and $m_{ll} \ge $ 116 GeV.
The total number of data points is 21.
The binning and the systematic uncertainties are determined from Refs.~\cite{ATLAS:2016gic,AbdulKhalek:2018rok}.

$\bullet$~The distributions for W and Z boson production in the forward region at $\sqrt{s} = $14 TeV, covered by the LHCb experiment, are generated according to the following cuts: 
$p_T^{l}\ge$ 20 GeV, 2.0 $\leq \eta^{l} \leq$ 4.5.
An additional requirement for Z production is that 60 GeV $\leq m_{ll} \leq$ 120 GeV.
The total number of data points is 90.
The binning and the systematic uncertainties are determined from Refs.~\cite{LHCb:2015mad,AbdulKhalek:2018rok}.

\begin{figure}[htbp]
  \centering
  \subcaptionbox{}[7.7cm]
    {\includegraphics[width=7.7cm]{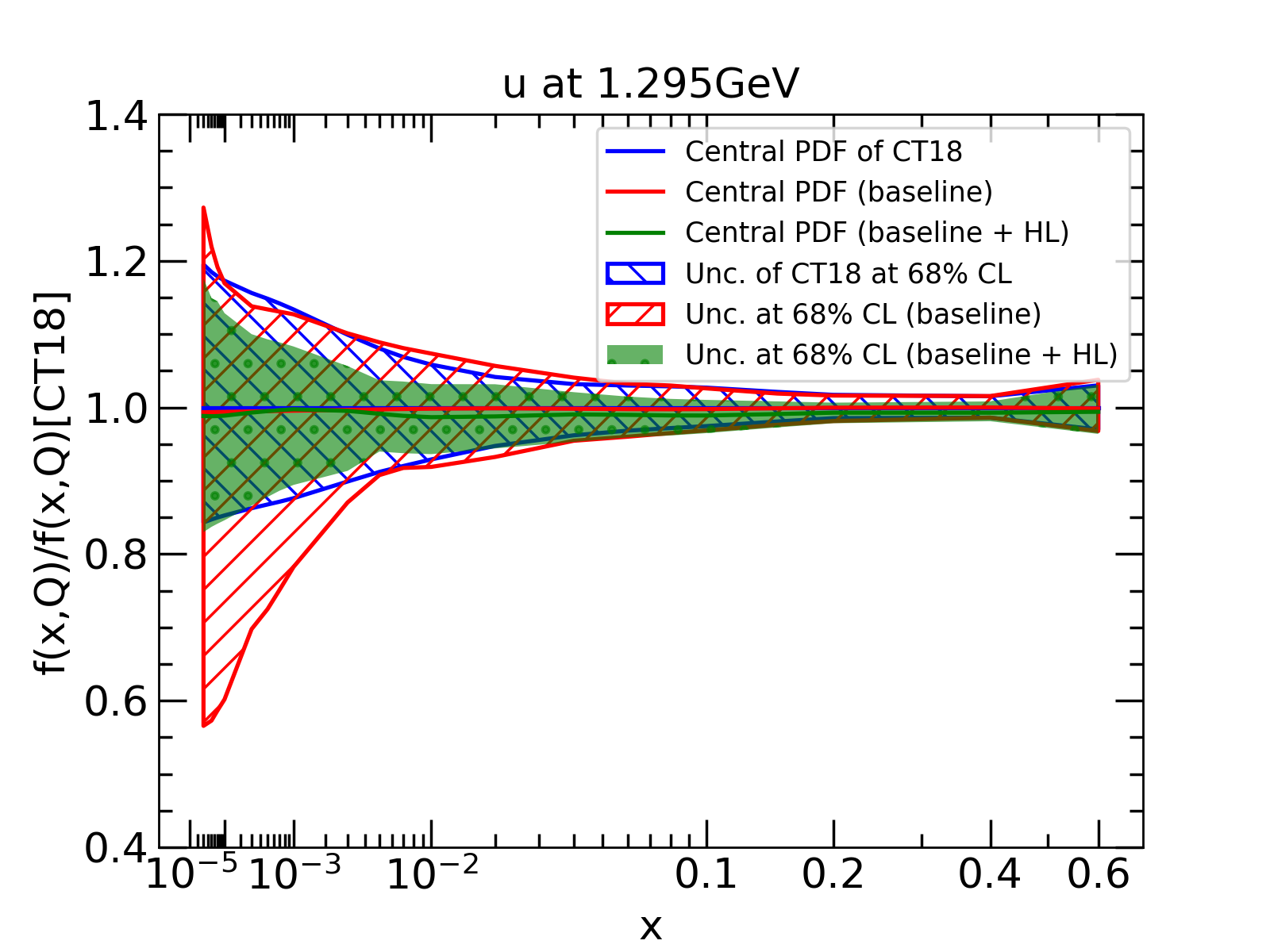}}
  \subcaptionbox{}[7.7cm]
    {\includegraphics[width=7.7cm]{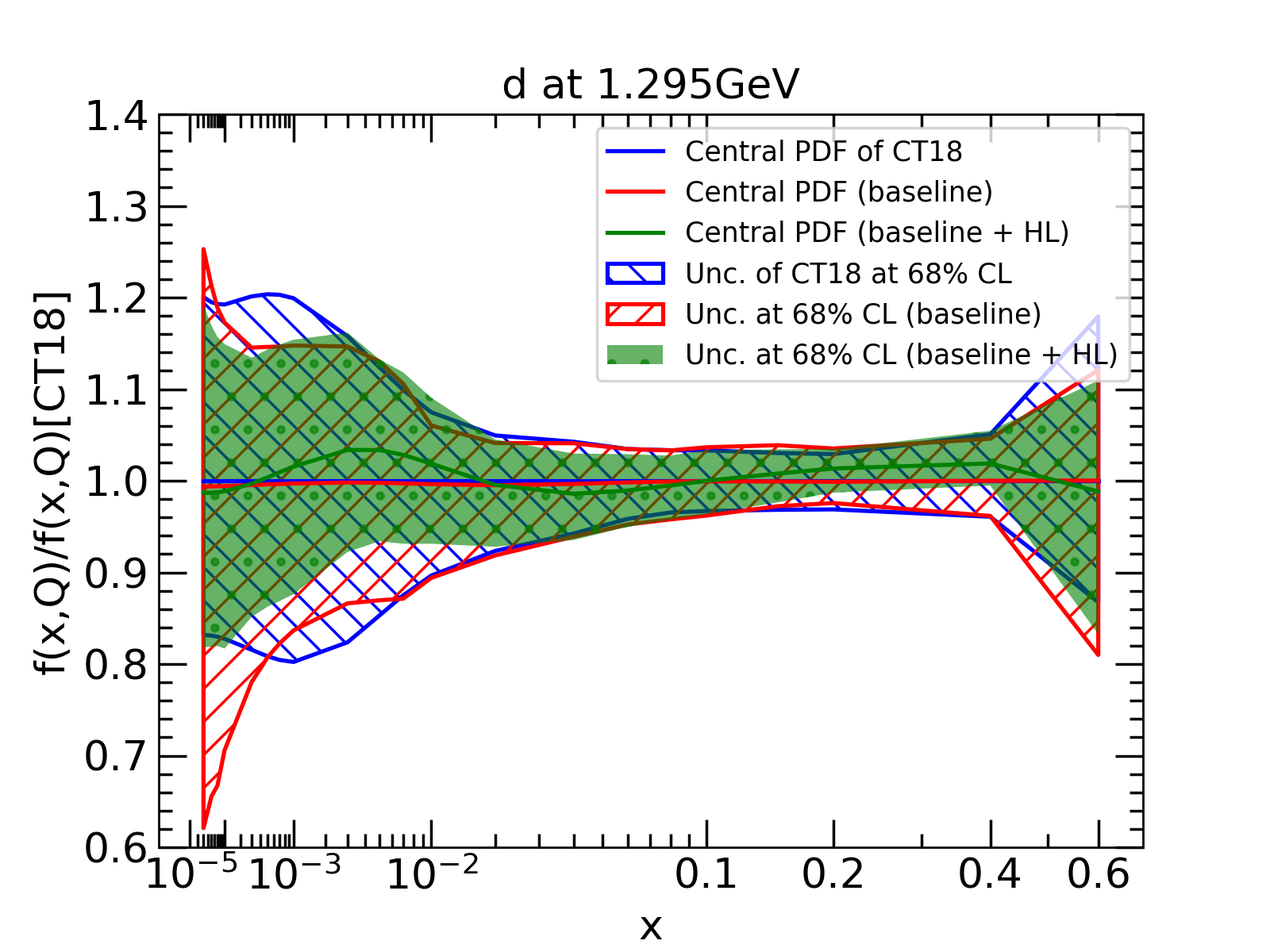}}
  \subcaptionbox{}[7.7cm]
    {\includegraphics[width=7.7cm]{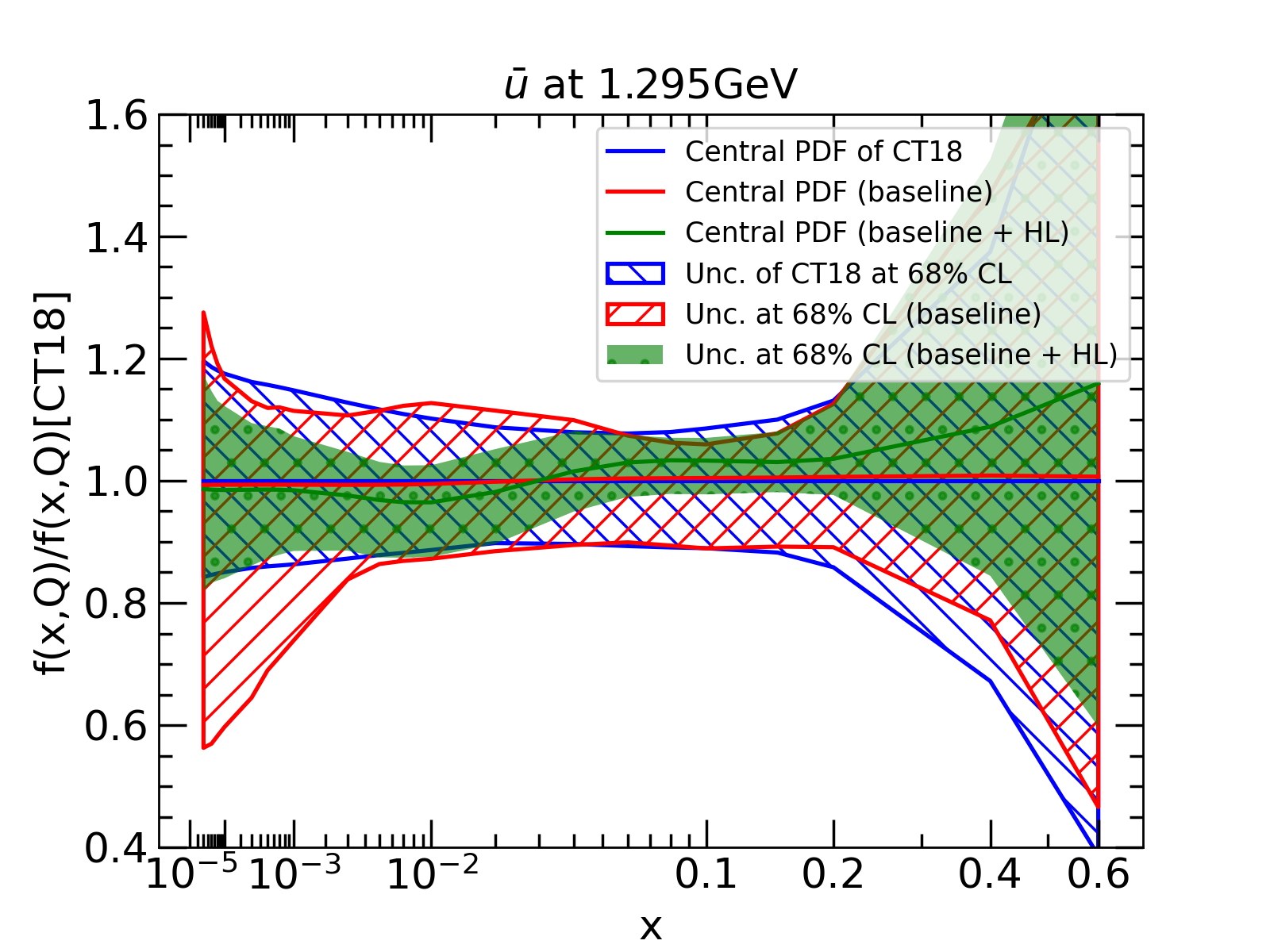}}
  \subcaptionbox{}[7.7cm]
    {\includegraphics[width=7.7cm]{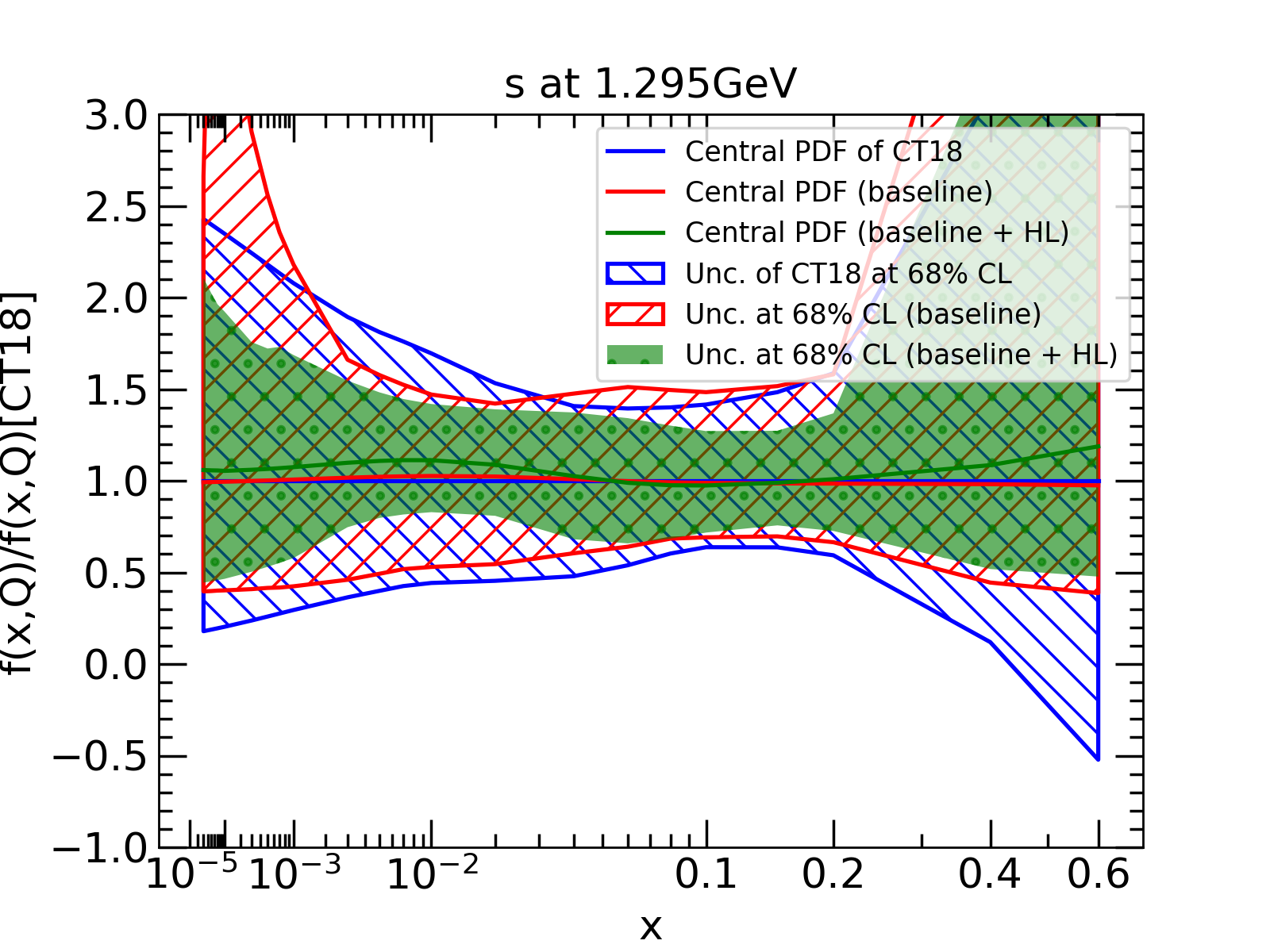}}
  \caption{The parton distribution functions at $Q = 1.295$ GeV for $u$, $d$, $\bar{u}$ and $s$. 
The red and the green solid lines represent the central values without and with HL-LHC pseudo-data respectively,
and the red and the green hatched areas represent the respective uncertainties at 68\% CL.
The results are normalized to the central value of CT18 NNLO (blue solid line).}
\label{Fig:HL_LM_total}
\end{figure}

We include those pseudo-data in the CT18 global fit and quantify their impact on PDFs.
In Fig.~\ref{Fig:HL_LM_total} we show a comparison of the PDFs with and without HL-LHC pseudo-data,
together with the published Hessian set of CT18.
All results are normalized to the central value of CT18.
The PDF uncertainties are shown through hatched areas with
relevant colors.
In Fig.~\ref{Fig:HL_LM_total}, a significant reduction in PDF uncertainties can be found in all cases
once including the pseudo-data, especially for sea quarks.
In the upper-left panel, the PDF uncertainties are reduced by almost a factor of 2, from about 30\% to about 15\%, at small-$x$.
Similar improvements can also be observed in $d$, $\bar{u}$ and $s$-quark PDFs.
That is because HL-LHC pseudo-data contribute a great improvement in statistics, and cover the kinematic regions where PDFs are not determined well.
Specifically, the process of high-mass Drell-Yan is directly sensitive to sea quarks at large-$x$,
and the process of forward W/Z production constrains the $s$-quark PDF at both small-$x$ and large-$x$.
In the lower-right panel, we find that the HL-LHC gives about 30\% reduction on PDF uncertainties of $s$-quark  in the regions of $x \sim 0.01$ and $x \sim 0.1$. 
This result highlights the importance of the process of forward W/Z production.

\begin{figure}[htbp]
  \centering
  \subcaptionbox{}[7.7cm] 
    {\includegraphics[width=7.7cm]{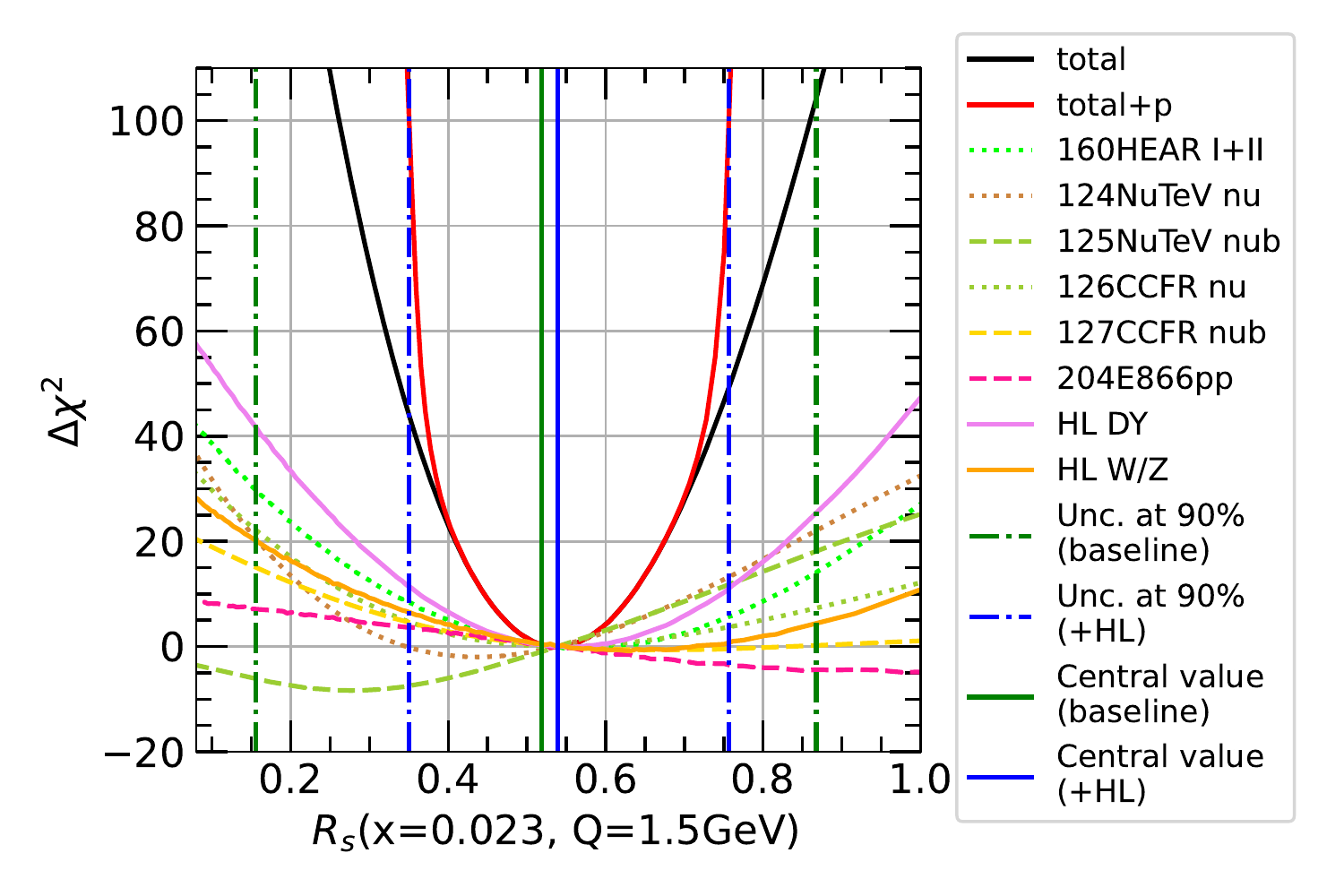}}
  \subcaptionbox{}[7.7cm]
    {\includegraphics[width=7.7cm]{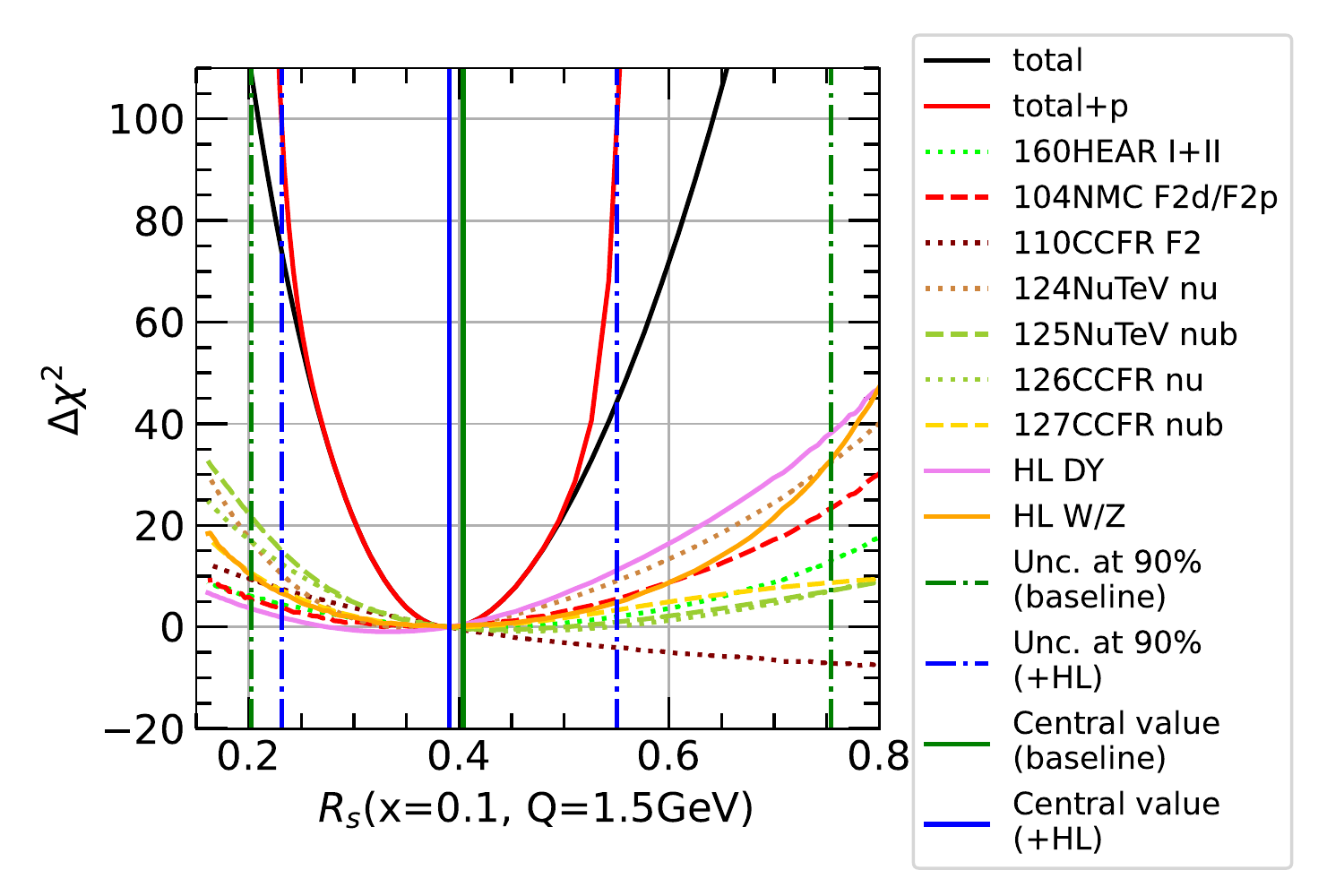}}
  \subcaptionbox{}[7.7cm]
    {\includegraphics[width=7.7cm]{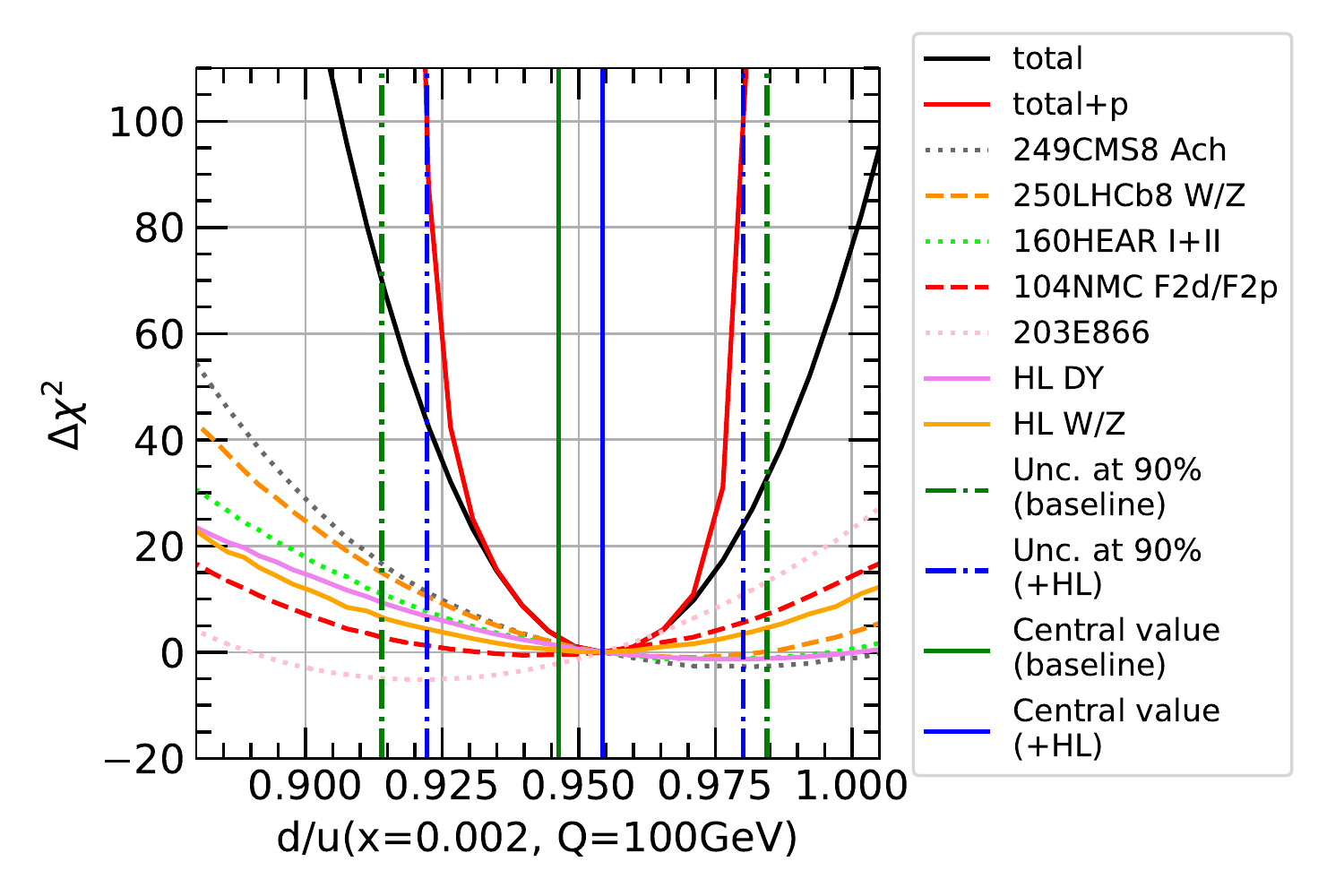}}
  \subcaptionbox{}[7.7cm]
    {\includegraphics[width=7.7cm]{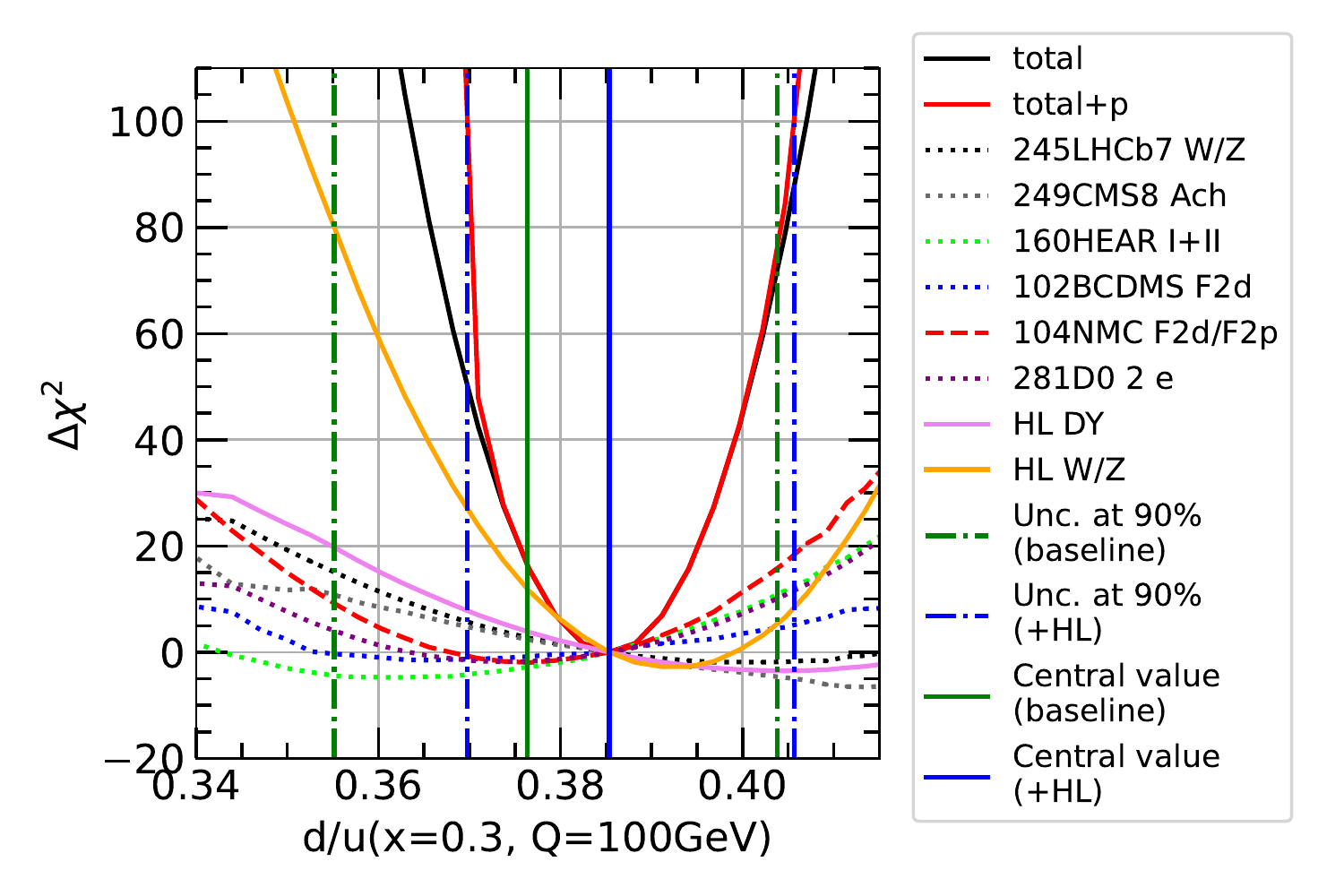}}
  \caption{LM scans on the $R_s$ at $Q = 1.5$ GeV and $x = 0.023$ or 0.1 (upper panels), and LM scans on the $d/u$ at $Q = 100$ GeV and $x = 0.002$ or 0.3 (lower panels).
  The blue and the green vertical solid (dot-dash) lines represent the central values (uncertainties) with and without HL-LHC pseudo-data respectively.
  }
  \label{Fig:LM_HL_ud}
\end{figure}

We show the results of LM scans on $R_s$ and $u/d$ ratio in Fig.~\ref{Fig:LM_HL_ud}.
We find measurements of high-mass Drell-Yan process and forward W/Z production process at HL-LHC give strong constraints on $R_s$ and $u/d$ ratio at both small-$x$ and large-$x$. 
The PDF uncertainties of $R_s$ and $u/d$ are significantly reduced after the inclusion of pseudo-data.

In the upper-left panel of Fig.~\ref{Fig:LM_HL_ud} for $R_s$ at $x = 0.023$ and $Q = 1.5$ GeV, the constraints from HL-LHC pseudo-data predominate as expected. In addition to that, constraints from NuTeV dimuon, CCFR dimuon and HERA inclusive DIS data also play important roles.
The fit without pseudo-data predicts a result of $R_s$ = $0.53^{+0.33}_{-0.38}$ at 90\% CL, while fit including pseudo-data gives $R_s$ = $0.54^{+0.22}_{-0.19}$.
After the inclusion of pseudo-data, the PDF uncertainties are reduced by almost 50\%.
As $x$ increases to 0.1 in the upper-right panel, HL-LHC forward W/Z data becomes more important.
Fit without pseudo-data gives a result of $R_s$ = $0.40^{+0.35}_{-0.20}$, while fit including pseudo-data gives $R_s$ = $0.39^{+0.16}_{-0.16}$.

For $d/u$ at $x = 0.002$ and $Q = 100$ GeV, in the lower-left panel, the most strong constraints originate from HL-LHC pseudo-data together with LHC W and Z boson data and the fixed target experiments E866 and NMC.
The results of $d/u$ are $0.946^{+0.038}_{-0.032}$ and $0.954^{+0.026}_{-0.032}$ corresponding to fit without and with pseudo-data respectively.
After the inclusion of pseudo-data, the PDF uncertainties are reduced by almost 30\%.
In the lower-right panel for $x = 0.3$ and $Q = 100$ GeV, pseudo-data predominate over the other experimental data sets.
Fit without and with pseudo-data give $d/u$ = $0.376^{+0.028}_{-0.021}$, and $d/u$ = $0.385^{+0.020}_{-0.016}$ respectively.
PDF uncertainties are reduced by almost 25\% in this case.
Both of the two HL-LHC processes prefer a larger $d/u$, and their inclusion leads to an increase of the central value.

\begin{figure}[htbp]
  \centering
  \subcaptionbox{CT18NNLO + HL-LHC}[7.7cm]
    {\includegraphics[width=7.7cm]{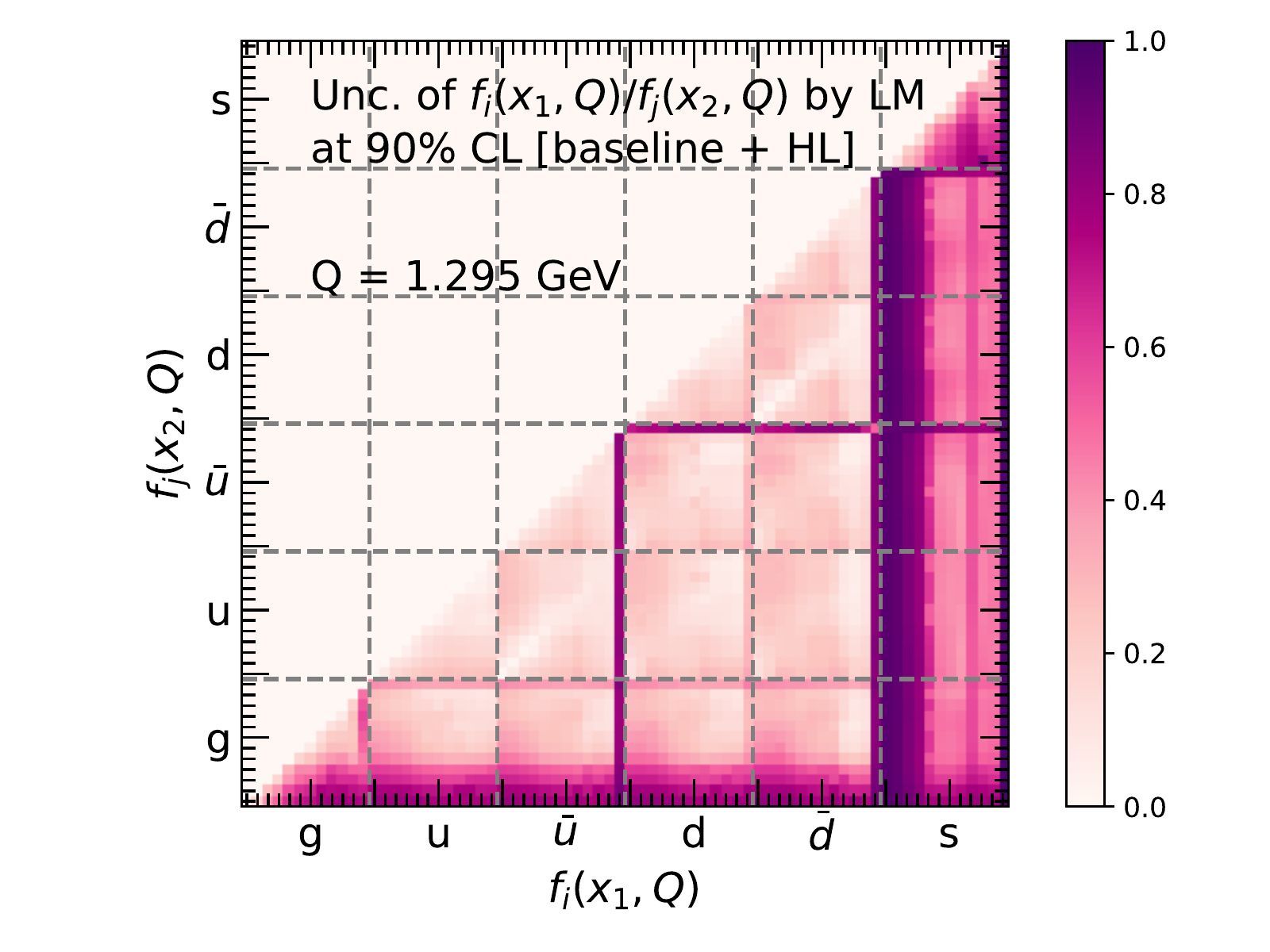}}
  \subcaptionbox{Reduction}[7.7cm]
    {\includegraphics[width=7.7cm]{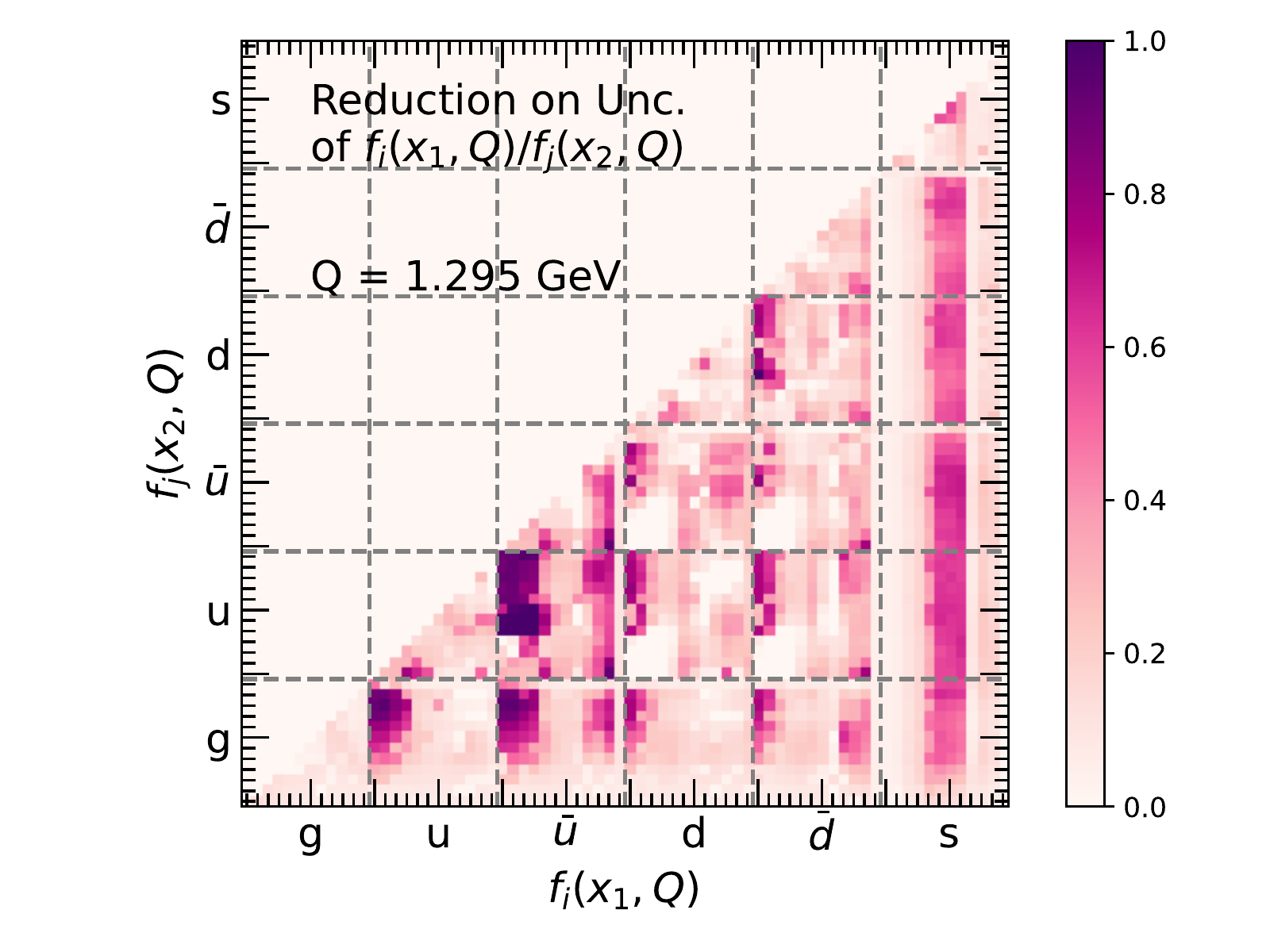}}
  \caption{The relative uncertainties of $R_f = f_{i}(x_1,Q)/f_{j}(x_2,Q)$ determined with the LM method at 90\% CL are shown in panel (a), where $Q = 1.295$ GeV.
  In panel (b) we show the reduction factors on the relative uncertainties of $R_{f}$ between with and without HL-LHC pseudo-data.}
  \label{Fig:HL_delta_r}
\end{figure}

In Fig.~\ref{Fig:HL_delta_r}, we show the results for the general PDF ratio $R_{f}$ as defined in Eq.~(\ref{eq:r}).
The uncertainties of $R_f$ are also determined with the LM method.
In panel (a), we show the relative uncertainties of $R_{f}$ at 90\% CL, $\Delta R_{f}$, from the fit with inclusion of HL-LHC pseudo-data.
To compare with the results from the fit without HL-LHC pseudo-data, a reduction factor of $\Delta R_{f}$,
\begin{equation}
    y_{red} =  2\dfrac{\Delta R_{f}^{base}-\Delta R_{f}^{base+HL}}{\Delta R_{f}^{base} +\Delta R_{f}^{base+HL}},
\end{equation}
is shown in panel (b).
We find that the relative uncertainties of $R_{f}$ have a noticeable reduction in general.
For the case of the $u$ and $\bar{u}$-quark PDFs as the numerator, we find that the HL-LHC gives about 80\% reduction on relative uncertainties in the region of $x_1 \lesssim 0.001$, which is because the HL-LHC pseudo-data give strong constraints on $u$ and $\bar{u}$-quark PDFs in this region as shown in Fig.~\ref{Fig:HL_LM_total}.
In addition, for the case of the $s$-quark PDFs as the numerator, we find that the HL-LHC gives about 50\% reduction on relative uncertainties in the region of $x_1\sim 0.01$.
However, for the $f_{d}(x_1 , Q)/f_{u}(x_2 , Q)$, $f_{d}(x_1 , Q)/f_{\bar{u}}(x_2 , Q)$, $f_{\bar{d}}(x_1 , Q)/f_{u}(x_2 , Q)$ and $f_{\bar{d}}(x_1 , Q)/f_{\bar{u}}(x_2 , Q)$, we find that the reduction factors on relative uncertainties are minor in the region of $x_1 \lesssim 1\times 10^{-3}$ and $x_2 \lesssim 1\times 10^{-3}$.
That is because the correlation between $u$-quark PDFs and $d$-quark PDFs at small-$x$ that originates from the parametrization form of PDFs.
Besides, for the ratios of gluon PDFs $f_{g}(x_1 , Q)/f_{g}(x_2 , Q)$, the uncertainties are reduced by only a few percent, which is expected due to the weak correlations between the two HL-LHC processes and gluon PDFs.

\subsection{Constraint on new physics with the global fit}\label{subsec:newp}

PDFs and their uncertainties play important roles in the indirect searches for new physics beyond the SM.
In this case, the scale of the new physics can be well beyond the typical scale of hard scatterings, and its effects can be formally described in the framework of the SMEFT.
PDFs are determined by fitting to a variety of experimental data under the assumption of the SM.
This leads to a problem that the degeneracy of PDF variations and the new physics contributions cannot be identified.
Therefore, to assess and furthermore constrain the new physics, a joint global fit including both PDFs and model parameters of new physics should be performed.
In this paper, we only consider one dimension-six operator, namely the lepton-quark contact interactions, to model the BSM effects in the SMEFT framework,
\begin{align}
\mathcal{L}_{\mathrm{SMEFT}}&=\mathcal{L}_{\mathrm{SM}}+  \sum_{i,j}\dfrac{c_{ij}}{\Lambda^2}(\bar{q_i} \gamma_{\mu} q_i)(\bar{l_j} \gamma^{\mu} l_j) \nonumber \\
&=\mathcal{L}_{\mathrm{SM}}+ \dfrac{\tilde c}{\Lambda^2}\sum_{i,j} e_{q_i} e_{l_j}(\bar{q_i} \gamma_{\mu} q_i)(\bar{l_j} \gamma^{\mu} l_j),
\end{align}
where $c_{ij}$ is the Wilson coefficient, $l_j$ and $q_i$ represent fields of charged leptons and quarks of flavor $j$ and $i$ respectively,
and $e_{q_i(l_j)}$ are the corresponding electric charges.
We assume the new interactions being vector-current type and have a flavor structure
similar to the QED coupling for simplicity.
Thus the contributions from the new physics are parametrized by a single variable of the
effective Wilson coeﬀicient $\tilde{c}$ that is normalized to the QED coupling.

In the case of data sets of the CT18 global fit, the DIS and the Drell-Yan processes receive contributions from this operator.
Processes with relatively large $Q^2$ are especially sensitive to BSM effects, where $Q$ is the momentum transfer.
Most of the data of Drell-Yan process included in the CT18 analysis are collected near the Z-pole region, which is less sensitive to new physics.
Hence, we only consider the HERA DIS process due to its large $Q^2$.
The amplitude of SM contributions from QED interactions is proportional to $1/Q^2$, and the amplitude of the BSM contributions is proportional to $\tilde{c}/\Lambda^{2}$
~\footnote{The weak interactions from $Z$ boson induce a different energy dependence of $1/(Q^2+M_Z^2)$ which we neglect here
for simplicity.}.
Hence, the total cross section including the BSM effects can be written as:
\begin{equation}
\sigma_{\mathrm{total}} = (1+\frac{\tilde{c}}{\Lambda^2}Q^2)^2\times\sigma_{\mathrm{DIS}} .
\end{equation}

\begin{figure}[htbp]
  \centering
  \includegraphics[width=0.5\textwidth,clip]{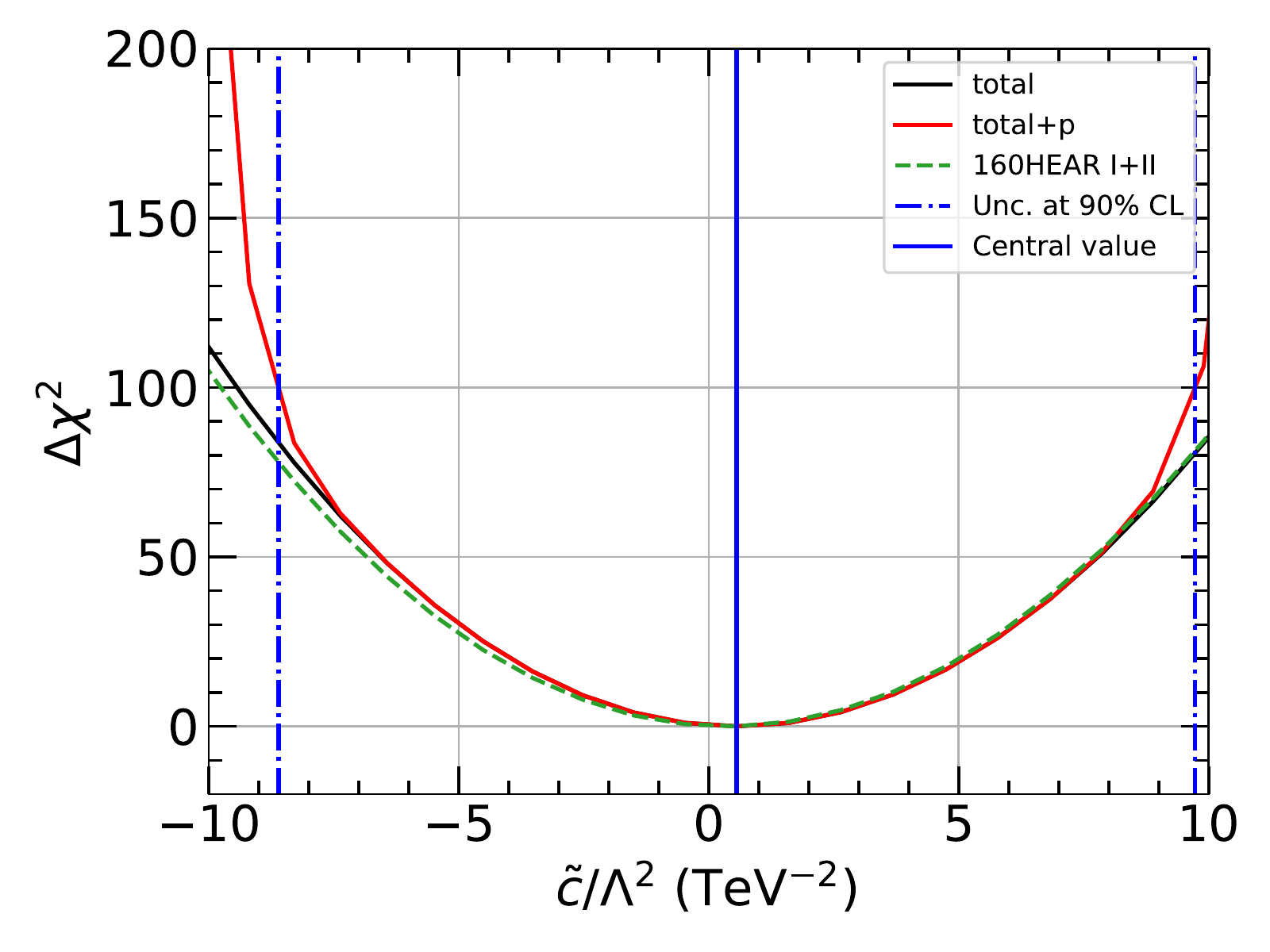}
  \hfill
  \caption{LM scans on $\tilde{c}/\Lambda^2$. The blue vertical solid line represents the central value of $\tilde{c}/\Lambda^2$, and the blue vertical dot-dash lines represent the uncertainties at 90\% CL.}
  \label{Fig:newp_r}
\end{figure}

A new NNs is built by adding the parameter $\tilde{c}/\Lambda^2$ into the input layer.
An association between the 29 variables \{$a_i, \tilde{c}/\Lambda^2$\} and $\chi^2$ is constructed.
With the new NNs, $\chi^2$ is recalculated and the results of LM scans on $\tilde{c}/\Lambda^2$ are shown in Fig.~\ref{Fig:newp_r}.
HERA inclusive DIS data give the dominant constraints as expected.
The LM scans predict a result of $\tilde{c}/\Lambda^2 = 0.56^{+9.16}_{-9.16}$ $\mathrm{TeV^{-2}}$ at 90\% CL, which is consistent with the SM.
The interplay between PDFs and BSM effects in the framework of the SMEFT has been studied in previous works~\cite{Carrazza:2019sec,Greljo:2021kvv,Iranipour:2022iak}.
A simultaneous determination of the PDFs and BSM effects from DIS data based on the NNPDF framework was presented in Ref.~\cite{Carrazza:2019sec}.
The Wilson coefficients of the lepton-quark contact interactions (i.e. $l$-$u$, $l$-$d$, $l$-$s$ and $l$-$c$ contact interactions) are constrained by the HERA inclusive DIS data.
The most stringent bounds are obtained for $u$-quark, followed by $d$-quark, and then $c$-quark and $s$-quark.
The constraint on the Wilson coefficients for $u$-quark converted to $\tilde{c}/\Lambda^2$ is [-6.5 $\mathrm{TeV^{-2}}$, 39.2 $\mathrm{TeV^{-2}}$] at 90\% CL.
In Ref.~\cite{Greljo:2021kvv}, the BSM effects are constrained by the high-mass Drell-Yan data.
The result converted to $\tilde{c}/\Lambda^2$ is [-6.5 $\mathrm{TeV^{-2}}$, 57.8 $\mathrm{TeV^{-2}}$] at 95\% CL.

\begin{figure}[htbp]
  \centering
  \subcaptionbox{}[7.7cm]
    {\includegraphics[width=7.7cm]{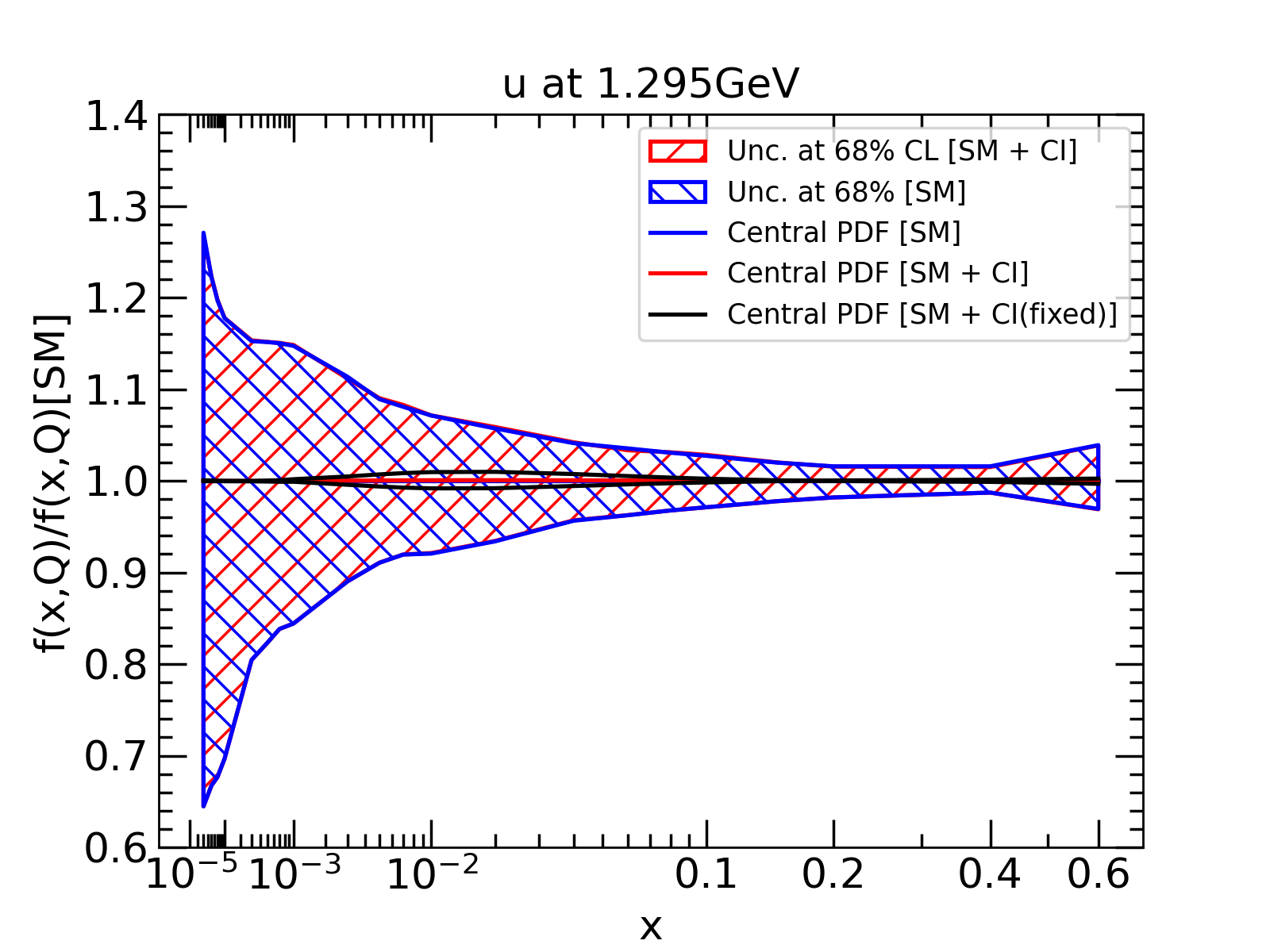}}
  \subcaptionbox{}[7.7cm]
    {\includegraphics[width=7.7cm]{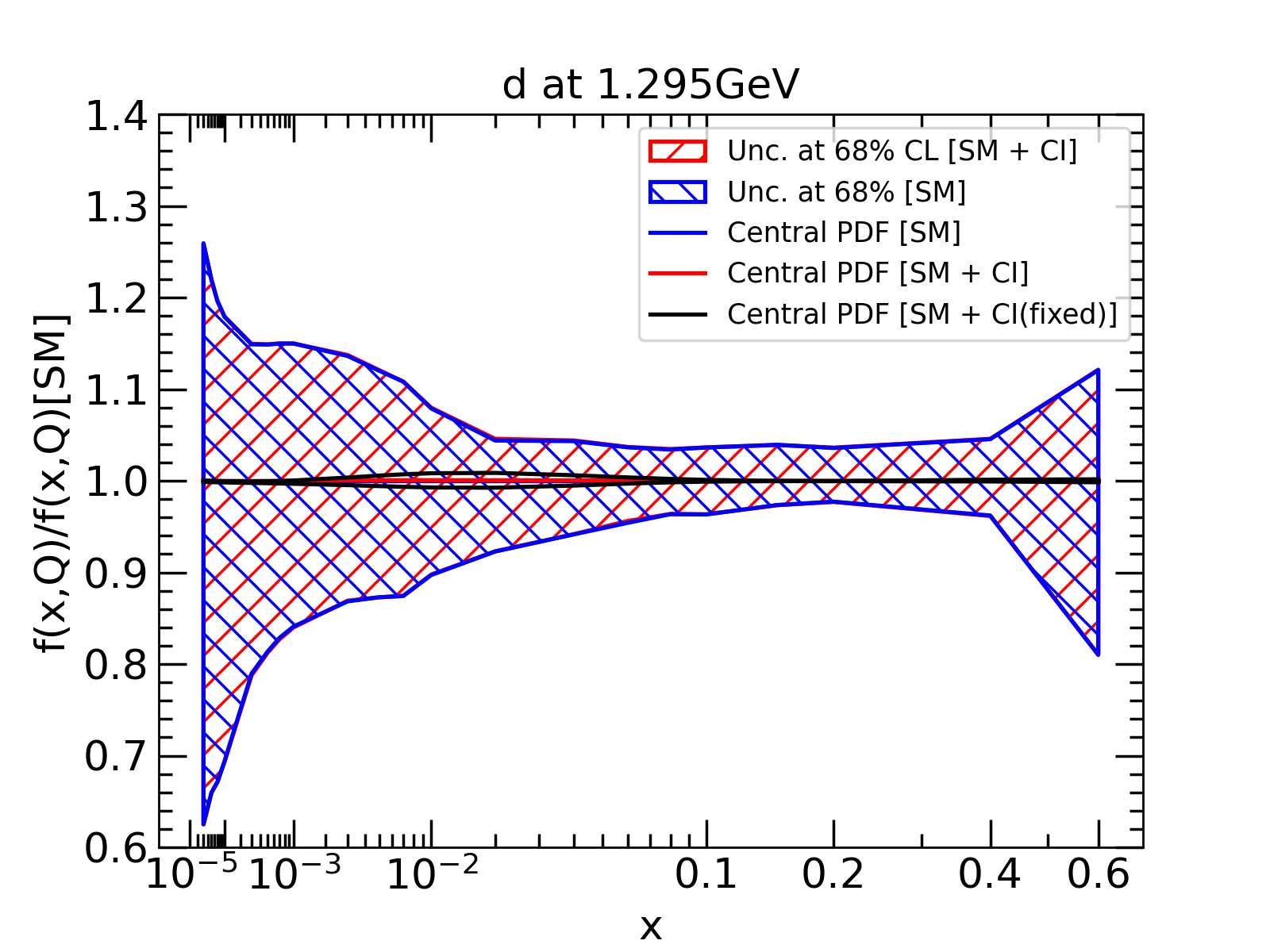}}
  \subcaptionbox{}[7.7cm]
    {\includegraphics[width=7.7cm]{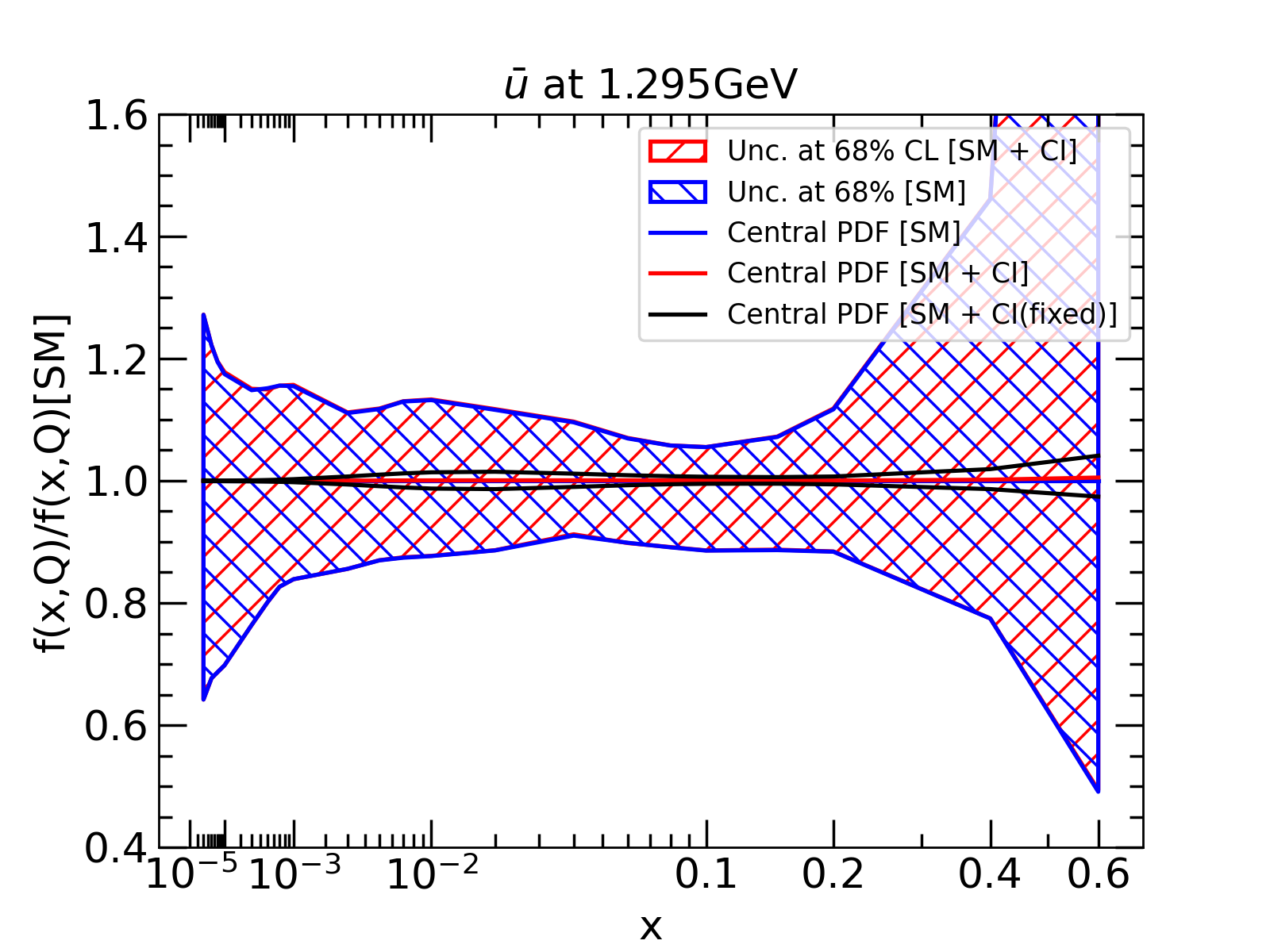}}
  \subcaptionbox{}[7.7cm]
    {\includegraphics[width=7.7cm]{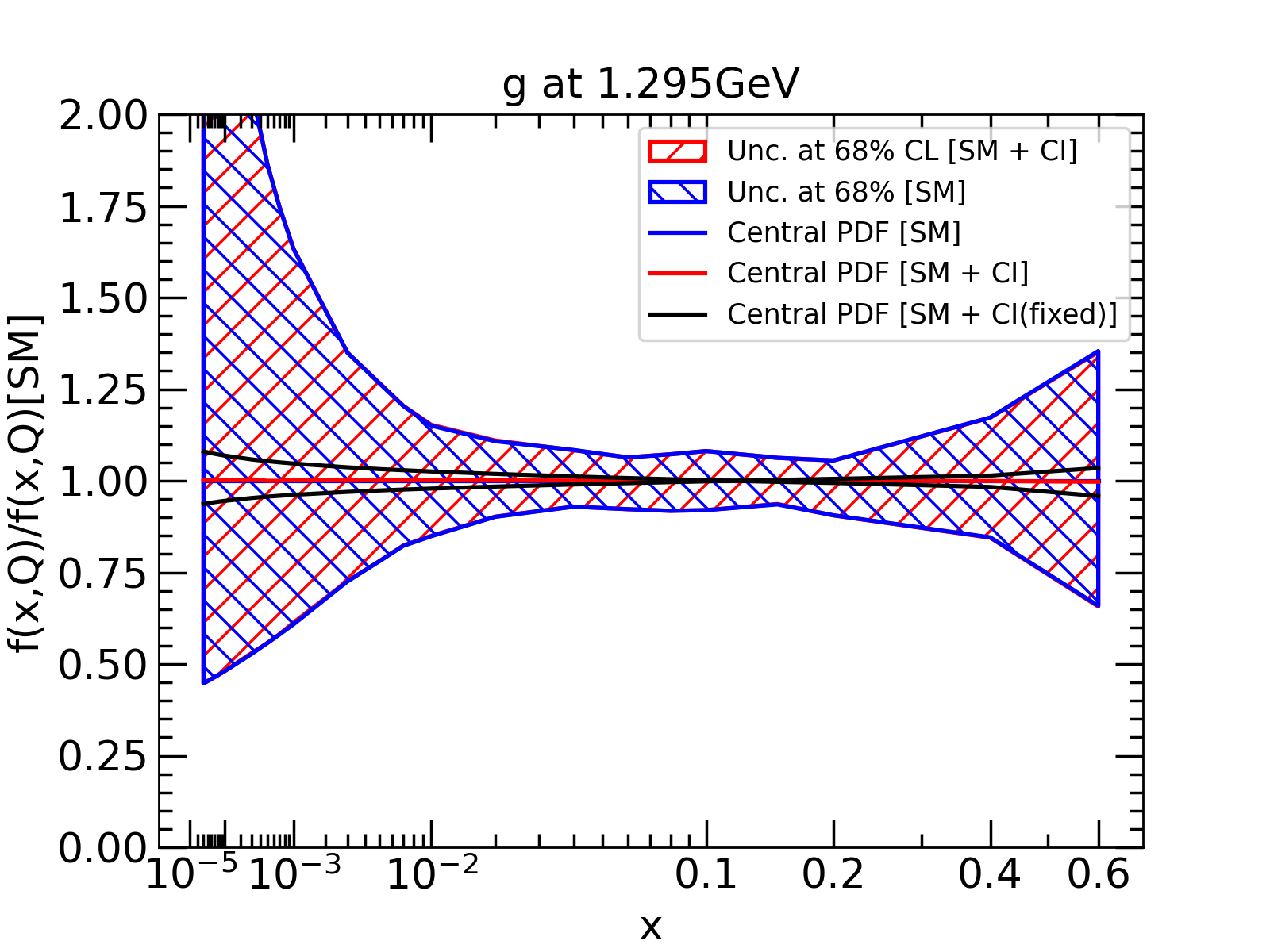}}
  \caption{The parton distribution functions at $Q = 1.295$ GeV for $u$, $d$, $\bar{u}$ and $g$. 
The blue and the red solid lines represent the central values determined by fitting without and with the new physics contributions respectively, and the blue and the red hatched areas represent the respective uncertainties at 68\% CL.
The black solid lines represent the PDFs when $\tilde{c}/\Lambda^{2}$ is fixed at -8.60 or 9.72 TeV$^{-2}$.}
\label{Fig:newp_total}
\end{figure}

In Fig.~\ref{Fig:newp_total},
we compare $u$, $d$, $\bar{u}$ and $g$ PDFs at $Q$ = 1.295 GeV determined by fitting with and without the new physics contributions.
The PDF uncertainties are shown through hatched areas with relevant colors.
The PDFs from two fits are almost indistinguishable for both central value and the uncertainties.
In addition, we find that the central value is slightly changed if the $\tilde{c}/\Lambda^2$ is fixed at -8.60 or 9.72 TeV$^{-2}$, as represented by the two black solid lines.
Specifically, in the upper-left panel, a shift as large as 2\% can be observed in the region of $x \sim 0.02$.
Similar shifts can also be observed in panel (b) and panel (c).
Besides, in the lower-right panel, a shift as large as 10\% can be observed at both small-$x$ ($\sim 10^{-4}$) and large-$x$ ($\sim 0.6$).
These shifts on PDFs are required to compensate for the contributions from the new physics on DIS cross sections.
Our approach can be extended to include more EFT operators from new physics which we leave for
future studies.

\section{Conclusion}\label{sec:conc}

Better understanding on parton distributions is essential for precision physics
at hadron colliders, as well as for study of QCD. 
Nowadays the analysis of PDFs requires calculations of the
log-likelihood functions $\chi^2$ from thousands of experimental data points, and
scans of multi-dimensional parameter space with tens of degrees of freedom.
Such analyses will benefit from development of new methods 
and improvement of computing efficiencies, for instance by various interpolation approaches.
In this paper we propose a new approach of using Neural Networks
and machine learning techniques to model the dependence of the $\chi^2$ or any
physics quantities on the PDFs.
We demonstrate the high accuracy of our approach through detailed comparisons in the PDF
parameter space of interest, taking the CT18 NNLO analysis as an example.
Importantly, compared with direct calculations the computational cost on calculating $\chi^2$
are reduced by several orders of magnitude.
The improvement ensures efficient scans of the full PDF parameter space and is desirable for
the determination of PDF uncertainties.

Based on our NNs, we perform a series of LM scans to reevaluate PDF uncertainties in the
CT18 NNLO analysis, and to understand the interplay between different data sets.
The LM method is generally more reliable through a scan of the $\chi^2$ along
the trajectory of constrained minimum of the physics quantity studied.
Our new approach renders such extensive scans almost costless and ensures the possibility of detailed
comparisons of PDF uncertainties determined from the LM and Hessian
method.
We first perform LM scans on PDF values and ratios at various $x$ and $Q$ values, and
find the results from the LM method and the Hessian method agree well in general.
However, a notable difference can be observed in the small and the large-$x$ regions.
Since the quadratic approximation fails in the region where PDF uncertainties
are large, and the results from the LM method are more reliable. 
Besides, we perform LM scans on the production cross sections of the Higgs boson pair
and the top-quark pair in association with a Higgs boson, at the LHC or future colliders,
as well as two dimensional scans on a pair of PDFs or cross sections.
Furthermore, using LM scans we study the impact of individual data sets in the
CT18 NNLO analysis by subtracting and adding back one data set at a time.
We show further applications of our approach on several extensions of the CT18 NNLO analysis.
Especially, we study the impact of the NOMAD dimuon data on constraining the strange-quark PDFs.
Theoretical predictions are calculated in the S-ACOT-$\chi$ general mass scheme up to NNLO,
based on which the NNs are constructed and LM scans are performed.
We find that the NOMAD data place stringent constraints on the strange-quark PDFs at intermediate and large-$x$ regions.
At $x \sim 0.05$, for example, the PDF uncertainties of the strange quark are reduced by almost a factor of 2.
An upward shift of more than 15\% in the strange-quark PDF as well as slight downward
shift in the $u$ and $d$-quark PDFs are also observed in most regions.
We show the interplay of the NOMAD data and other data sets in the CT18
by detailed LM scans on $R_s$ and $s$-quark PDFs at different scales and $x$ values.
The global fit with NOMAD data predicts $R_s(x=0.023,Q=1.5\rm{GeV})=0.70^{+0.40}_{-0.17}$ at 90\% CL and
a slight tension between NOMAD and NuTeV data is observed.
We also present a series of variant fits for clarifications on the impact of different theory predictions
and of different choices of the decay branching ratio of the charm quark.

Afterwards, we study the impact of two HL-LHC pseudo-data constructed in Ref.~\cite{AbdulKhalek:2018rok}, including
the high-mass Drell-Yan data and the forward W/Z production data.
We find potentially large reduction on PDF uncertainties of the sea quarks.
These results highlight the importance of HL-LHC measurements.
Besides, we performed a joint fit on both PDFs and effects of new physics beyond the SM.
We take the lepton-quark contact interactions as an example that are described by high dimensional
operators in the SMEFT.
We determine the effective Wilson coefficient to be $\tilde{c}/\Lambda^2 = 0.56^{+9.16}_{-9.16}$ TeV$^{-2}$ at 90\% CL
as mostly constrained by the HERA inclusive DIS data.
Foreseen extensions of the study would be to include more SMEFT operators in the joint fit
that is under investigation.

\begin{acknowledgments}
This work was sponsored by the National Natural
Science Foundation of China under the Grant No. 11875189 and No.11835005.
JG would like to thank members of CTEQ-TEA collaboration for helpful discussions
and proofreading of the manuscript.
\end{acknowledgments}

\begin{appendix}

\section{More on the Neural Network approach}\label{sec:para}

In this appendix we collect various details of the NN approach, including
on the architectures and parametrization dependence, the generation of training and test samples,
and the performances in terms of computational cost.
One important feature of our approach is to use directly PDF values as
inputs to the NNs rather than the PDF parameter themselves.
That ensures a great flexibility of the functional space since we can
select PDF values at an arbitrary number of $x$ points.
In our current study with CT18 parametrization form, we select the $x$
grid consisting of 14 points for each PDF flavor with their values shown
in Table~\ref{tab:x_value}.
They are selected randomly with the only criteria being distributed evenly
in $\ln x$.
We explain briefly on the mathematical model behind our NNs.
The true dependence of our target function, for instance, the $\chi^2$, on the
PDF parameters is uniquely determined by the theory and experimental data
used in the global fit, which we denote as $A_{\rm TR}$.
On another hand if we exchange the PDF parameters by the PDF values at
discrete $x$ points, the mapping is not unique since we have input PDF
values far more than the number of PDF parameters.
Thus we arrive at a bunch of possible functions $\{A^*_{\rm TR}\}$ depending explicitly
on $\{I_k\}$ which is the PDF value at the $k_{th}$ node, satisfying
\begin{equation}
    \chi^2_{truth}=A_{{\rm TR}}(a)=A^*_{{\rm TR}}(\{I_k(a)\}).
\end{equation}
The purpose of our NNs is to construct an explicit function of $\{I_k\}$
depending on a set of tuneable parameters $t_{\beta}$.
By the training procedure we update $A_{\rm NN}$ iteratively until it
converges to the neighborhood of one of the truth function $A^*_{\rm TR}$ with
a choice on the parameters $\hat{t}_{\beta}$.
Finally we arrive at our approximation to the $\chi^2$ dependence
on the PDF parameters as
\begin{equation}
    \chi^2_{pred}=A_{{\rm NN}}(\{I_k(a)\};\, \{\hat{t}_{\beta}\}).
\end{equation}
As from above one expects that the outcome NN (or equivalently the solution $\hat{t}_{\beta}$) in general depends on the parametrization form of PDFs.
However, in practice one can approximate either PDFs or cross sections in
terms of interpolated functions on a dense $x$-grid with sufficient
accuracy, as implemented successfully in APPLgrid~\cite{Carli:2010rw}, FastNLO~\cite{Kluge:2006xs}, and FastKernal~\cite{Ball:2021leu}.
In that sense there may exist an almost universal solution for different
parametrization forms if one start with a sufficiently large number of
PDF inputs.
We leave that for future investigations.

\begin{figure}[htbp]
  \centering
    \subcaptionbox{}[7.7cm] 
    {\includegraphics[width=7.7cm]{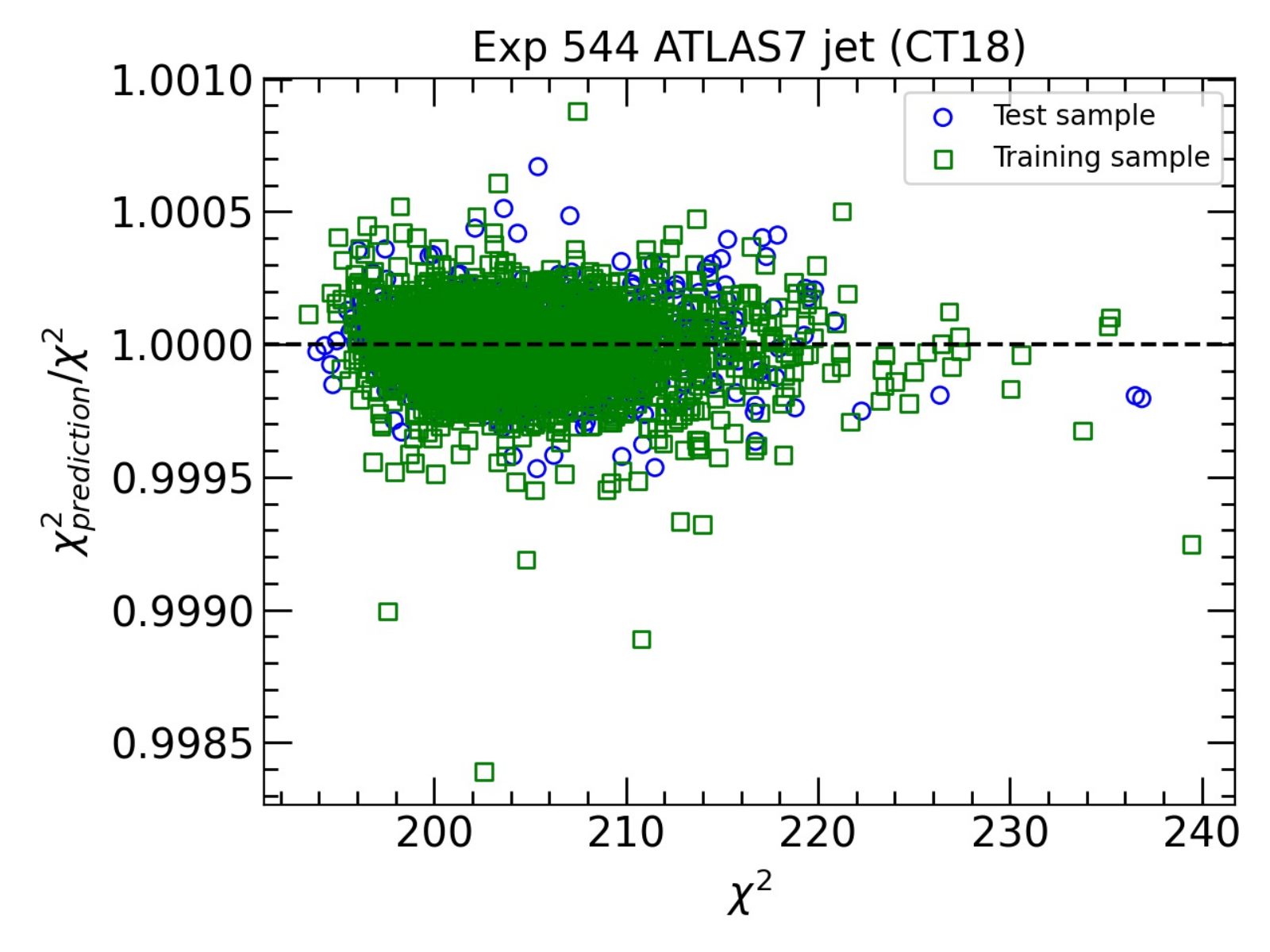}}
  \subcaptionbox{}[7.7cm] 
    {\includegraphics[width=7.7cm]{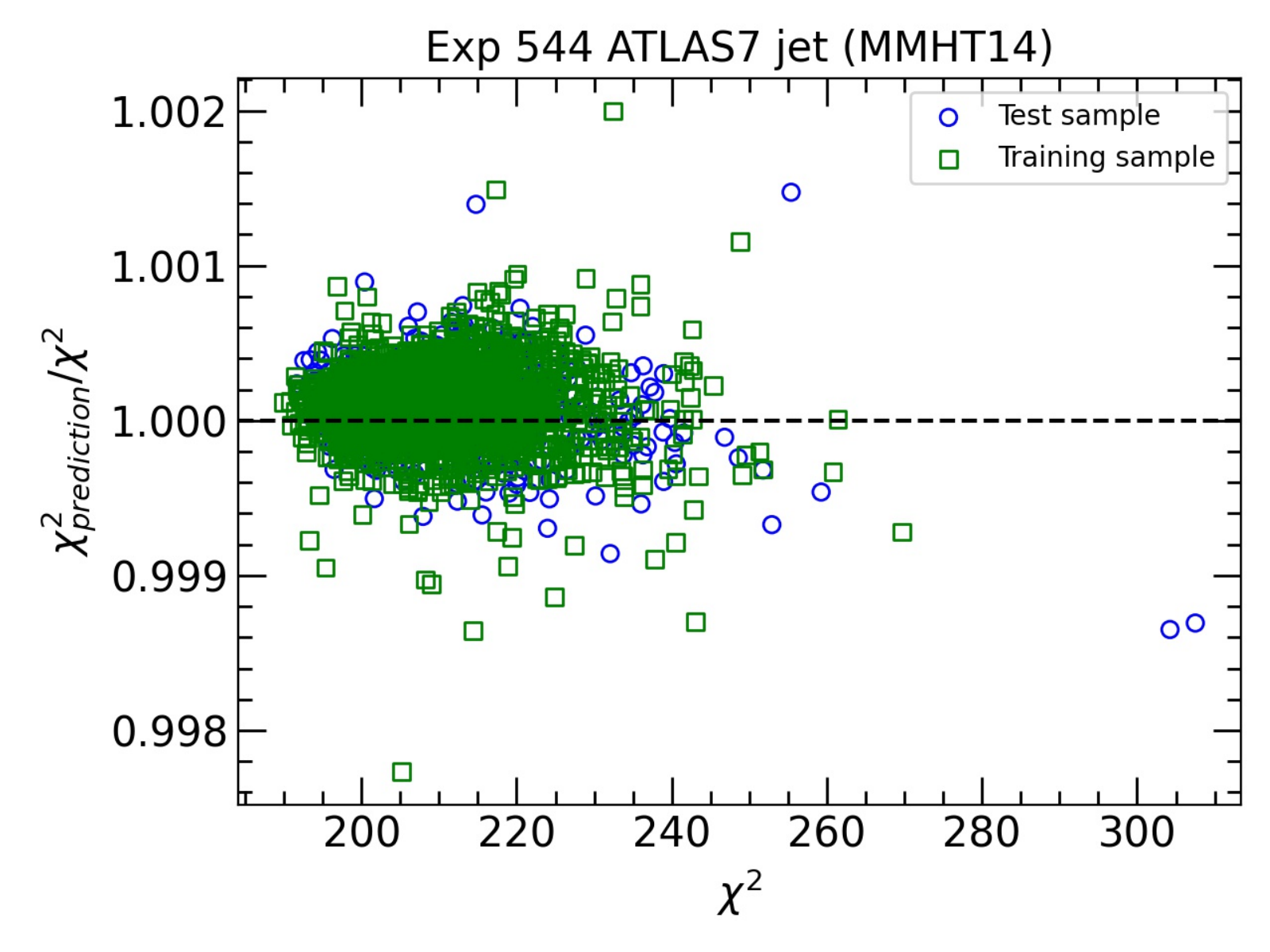}}
  \subcaptionbox{}[7.7cm]
    {\includegraphics[width=7.7cm]{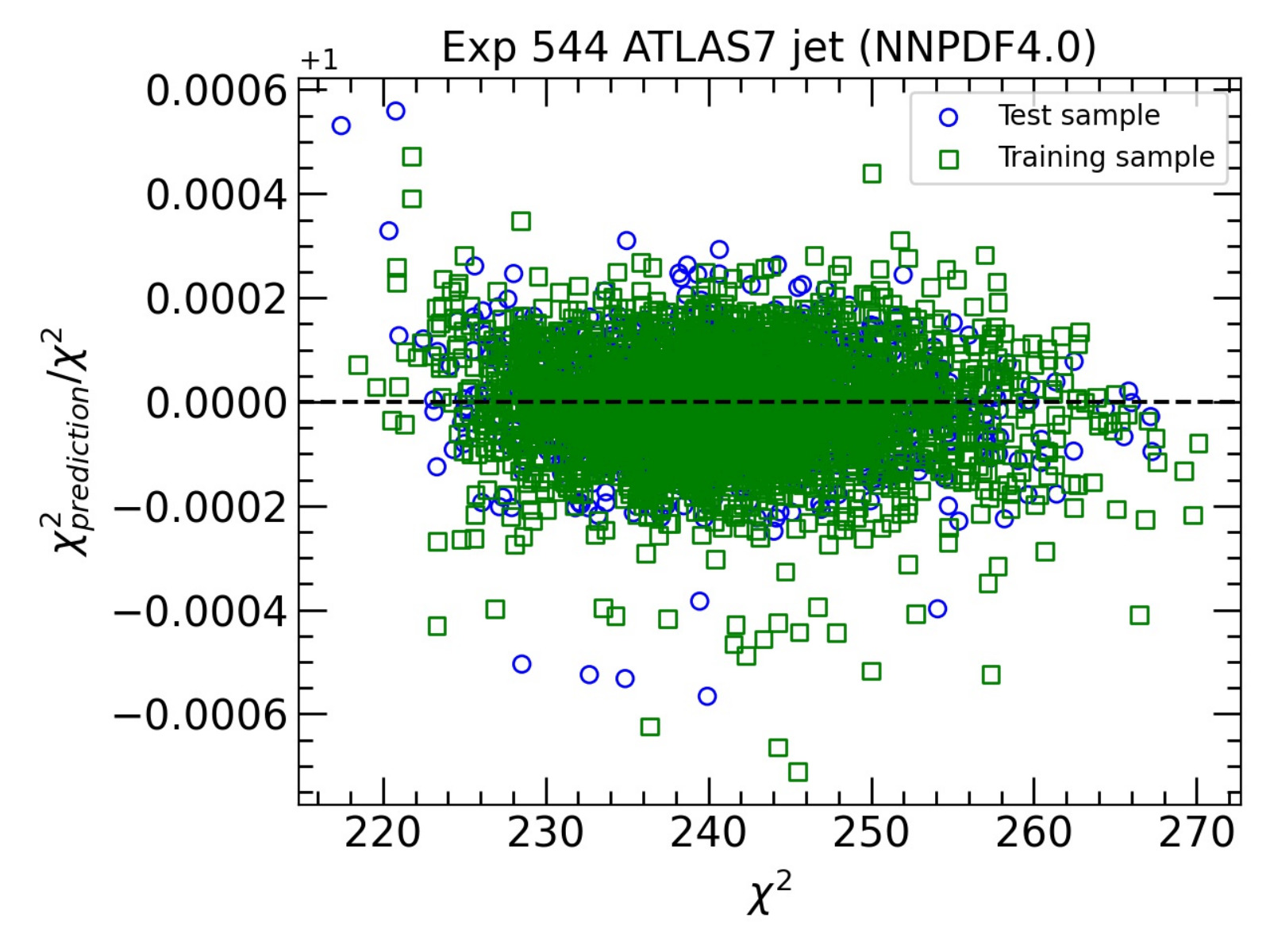}}
   \subcaptionbox{}[7.7cm]
       {\includegraphics[width=7.7cm]{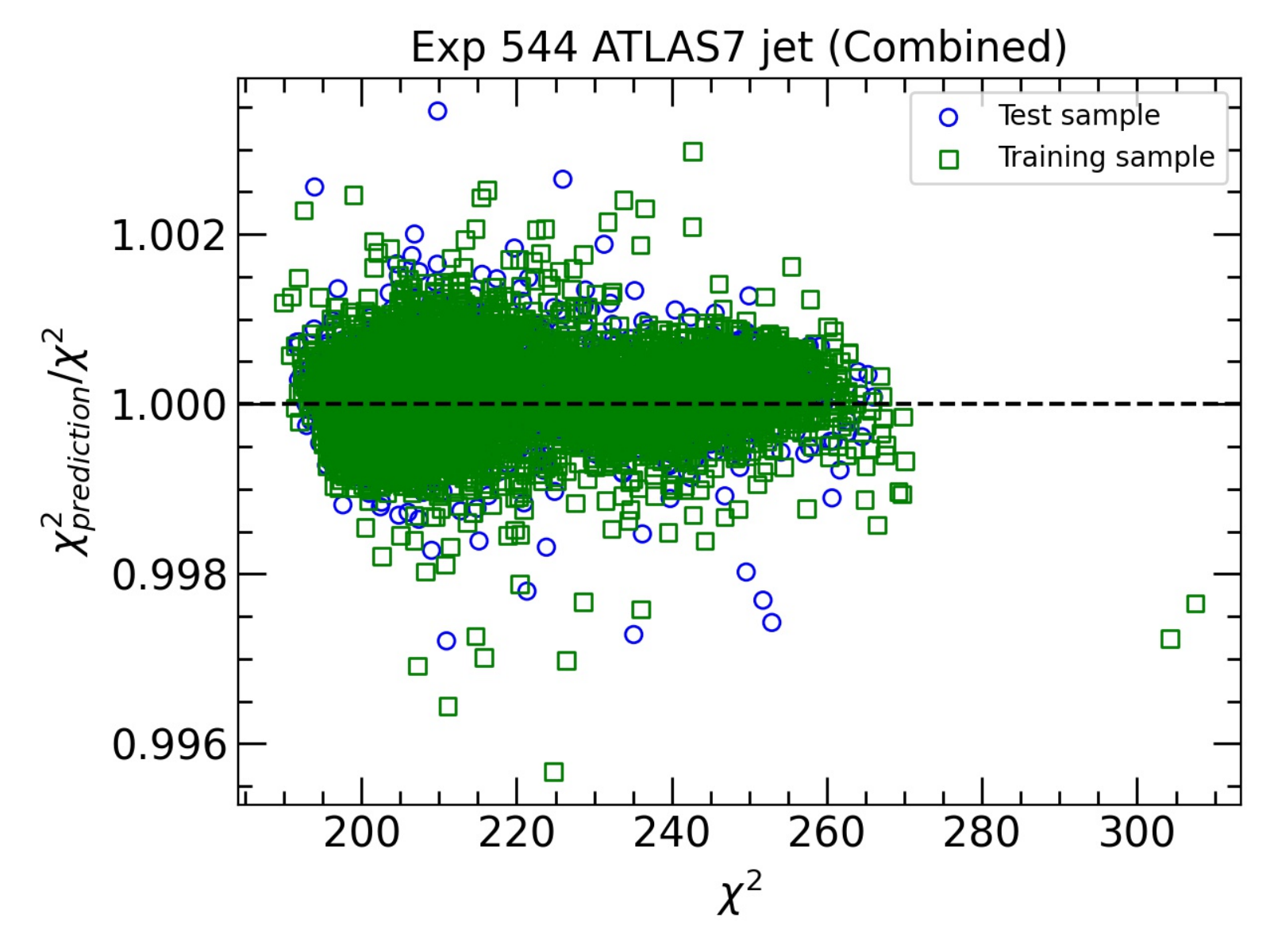}}
  \caption{The predictions to truths ratios of $\chi^2$ for
  measurement of the inclusive jet cross sections at $\sqrt{s} =$ 7 TeV by ATLAS
  when training the NNs to individual PDF parametrization forms including CT18,
  MSHT14 and NNPDF4.0, or an ensemble of PDF replicas of the three.}
  \label{Fig:paradep}
\end{figure}

\begin{table}[htpb]
  \centering
  \resizebox{\textwidth}{13mm}{
  \begin{tabular}{c|ccccccc}
  \hline
   & 1 & 2 & 3 & 4 & 5 & 6 & 7\\
  \hline
  $x$ & 3.30$\times 10^{-5}$ & 3.73$\times 10^{-4}$ & 3.50$\times 10^{-3}$ & 1.24$\times 10^{-2}$ & 2.48$\times 10^{-2}$ & 4.34$\times 10^{-2}$ & 8.59$\times 10^{-2}$ \\
  \hline
  \hline
   & 8 & 9 & 10 & 11 & 12 & 13 & 14 \\
  \hline
  $x$ & 0.118 & 0.167 & 0.206 & 0.302 & 0.406 & 0.637 & 0.831\\
  \hline
  \end{tabular}}
  \caption{The $x$ values we choose for the training and test samples of NNs.} 
  \label{tab:x_value}
\end{table}
The 8000 PDF replicas used for training and test are generated through a randomly sampling of
the PDF parameters defined in Eq.~(\ref{eq:para}) with the help of CT18 NNLO
Hessian PDF set.
Each replica or PDF parameters ${a^{rep}_i}$ is determined by 28 randomly distributed
variables ${r_j}$, namely
\begin{equation}
a_{i}^{rep}=a_{i}^{0}+\sum_{j=1}^{28}r_{j} (a_{i}^{2 j-1}- a_{i}^{2 j})/2,
\end{equation}
where \{$a_i^0$\} represent the $i_{th}$ PDF parameter of the central PDF of CT18 NNLO,
\{$a_i^{2j-1}$\} and \{$a_i^{2j}$\} represent the $i_{th}$ PDF parameter of the error PDFs
in the plus and the minus direction of the $i_{th}$ eigenvector respectively.
For each $r_j$ we use a Gaussian sampling with mean value 0 and variance $1/\sqrt{28}$
which ensures coverage of the PDF parameter space with average increase of global $\chi^2$ of a
few hundred units comparing to CT18 best fit as shown in Fig.~\ref{Fig:NN_result_chi}.
We note that performances of the trained NNs are not sensitive to the choice of training
samples as far as we are within or close to the uncertainty range of CT18.
We further test performance of our NN approach with alternative PDF parametrization
forms taking the target function of $\chi^2$ of the ATLAS 7 TeV jet data as an example.
We have chosen MMHT2014~\cite{Harland-Lang:2014zoa}, NNPDF3.1~\cite{NNPDF:2017mvq}
and NNPDF4.0~\cite{Ball:2021leu} NNLO PDFs with 4000 MC PDF replicas each
generated from the corresponding Hessian PDF sets with LHAPDF6~\cite{Buckley:2014ana}
\footnote{We have not used the native MC replicas of NNPDF since the numbers of replicas
are limited to be 1000 in that case.}.
We use the same architecture as used for CT18 except for extensions to include 9 PDF
flavors, namely with $\bar s$, $c$, and $b$-quark PDFs in addition.
The NNs have been trained and tested for each individual parametrization form with
the corresponding MC replicas.
We find very good performance of the NNs in cases of MMHT2014 and NNPDF4.0 as
shown in Fig.~\ref{Fig:paradep}, similar to the case of CT18.
Interestingly, we find performance of the same architecture is much better for the
parametrization form of NNPDF4.0 than NNPDF3.1, possibly due to the smooth conditions
applied in NNPDF4.0~\cite{Ball:2021leu}.
We also try to train the NNs with an ensemble of PDF replicas, 12000 replicas in total,
from CT18, MMHT14 and NNPDF4.0.
The accuracy of the trained NNs is only marginally worse than the NNs trained
to individual parametrization forms.
That hints the possibility of a universal
NN to accommodate for a variety of smooth PDF parametrizations, as discussed at
earlier this section.
Finally we summarize the performances of our NNs in terms of computational cost in Table~\ref{tab:time}
comparing to the traditional approaches.
Note that we have not included the time cost for the process of training of the NNs since we
do not need to repeat it in later scans of the PDF parameters.
In Table~\ref{tab:time} the numbers indicate the time cost on a single CPU-core (2.4 GHz)
of calculating the target functions for a single point in the PDF parameter space.
For $\chi^2$ the cost includes those for the calculations of the needed cross sections (taking
10 points per data set as an example) and for the multiplications with covariance matrix. 
In the conventional approach the computing efficiency varies significantly, e.g., for the $\chi^2$,
depending on the number of data points, the perturbative order of the theory calculations,
and importantly whether the fast interpolation algorithms are used or not.
Thus included numbers only represent typical average cost in the CT18 NNLO
analysis for a direct calculation or using fast interpolations (shown in parenthesis).
The fast interpolation method for calculations of a single cross section
involves more PDF values on a dense grid and thus is slower than the NN approaches.
Nevertheless, the NN approaches lead to significant improvement in general and ensure
efficient scans of the PDF parameter space with much less cost. 
The NNs was programmed with PYTHON2.7, and we expect further reduction of
the computational cost if transferred into more efficient programming languages like Fortran/C++. 
We are planning to provide open source access for the NN framework used together
with trained NNs for various target functions of CT18 in the near future.

\begin{table}[htpb]
  \centering
  \begin{tabular}{l|ccc}
  \hline
  \diagbox{method}{cost}{target} & $\chi^2$ & $\sigma$ & $f(x,Q)$ \\
  \hline
  NNs & 0.70 ms & 0.41 ms & 0.37 ms\\
  traditional & $10^7$(200) ms & $10^6$(20) ms & 20(2) ms\\
  \hline
  \end{tabular}
  \caption{Comparison on computational cost between NNs and traditional methods. Numbers in
  parenthesis represent cases if fast interpolations on PDFs are used.}
  \label{tab:time}
\end{table}

\section{Hessian PDF set}\label{sec:nnhes}
We further generate a Hessian PDF set based on the $\chi^2$ profile obtained
with the NN approaches.
The Hessian error matrix on the PDF parameters is calculated using a numeric method
of finite difference.
We use an iterative algorithm on diagonalization of the Hessian matrix that is developed
in Ref.~\cite{Pumplin:2000vx,Pumplin:2001ct} and used in later CTEQ analyses.
The iterative procedure greatly improves the performance of Hessian approximation in the
case of large number of free parameters (28 here) and in the existence of flat directions.
Once all orthogonal eigenvectors are determined, two error PDFs are generated
for each eigenvector by scanning along the plus and the minus directions and looking
for solutions with $\Delta \chi^2+P=100$ (for 90\% CL).

\begin{figure}[htbp]
  \centering
  \subcaptionbox{}[7.1cm]
    {\includegraphics[width=7.1cm]{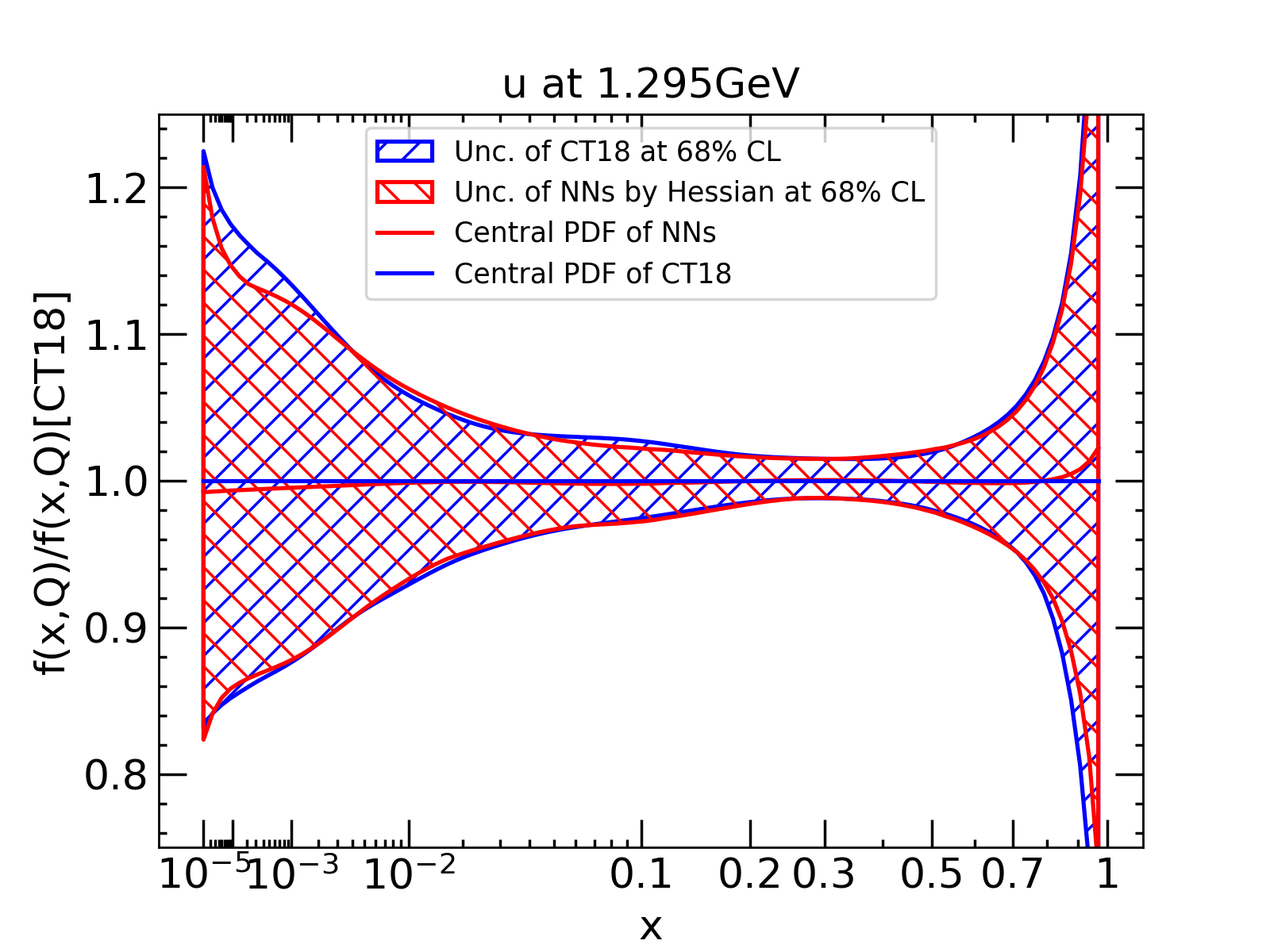}}
  \subcaptionbox{}[7.1cm]
    {\includegraphics[width=7.1cm]{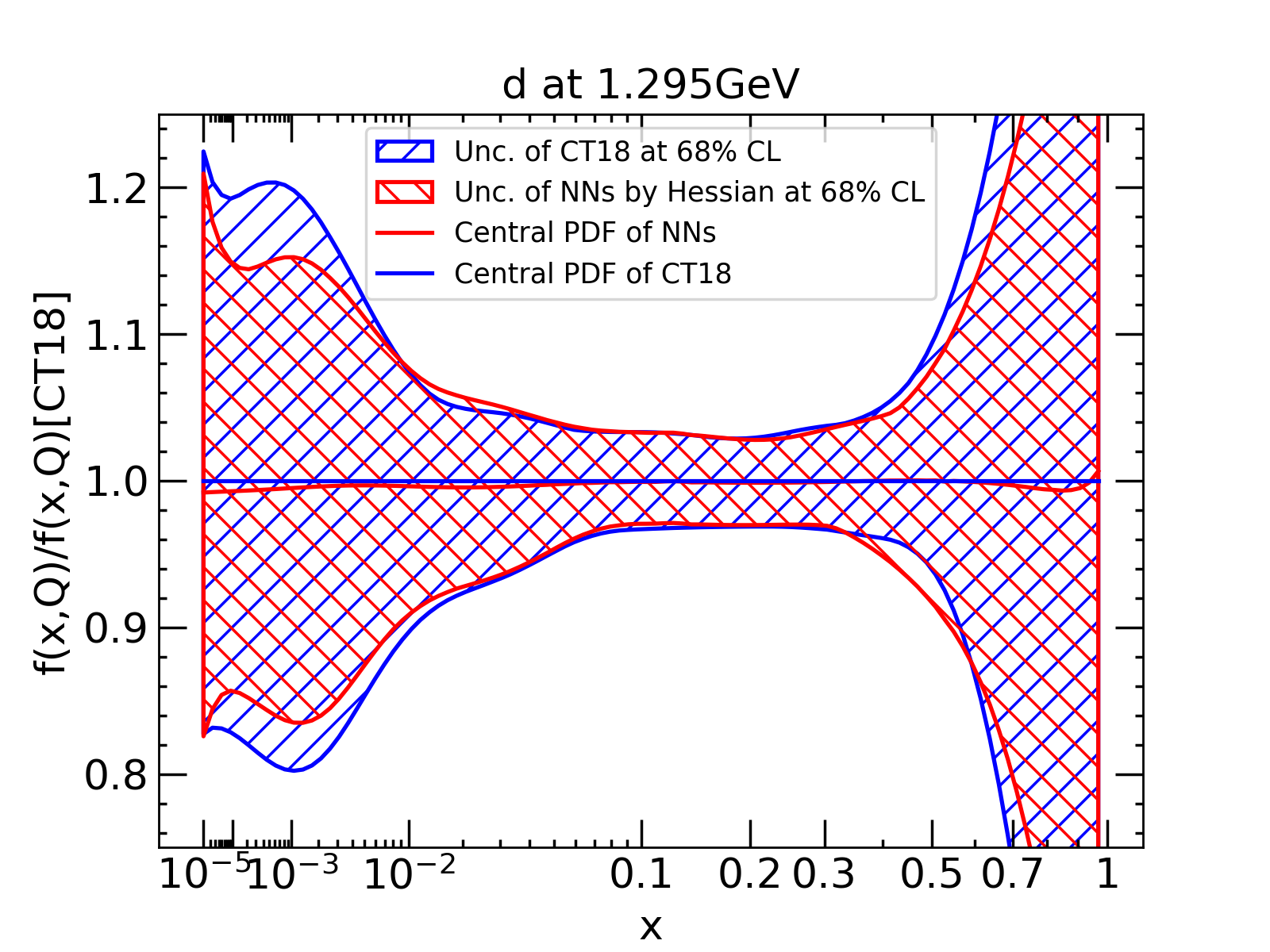}}
  \subcaptionbox{}[7.1cm]
    {\includegraphics[width=7.1cm]{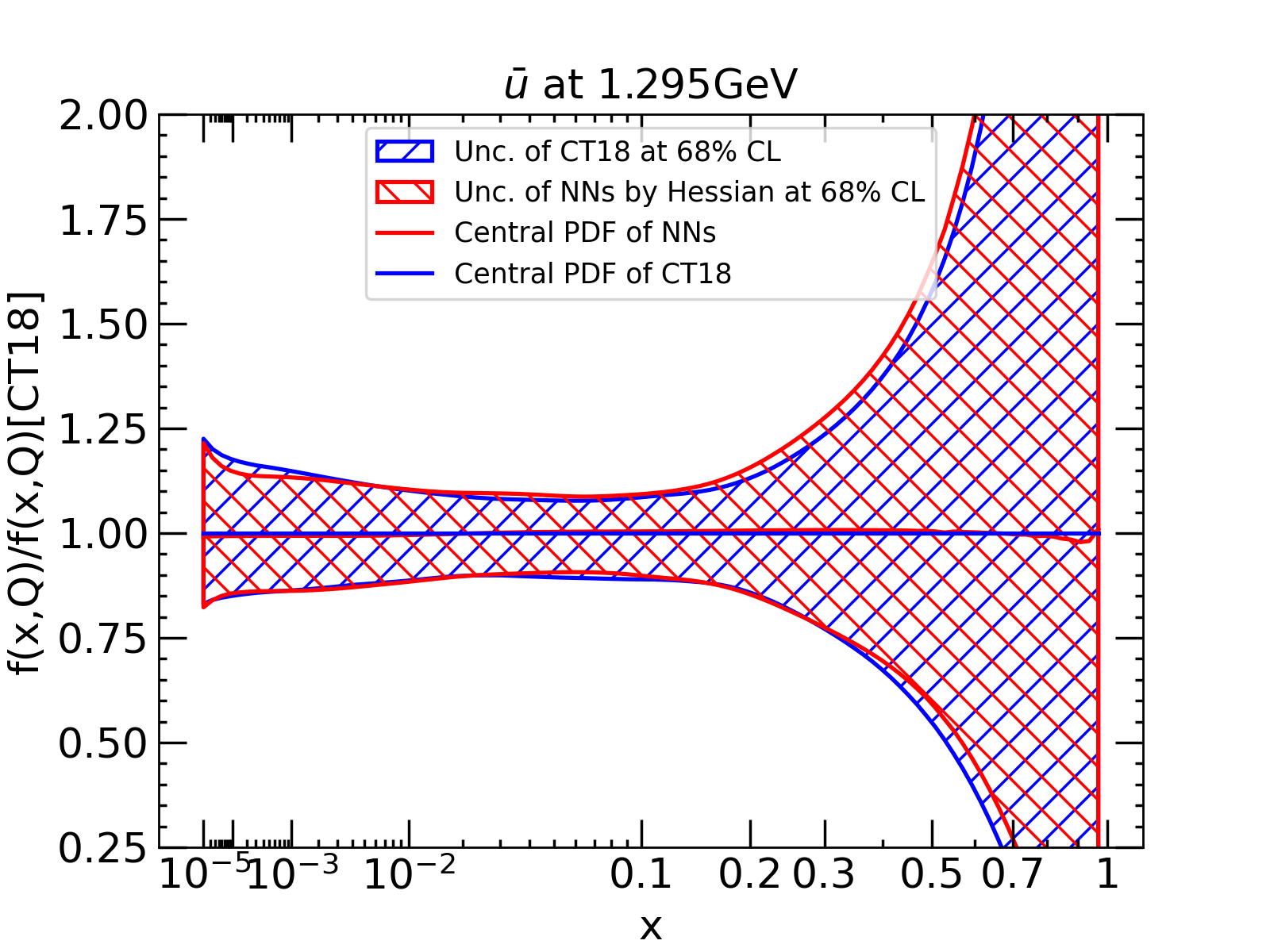}}
  \subcaptionbox{}[7.1cm]
    {\includegraphics[width=7.1cm]{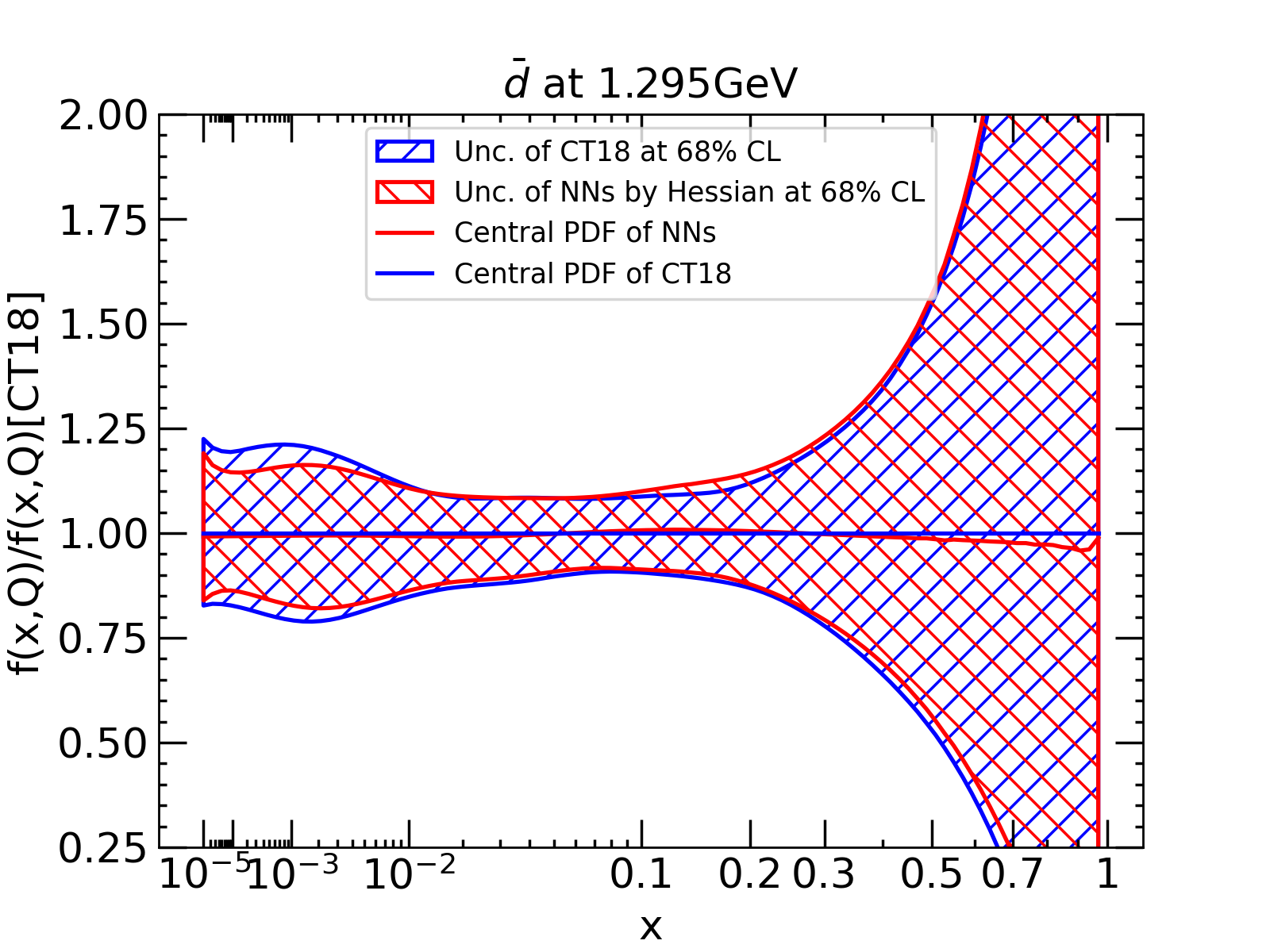}}
  \subcaptionbox{}[7.1cm]
    {\includegraphics[width=7.1cm]{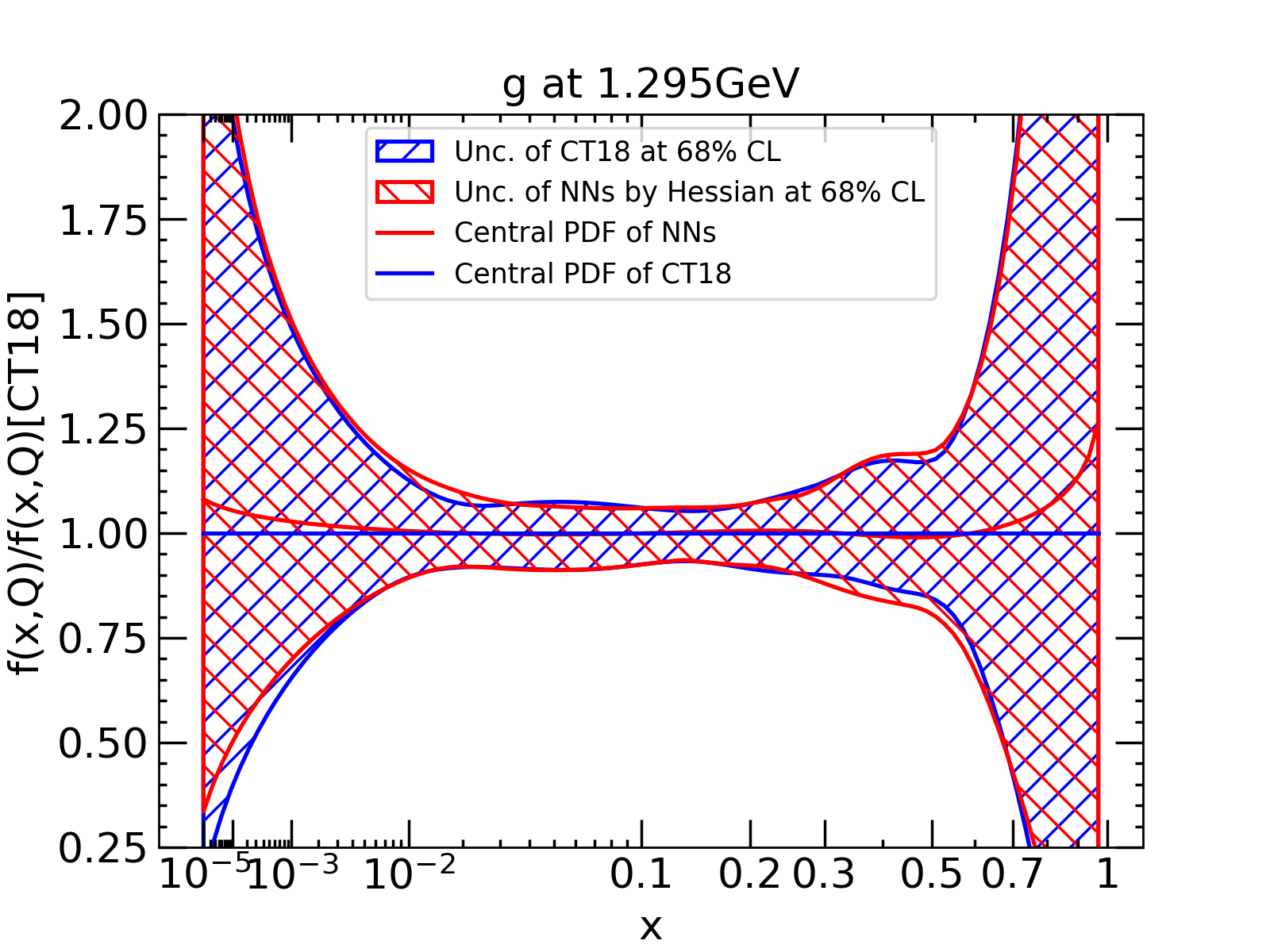}}
  \subcaptionbox{}[7.1cm]
    {\includegraphics[width=7.1cm]{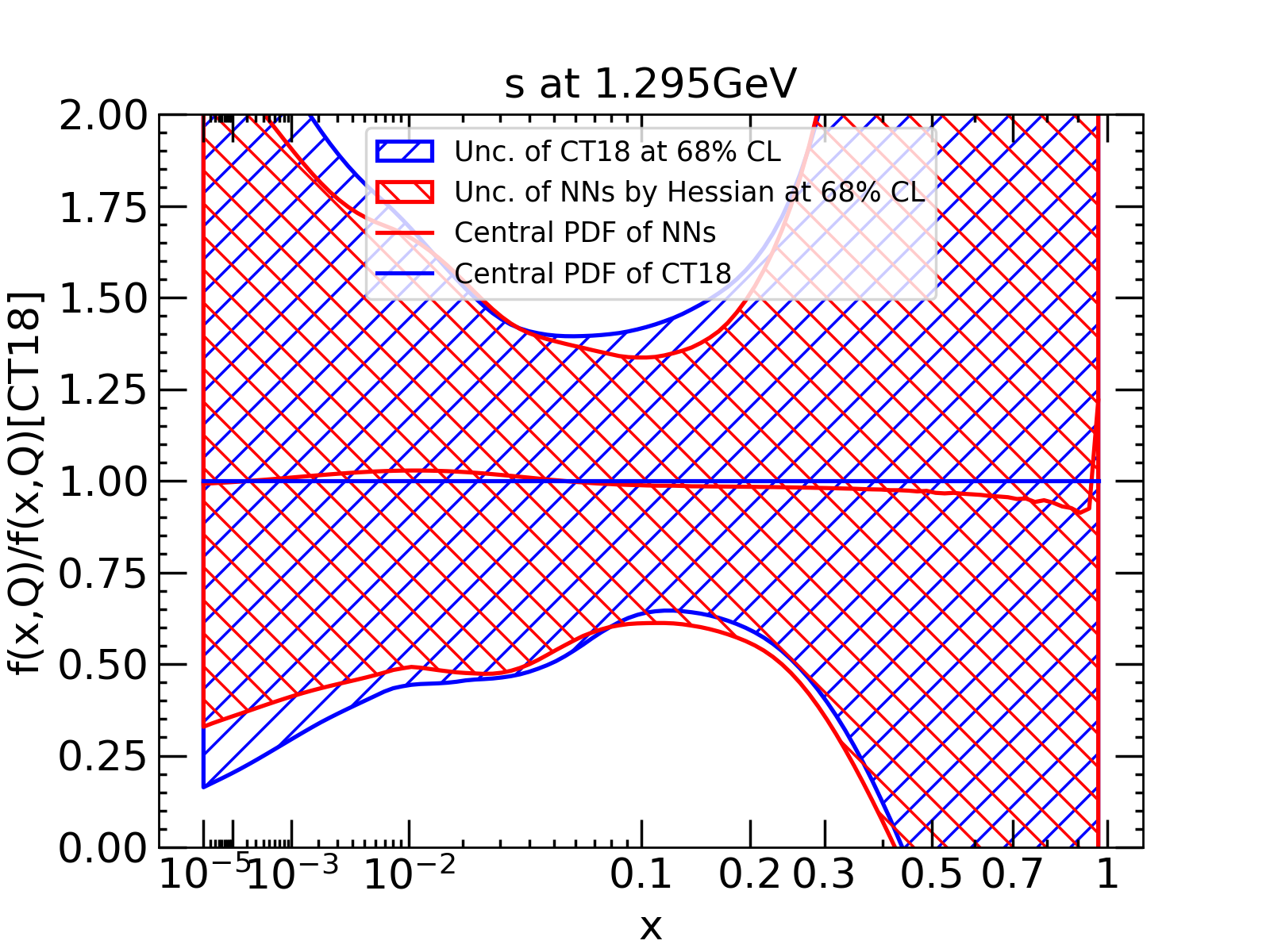}}
  \caption{The parton distribution functions at $Q = 1.295$ GeV for $u$, $d$, $\bar{u}$, $\bar{d}$, $g$ and $s$. 
The blue and the red solid lines represent the central values of CT18NNLO and NNs respectively. 
The blue and the red hatched areas represent the uncertainties of CT18NNLO and final PDFs in this paper at 68\% CL respectively.}
\label{Fig:hessian_total}
\end{figure}

We compare the PDF uncertainties at 68\% CL from the Hessian PDF set to the
published CT18 NNLO PDFs in Fig.~\ref{Fig:hessian_total}.
We find very good agreements between predictions of the two Hessian PDF sets in general.
However, some notable differences can be seen for $d$-valence and gluon PDFs at
large-$x$ ($\sim 0.4$) as well as for sea quarks at $x\lesssim 10^{-3}$.
There are two reasons that lead to the differences in the new Hessian set and the CT18
set.
First as mentioned earlier in the global fit presented in this paper the
NNLO K-factors used for predictions of the Drell-Yan data have been updated comparing
to those used in the CT18 analysis.
Besides, when calculating the Hessian error matrix numerically we use a step size
of $\Delta \chi^2=10$ on sampling of the PDF parameters while a value of $\sim 1$
is used in the CT18 analysis.
The dependence on choices of this step size reflect one intrinsic uncertainty
of the Hessian approaches~\cite{Pumplin:2000vx}.

\section{Variant fits with NOMAD data}\label{sec:vano}

In this appendix we present a series of global fit with inclusion of the NOMAD data
and with different theories or different choices of decay branching ratios of charm
quark to muon.
In Sect.~\ref{subsec:nomad} when comparing to the global fit without NOMAD data, we use
NLO cross sections (but with NNLO PDFs) for NuTeV and CCFR dimuon data to be consistent
with the CT18 analysis.
Thus the changes observed after including the NOMAD data can be due to both the 
NOMAD data or the changes of theories for the other two dimuon data.
Now we further consider different choices of the theory predictions, namely either
calculated at NLO or NNLO in QCD, to disentangle their effects.
Furthermore, to compare with dimuon data, one has to convert the cross sections
of charm-quark production to production of dimuon which relies on the input of
inclusive semileptonic branching ratio Br($c\to \mu$).
Since NOMAD dimuon data extend down to $E_{\nu} \sim$ 6 GeV the energy dependence of Br($c\to \mu$)
is taken into account in NOMAD analysis, with a parametrization form
\begin{equation}
Br(c\to \mu) = \dfrac{a}{1+b/E_{\nu}},
\end{equation}
where $a$ and $b$ are free parameters.
In the NOMAD paper it suggests values of $a = 0.094\pm 0.010$ and $b = 6.6 \pm 3.9$ GeV as
measured by the E531 experiment~\cite{FermilabE531:1988hzw}.
The uncertainties on parameters $a$ and $b$ will propagate into the unfolded
charm-quark cross sections and are treated as additional correlated systematic errors that
are summarized in Table~\ref{tab:nomadsys} for distribution in Bjorken-$x$.
Other correlated systematic uncertainties for NOMAD data can be found in Ref.~\cite{NOMAD:2013hbk}.  
On the other hand, for NuTeV and CCFR data, since the neutrino energies are sufficiently high,
a constant value of Br($c\to \mu$)=$0.099\pm 0.010$ has been suggested~\cite{osti_879078} and is used in
the CT18 analysis.
The central value is slightly higher than the parameter $a$ used in our nominal fit of NOMAD data.
Thus we perform variant fits using $a=0.099\pm 0.010$ for NOMAD data to further investigate the impact
of this overall normalization on the outcome PDFs.

\begin{figure}[htbp]
  \centering
  \includegraphics[width=0.8\textwidth,clip]{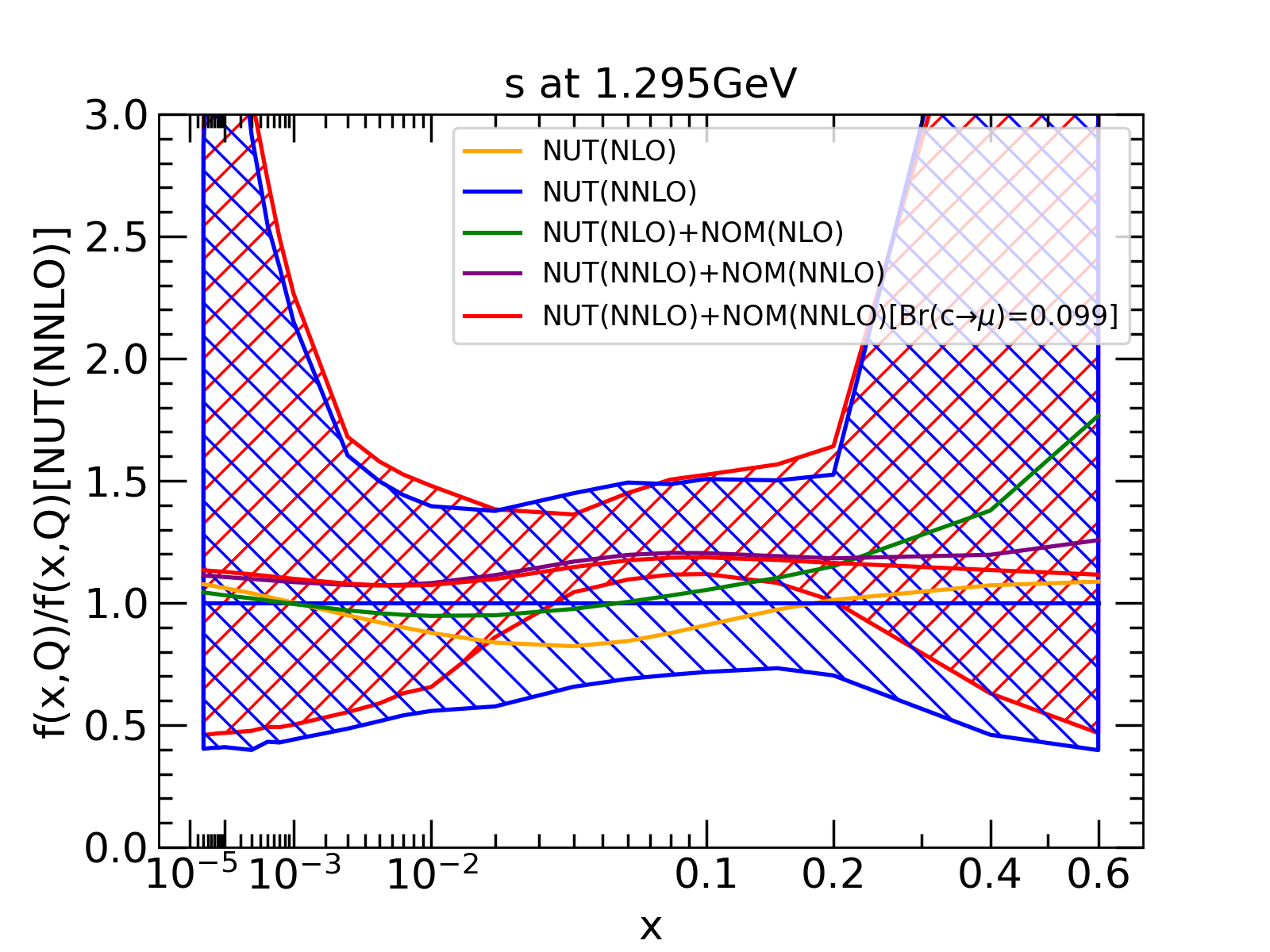}
  \hfill
  \caption{Strange-quark PDF at $Q = 1.295$ GeV from LM scans in global fits with various conditions.
  The PDF uncertainties are shown for 68\% CL.}
  \label{Fig:nomad_appendix}
\end{figure}

In Fig.~\ref{Fig:nomad_appendix} we compare the strange-quark PDFs at 1.295 GeV from all variant fits.
We show the PDF uncertainties at 68\% CL from LM scans
for fits with and without NOMAD data and using NNLO predictions from dimuon production consistently.
That can be compared with Fig.~\ref{Fig:LM_total_xnnlo} where NLO predictions are used
in fit without NOMAD data.
We also present central PDFs obtained with NLO predictions or with higher branching ratio for
NOMAD data.
We find including the NNLO corrections leads to a moderate increase of the strange-quark PDF, which is
in consistent with the conclusions in Ref.~\cite{Hou:2019efy}.
The inclusion of NOMAD data results in about 20\% enhancement of the strange-quark PDF at
$x$ around 0.05 and a significant reduction of the PDF uncertainties.
Changing to $a=0.099\pm 0.010$ for NOMAD only induces a minor reduction of the strange-quark PDF. 
We further summarize the total or individual $\chi^2$ of all variant fits together with
predictions on $R_s (x = 0.023, Q = 1.5 {\rm GeV})$ with uncertainties at 68\% CL in
Table~\ref{tab:nomad_compare_chi2}.
By comparison with the $\chi^2$ we find the NNLO predictions in general lead to a slightly worse fit
with increase on $\chi^2$ of a few units.
However, in all cases the global fit can describe well various dimuon data as can be
seen from the $\chi^2$ per number of degree of freedoms.
When comparing fits in the last two rows we find using a consistent branching ratio in
different data sets results in a better fit and reduced PDF uncertainties.
\begin{table}[h]
\begin{center}
\small
\begin{tabular}{||c|c|c|c|c||}
\hline
%----------------------------------------------------------------------------------------------------------
       $x_{\rm Bj}$      & Bin center &  $\sigma_{\mu\mu}/\sigma_{cc} \pm \delta^{stat} \pm \delta^{syst}$ ($10^{-3}$) & $\delta^{a}$, \% & $\delta^{b}$, \% \\
%----------------------------------------------------------------------------------------------------------
\hline
    0.0000 -   0.0336 &    0.0267 & 13.383 $\pm$ 0.441 $\pm$ 0.289             &   10.6    &  5.3    \\
    0.0336 -   0.0511 &    0.0440 & 11.245 $\pm$ 0.380 $\pm$ 0.210             &   10.6    &  6.8    \\
    0.0511 -   0.0672 &    0.0598 &  ~9.991 $\pm$ 0.347 $\pm$ 0.201             &   10.6    &   7.7    \\
    0.0672 -   0.0836 &    0.0756 &  ~9.141 $\pm$ 0.324 $\pm$ 0.189             &   10.6    &   8.3    \\
    0.0836 -   0.1000  &    0.0917 &  ~8.198 $\pm$ 0.297 $\pm$ 0.169             &   10.6    &   8.8    \\
    0.1000  -   0.1246  &    0.1122  &  ~7.176 $\pm$ 0.225 $\pm$ 0.144             &   10.6    &  9.0    \\
    0.1246  -   0.1535  &    0.1389  &  ~6.229 $\pm$ 0.195 $\pm$ 0.118             &   10.6    &  9.4    \\
    0.1535  -   0.1870  &    0.1699  &  ~5.427 $\pm$ 0.171 $\pm$ 0.106             &   10.6    &  9.6    \\
    0.1870  -   0.2277  &    0.2066  &  ~4.837 $\pm$ 0.151 $\pm$ 0.093             &   10.6    &  9.9    \\
    0.2277  -   0.2800  &    0.2524  &  ~4.235 $\pm$ 0.133 $\pm$ 0.083             &   10.6    &  10.0    \\
    0.2800  -   0.3590  &    0.3165  &  ~3.595 $\pm$ 0.113 $\pm$ 0.072             &   10.6    &  10.0    \\
    0.3590  -   0.4583  &    0.4036  &  ~2.955 $\pm$ 0.111 $\pm$ 0.062             &   10.6    &  10.1    \\
    0.4583  -   0.5838  &    0.5116  &  ~2.355 $\pm$ 0.120 $\pm$ 0.055             &   10.6    &  9.9    \\
    0.5838  -   0.7500  &    0.6465  &  ~1.607 $\pm$ 0.150 $\pm$ 0.047             &   10.6    &  9.4    \\
%----------------------------------------------------------------------------------------------------------
\hline
\end{tabular}
\normalsize
\caption {NOMAD measurements on the Bjorken-$x$ distribution of dimuon to inclusive CC cross section ratio,
including the binning, central values, statistical and total systematic uncertainties. 
The last two columns show the additional correlated systematic uncertainties in percentages, if converting back to
production cross sections of charm-quark, due to input parameter $a$ and $b$ respectively.
These additional errors are derived based on theoretical cross sections at NLO with CT18 NNLO PDFs.
}   
\label{tab:nomadsys}
\end{center}
\end{table}

\begin{table}[h]
\begin{center}
\small
\begin{tabular}{||c|c|c|c|c|c|c|c||}
\hline
%----------------------------------------------------------------------------------------------------------
data sets & \makecell*[c]{$\chi^2_{total}$\\(3671/3683)} & \makecell*[c]{$\chi^2_{nomad}$\\(12)} & \makecell*[c]{$\chi^2_{124}$\\(38)} & \makecell*[c]{$\chi^2_{125}$\\(33)}  & \makecell*[c]{$\chi^2_{126}$\\(40)} & \makecell*[c]{$\chi^2_{127}$\\(38)} & $R_s$(0.023,1.5GeV) \\
%----------------------------------------------------------------------------------------------------------
\hline
NUT (NLO) & 4272.64 & - & 18.83 & 39.55 & 29.89 & 19.42 & $\mathrm{0.518^{+0.349}_{-0.363}}$\\
\hline
NUT (NNLO) & 4268.77 & - & 21.44 & 32.84 & 34.06 & 22.54 & $\mathrm{0.616^{+0.441}_{-0.377}}$\\
\hline
\makecell*[c]{NUT (NLO) +\\NOM (NLO)} & 4286.28 & 8.39 & 24.81 & 41.95 & 29.30 & 18.79 & $\mathrm{0.593^{+0.256}_{-0.155}}$\\
\hline
\makecell*[c]{NUT (NNLO) +\\NOM (NNLO)}& 4291.47 & 14.47 & 28.13 & 34.26 & 34.21 & 22.15 & $\mathrm{0.695^{+0.384}_{-0.169}}$ \\
\hline
\makecell*[c]Br(c$\to \mu$)=0.099& 4289.61 & 13.79 & 27.38 & 34.06 & 34.10 & 22.13 & $\mathrm{0.685^{+0.290}_{-0.174}}$\\
\hline
%----------------------------------------------------------------------------------------------------------
\hline
\end{tabular}
\normalsize
\caption {Total $\chi^2$ and individual $\chi^2$ of dimuon data for global fits with various conditions.
Numbers in the first row indicate the total number of data points. The last column includes predictions
on $R_s (x = 0.023, Q = 1.5 {\rm GeV})$ with uncertainties at 90\% CL.}   
\label{tab:nomad_compare_chi2}
\end{center}
\end{table}

\end{appendix}
\bibliography{pdfnn}
\bibliographystyle{jhep}

\end{document}